\documentclass[a4paper,
portrait,%
11pt]{book}

\usepackage[english]{minitoc,varioref}
\usepackage{makeidx}
\usepackage{hyperref}
\usepackage{mdframed}
\usepackage{xcolor}
\usepackage{tikz}
\usetikzlibrary{positioning}
\tikzset{>=stealth}
\usepackage{pstricks}
\usepackage[paperwidth=9.5in, paperheight=14in, bindingoffset=0.1in,
marginpar=4cm,outer=2in,marginparsep=0.3in]{geometry}
\usepackage[Bjornstrup]{fncychap}

\usepackage[most]{tcolorbox}

\tcbset{
    frame code={}
    center title,
    left=10pt,
    right=10pt,
    top=10pt,
    bottom=10pt,
    colback=gray!40,
    width=\dimexpr\textwidth\relax,
    enlarge left by=0mm,
    boxsep=5pt
    }

\usepackage[utf8]{inputenc} 
\usepackage[T1]{fontenc} 

\nomtcrule

\usepackage{graphicx}
\usepackage{amsfonts}
\usepackage{amssymb}
\usepackage{bbold}
\usepackage{slashed} 
\usepackage{color}
\usepackage{hyperref}
\usepackage{wrapfig}
\usepackage{booktabs}
\usepackage{tabularx}
\usepackage{amsmath}
\usepackage{mathtools}
\def\eq#1{{Eq.~(\ref{#1})}}

\def\Re{\mbox{Re}\,}

\def\ltap{\ \raisebox{-.4ex}{\rlap{$\sim$}} \raisebox{.4ex}{$<$}\ }
\def\alt{\ \raisebox{-.4ex}{\rlap{$\sim$}} \raisebox{.4ex}{$<$}\ }
\def\gtap{\ \raisebox{-.4ex}{\rlap{$\sim$}} \raisebox{.4ex}{$>$}\ }
\def\agt{\ \raisebox{-.4ex}{\rlap{$\sim$}} \raisebox{.4ex}{$>$}\ }
\definecolor{titlepagecolor}{cmyk}{1,.60,0,.40}
\definecolor{oucrimsonred}{rgb}{0.6, 0.0, 0.0}
\definecolor{persianblue}{rgb}{0.11, 0.22, 0.73}
\definecolor{forestgreen}{rgb}{0.13,0.35,0.13}
 \hypersetup{colorlinks, citecolor=forestgreen, linkcolor=persianblue, urlcolor=oucrimsonred}
\newcommand{\be}{\begin{equation}}
\newcommand{\ee}{\end{equation}}
\newcommand{\bea}{\begin{eqnarray}}
\newcommand{\eea}{\end{eqnarray}}
\newcommand{\nn}{\nonumber}

\newcommand{\com}[1]{}
\newcommand{\ii}[1]{\mathrm{i}}
\newcommand{\uu}{\scriptscriptstyle U}
\newcommand{\dd}{\scriptscriptstyle D}

\newcommand{\D}{\scriptscriptstyle D}
\newcommand{\BR}{\mbox{BR\,}}
\newcommand{\bc}{\begin{center}}
\newcommand{\ec}{\end{center}}
\font\beeg=cmr17 scaled 1800
\newbox\ibox
\def\versal#1{\setbox\ibox=\hbox{{\beeg #1}~}%
	    \noindent\global\hangindent=\wd\ibox\global\hangafter-2%
	    \sc\smash{\llap {\lower 14pt \box\ibox}}}
\makeindex
\begin{document}
%
%
%
\begin{titlepage} 
	\rule{1pt}{\textheight} 
	\hspace{0.05\textwidth} 
	\parbox[b]{0.8\textwidth}{ 

\begin{tcolorbox}
 {\Huge\bfseries The Dark Photon }
\end{tcolorbox}
\vskip 7.0cm
\vskip 1.5cm
Marco Fabbrichesi\\
INFN, Sezione di Trieste, \\
Via  A. Valerio 2, 34127 Trieste, Italy\\
email: \texttt{marco.fabbrichesi@ts.infn.it}\\
\vskip 1cm
Emidio Gabrielli\\
Dipartimento di Fisica Teorica, Universit\`a di Trieste, \\
Strada Costiera 11, 34151 Trieste, Italy\\
and NICPB, R\"avala 10, 10143 Tallinn, Estonia\\
email: \texttt{emidio.gabrielli@cern.ch}\\
\vskip 1cm
Gaia Lanfranchi\\
INFN, Laboratori Nazionali di Frascati, \\
Via E. Fermi 40, 00044 Frascati, Roma,  Italy\\
email:\texttt{gaia.lanfranchi@lnf.infn.it}
\vskip 3.0cm	
}

\end{titlepage}

\newpage
\vspace*{5cm}
\begin{flushleft}
{\Large \bfseries Abstract}
\end{flushleft}	
\vspace*{0.2cm} 

\begin{tcolorbox}

 \noindent
 {\versal  The dark photon is a new  gauge boson} whose existence has been conjectured. It is dark because it arises from  a symmetry of  a hypothetical dark sector comprising particles completely neutral under the Standard Model interactions. 
Dark though it is, this new gauge boson can be detected 
because of its kinetic mixing with the ordinary, visible photon.  
We review its physics  from  the theoretical and the experimental point of view. We discuss the difference between the massive and the massless case. 
We explain how the   dark photon  enters laboratory, astrophysical and cosmological observations as well as  dark matter physics. We survey the  current and future experimental limits on the parameters of the massless and massive dark photons   together with  the related bounds on milli-charged  fermions.
\end{tcolorbox}

\newpage

%
\tableofcontents
%
%
\chapter{Introduction}
\label{sec:intro}

\vspace*{1cm}
{\versal New particles beyond the Standard Model}  have always been thought   to be charged under  at least some of the same gauge interactions of  ordinary  particles. Although this assumption has driven the theoretical speculations as well as the  experimental searches of the last 50 years, it has also been  increasingly challenged by  the  negative results of all these searches---and the mounting frustration for the  failure to  discover   any of these hypothetical new particles. 

As the  hope of a breakthrough along these lines is waning, interest in a \textit{dark sector}---dark because not charged under the Standard Model gauge groups---is growing: Maybe no new particles have been seen simply because they do not interact  through the Standard Model gauge interactions.

The dark sector is assumed to exist as a world parallel to our own. It may contain few or many states, and these  can be fermions or scalars or both, depending on the model. Dark matter proper---the existence of which is deemed necessary to explain astrophysical data---is found among these states.  Its   relic density can be computed and constrained by observational data. In addition, the dark states can interact; their interactions can be Yukawa-like or mediated by dark gauge bosons or both depending on the model.  

If the dark and the visible sectors were to interact only gravitationally---which they cannot avoid---there would be little hope of  observing   in the laboratory  particles belonging to the dark sector. A similar problem exists for dark matter: Although its presence is motivated by gravitational physics, it is searched mostly through  its putative weak interactions---as in the direct- and indirect-detection searches of a weakly interacting massive particle. For the same reason, we must pin our hopes on assuming 
that dark and ordinary sectors  also  interact  through a \textit{portal}---as the current terminology has it---that is, through a sallow glimmer, in a manner that, though feeble, is (at least in principle) experimentally accessible. 

The portal may take various forms that can be classified by the type and dimension of its operators. The best motivated and most studied cases contain relevant operators taking different forms depending on the spin of the mediator: \textit{Vector} (spin 1), \textit{Neutrino} (spin 1/2),  \textit{Higgs} (scalar) and \textit{Axion} (pseudo-scalar).

Among these possible portals,  the vector portal  is the one where the interaction takes place because of the kinetic mixing between one dark and  one visible Abelian gauge boson (nonAbelian gauge bosons do not mix). The visible photon is  taken to be the  boson of the $U(1)$ gauge group  of electromagnetism---or, above the electroweak symmetry-breaking scale, of the hyper-charge---while the \textit{dark photon} comes to be identified as the boson of an extra $U(1)$ symmetry.

The names \textit{para}-~\cite{Holdom:1985ag}, \textit{hidden-sector}, \textit{secluded photon} and \textit{U-boson}~\cite{Fayet:1990wx} have also being used to indicate the same particle.  
The  idea of adding to the SM a new gauge boson similar to the photon was first considered in the context of supersymmetric theories in  \cite{Fayet:1980ad,Fayet:1980rr}. It was discussed more in general shortly after in~\cite{Okun:1982xi,Georgi:1983sy}.

Dark though it is, the dark photon  can be detected 
because of its kinetic mixing with the ordinary, visible photon. This kinetic mixing is always possible because the field strengths of two Abelian gauge fields can be multiplied together to give a dimension four operator. The existence of such an operator means that the two gauge bosons can go into each other as they propagate. This kinetic mixing provides  the portal linking the dark and visible sectors. It is this portal that makes possible to  detect the dark photon in the experiments. 

The  concept of a portal---which   at first blush might seem rather harmless---actually   represents  a radical departure from
what is  the main conceptual outcome of our study of particle physics, namely, the  gauge principle and
 idea that all interactions must be described by a gauge theory. The portal, and the new interactions that  it brings into the picture, adds   a significant  exception to this principle. Among the possible portals,   the vector case deviates  the least  from the gauge principle as it only introduces a mixing for the gauge bosons while the interaction to matter remains of the  gauge type (albeit with an un-quantized charge). Instead, the other kinds of portal imply a manifest new violation to the gauge principle, the other's being  the notable case of the Yukawa and self-interactions of the Higgs boson---which are themselves, exactly because of their not being gauge interactions,  the least understood 
 part of the Standard Model. 

There is an additional and  important reason to study the dark sector in general, and the dark photon in particular: The main motivation in introducing new-physics scenarios is to use them as a foil for the Standard Model in mapping possible experimental discrepancies. In the absence of clearly identified new states,  the many parameters, for instance, of the supersymmetric extensions to the Standard Model or even of the effective field theory approach to physics beyond the Standard Model, are  working against their usefulness. Instead, each dark sector can be reduced to few parameters---to wit, just two in the case of the dark photon---in terms of which the possible discrepancies with respect to the Standard Model are more effectively mapped in the experimental searches and the potential discovery more  discernible.

In this primer, we review the physics of this new gauge boson  from  the theoretical and the experimental point of view. 
We explain how the   dark photon  enters laboratory, astrophysical and cosmological observations as well as  dark matter physics. 

 As explained in detail in section~\ref{sec:mass}, there are actually two kinds of dark photons: The massless and the massive---whose  theoretical frameworks as well as  experimental signatures are quite distinct. They give rise to  dark sectors with different features; their characteristic physics  and experimental searches are best reviewed separately. The massive dark photon has been receiving so far most of the attention because it couples directly to the SM currents and is more readily accessible in the experimental searches. The massless dark photon arises from a  sound theoretical framework and, as we shall argue, provides, with respect to the massive case, a comparably rich, if perhaps more challenging, experimental target. 
 
 We look into the ultraviolet (UV) completion of models  of the dark photon in  section~\ref{sec:UV} to better understand the origin of their interactions with the SM particles. 

Section \ref{sec:dark} describes the interplay between the dark photon and dark matter and introduces many of the definitions used in the experimental searches. 

We survey the  current and future experimental limits on the parameters of the massless and massive dark photons   together with  the related bounds on milli-charged  fermions. We discuss all these  constrains for the massless case in section \ref{sec:massless} and for the massive case in section \ref{sec:massive}.  At the best of our knowledge, these two sections  provide the reader with a comprehensive review of the physics of the dark photon. 

We collect in three appendices a number of definitions and equations, which the reader may find useful to better follow  the discussion in the main text. 

In  the past few years a number of   reports  on the dark sector (and the massive dark photon within it)  have been published \cite{Hewett:2012ns,Essig:2013lka,Raggi:2015yfk,Deliyergiyev:2015oxa,Alekhin:2015byh,Curciarello:2016jbz,Alexander:2016aln,Beacham:2019nyx}.   The interested reader can therein find different points of view to complement the present review as well as
additional details on the other portals. A   previous discussion of the astrophysical, cosmological and other constraints for the massless dark photon can be found in  \cite{Dobrescu:2004wz}.

\section{Massless and massive dark photons}
\label{sec:mass}

The most general kinetic part of the Lagrangian of two Abelian gauge bosons, described by two gauge groups  $U(1)_a$ and $U(1)_b$, is given by
\be
{\cal L}_0=- \frac{1}{4} F_{a \mu\nu} F^{\mu\nu}_a- \frac{1}{4} F_{b \mu\nu} F^{\mu\nu}_b -\frac{\varepsilon} {2} F_{a \mu\nu} F^{\mu\nu}_b \, . \label{kinetic}
\ee

The gauge boson $A^{\mu}_b$ is taken to couple to  the current $J_{\mu}$ of ordinary SM matter, the other, $A^{\mu}_a$, to the current $J_{\mu}^\prime $, which is made of dark-sector  matter, to give the Lagrangian
\be
{\cal L}= e \, J_{\mu} A^{\mu}_b + e^\prime J_{ \mu}^\prime A^{\mu}_a \, , \label{int}
\ee
with $e$ and $e^\prime$ the respective coupling constants.

To discuss the physics arising from the Lagrangians in \eq{kinetic} and \eq{int}, it is  useful to identify from the very beginning two kinds of dark photons: 
\begin{itemize}
\item[-] the \underline{massless kind}, which, as we are about to show,  does not couple directly to any of the SM  currents and interacts  instead  with ordinary matter only through operators of dimension higher than four;
\item[-] the  \underline{massive kind}, which   couples to ordinary matter through a current (with arbitrary charge), that is, a renormalizable operator of dimension four. The massless limit of this case does not correspond to the massless case above.
\end{itemize}
Because of their different coupling to SM particles, the two kinds  are best  discussed  separately.

Let us first consider the massless case. 

As first discussed in \cite{Holdom:1985ag} in this case the classical Lagrangian can be diagonalized. What happens at the quantum level and how the mixing  manifests itself 
has been  analyzed in detail in \cite{delAguila:1995rb} for the unbroken gauge theory as well as the spontaneously broken case (see, also, the appendix of \cite{Feldman:2007wj} which we mostly follow).

 \begin{figure}[t!]
\begin{center}
\includegraphics[width=3in]{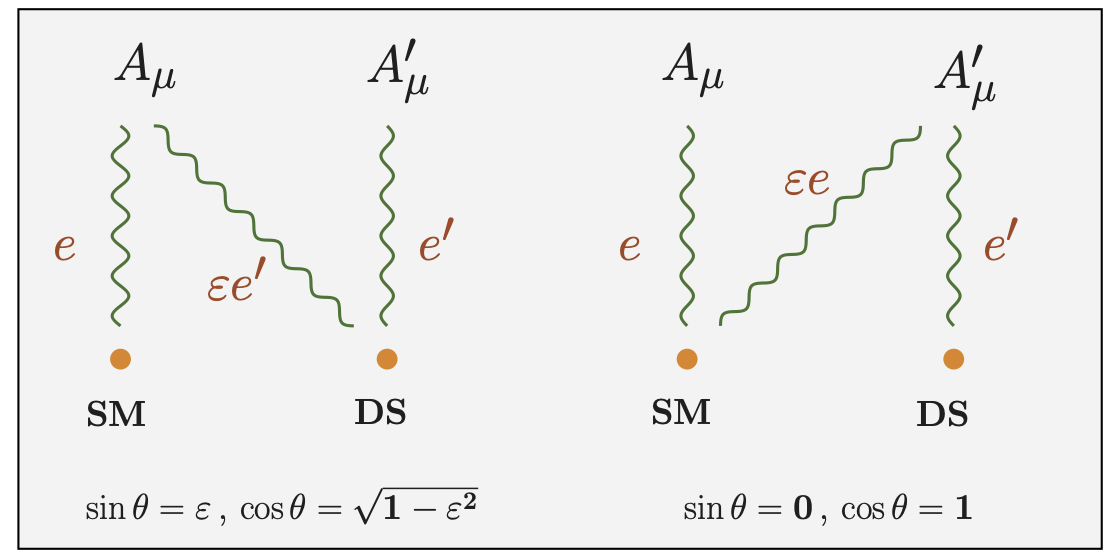}
\caption{\small Scheme of the coupling of the ordinary  ($A_\mu$) and dark ($A_\mu^\prime$) photon to the SM and dark-sector (DS) particles for the two choices of the angle $\theta$ discussed in the main text. $e$ and $e^\prime$ are the couplings of the ordinary and dark photons to their respective sectors.
\label{fig:mix} }
\end{center}
\end{figure}

The kinetic terms in \eq{kinetic} can be diagonalized by rotating the gauge fields as
\be
{ A^{\mu}_a \choose  A^{\mu}_b } = \left( \begin{array}{cc}{\displaystyle \frac{1}{\sqrt{1-\varepsilon^2}}} &  0 \\
{\displaystyle -\frac{\varepsilon}{\sqrt{1-\varepsilon^2}}} & 1  \end{array} \right)  \left( \begin{array}{cc} \cos \theta & -\sin \theta \\
\sin \theta & \cos \theta  \end{array} \right) 
{ A^{\prime \mu} \choose  A^{\mu} }  \, ,\label{rot}
\ee
where now we can identify $A^{ \mu}$ with the ordinary photon and $A^{\prime \mu}$ with the dark photon. The additional orthogonal rotation in \eq{rot} is always possible and introduces an  angle $\theta$ which is arbitrary as long as the gauge bosons are massless.

After the rotation in \eq{rot}, the interaction Lagrangian in \eq{int} becomes
\bea
{\cal L^\prime} &=& \left[ \frac{e^\prime  \cos \theta}{\sqrt{1 -\varepsilon^2}}  J_{\mu} ^\prime + e \left( \sin \theta -  \frac{\varepsilon  \cos \theta }{\sqrt{1 -\varepsilon^2}} \right)  J_{\mu} \right] A^{\prime \mu} \label{phys} \nn \\
&+ &\left[ -\frac{e^\prime \sin \theta}{\sqrt{1 -\varepsilon^2}}   J_{\mu} ^\prime + e \left( \cos \theta +  \frac{\varepsilon \sin \theta }{\sqrt{1 -\varepsilon^2}} \right)   J_{\mu} \right]  A^{\mu}. 
\eea
 By choosing $ \sin \theta =0\, (\cos \theta =1)$  (see right-side of Fig.~\ref{fig:mix}) we can have the ordinary photon $A_\mu$ coupled only to the ordinary current $J_\mu$ while the dark photon couples to both the ordinary and the dark current $J^\prime_\mu$, the former with strength $\varepsilon e/\sqrt{1 -\varepsilon^2}$ proportional to the mixing parameter $\varepsilon$. The Lagrangian is therefore:
\be
\boxed{
{\cal L^\prime} = \left[ \frac{e^\prime }{\sqrt{1 -\varepsilon^2}}  J_{\mu} ^\prime - \frac{e\varepsilon}{\sqrt{1 -\varepsilon^2}}  J_{\mu} \right] A^{\prime \mu} + e J_{\mu}  A^{\mu} \, . \label{Lmassive} 
}
\ee

\textit{Vice versa},  
with the choice $\sin \theta =\varepsilon$ and $\cos \theta =\sqrt{1 - \varepsilon^2}$ (see left-side of Fig.~\ref{fig:mix}), 
we have the opposite situation with the dark photon only coupled to the dark current and the ordinary photon to both currents, with strength $\varepsilon e/\sqrt{1 -\varepsilon^2}$ to the dark one. This latter coupling between the dark-sector matter  to the ordinary photon is called a \textit{milli-charge}. Its value is experimentally known to be  small \cite{Davidson:2000hf}. The dark photon sees ordinary matter only through the effect of operators  like the magnetic moment or the charge form factors (of dimension higher than four). This is the choice defining the massless dark photon proper:
\be
\boxed{
{\cal L^\prime} = e^\prime  J^{\prime}_{\mu} A^{\prime\mu} 
+ \left[ -\frac{e^\prime \varepsilon}{\sqrt{1 -\varepsilon^2}}   J_{\mu} ^\prime + \frac{e}{\sqrt{1 -\varepsilon^2}} J_{\mu} \right]  A^{\mu}  \label{Lmassless} 
}
\ee

If the gauge symmetry is spontaneously broken, the diagonalization of the mass terms locks the angle $\theta$ to the value required by the rotation of the gauge fields to the mass eigenstates and we cannot have that one of the two currents only couples to one of the two gauge bosons.

This is also the  case when the $U(1)$  gauge bosons acquire  a mass by means of the  Stueckelberg Lagrangian (see \cite{Ruegg:2003ps} for a review and the relevant references)
\be
{\cal L}_{Stu}=  -\frac{1}{2} M_a^2 A_{a \mu} A^{\mu}_a- \frac{1}{2} M_b^2 A_{b \mu} A^{\mu}_b- M_a M_b A_{a \mu} A^{\mu}_b \label{Stu}\, .
\ee
In this case, as in the spontaneously broken case, the angle $\theta$ is fixed and equal to
\be
\sin \theta =  \frac{\delta \sqrt{1 - \varepsilon^2}}{\sqrt{1-2 \delta \varepsilon + \delta^2}} \quad \cos \theta =   \frac{1 - \delta \varepsilon}{\sqrt{1-2 \delta \varepsilon + \delta^2}}
\ee
where $\delta = M_b/M_a$, and we have no longer the freedom of rotating the fields as in \eq{rot}. The Lagrangian in \eq{phys} is now
\bea
{\cal L^{\prime\prime}} &=&  \frac{1}{\sqrt{1 - 2 \delta \varepsilon + \delta^2}} 
\left[ \frac{e^\prime \left( 1 -\delta \varepsilon \right)}{\sqrt{1 -\varepsilon^2}}  J_{\mu} ^\prime 
+ \frac{e \left( \delta - \varepsilon \right)}{\sqrt{1 -\varepsilon^2}}   J_{\mu} \right] A^{\prime \mu}  \nn \\
&+ & \frac{1}{\sqrt{1 - 2 \delta \varepsilon + \delta^2}} 
\left[e   J_{\mu} - \delta e^\prime   J_{\mu} ^\prime   \right]  A^{\mu} \, .\label{massive}
\eea

The case of spontaneously broken symmetry can be distinguished from the Stueckelberg mass terms because the former will give rise to processes in which the dark photon is produced together with the dark Higgs boson, the vacuum expectation value of which hides the symmetry.

Whereas the Lagrangian in \eq{massive} is the most general, the simplest and most frequently discussed case consists in giving mass directly to only one of the $U(1)$ gauge bosons so that, for instance, $M_b=0$ in \eq{Stu}, the mass states are already diagonal. Even in this simple case, the mass term removes the freedom of choosing the angle $\theta$ in \eq{rot}. With this choice, $\delta=0$ in \eq{massive},  the ordinary photon couples only to ordinary matter and the massive dark photon is characterized by a direct coupling 
to the electromagnetic current of the the SM particles (in addition to that to dark-sector matter) and described by the Lagrangian 
\be
\mathcal{L} \supset -\frac{e \varepsilon}{ \sqrt{1 - \varepsilon^2}}   J_{\mu}  A^{\prime \mu}  \simeq -e\, \varepsilon\,  J_{\mu}  A^{\prime \mu}   \label{coupling} \, ,
\ee
as in \eq{Lmassive} above.  This is the choice defining the massive dark photon. The coupling of the massive dark photon to SM particles is not quantized---taking the arbitrary value $e \varepsilon$.
 Because of this direct  current-like coupling to ordinary matter, it is the spontaneously broken or Stueckelberg massive dark photon that is mostly discussed in the literature and considered in the experimental proposals.
 
Notice that the  massive dark photon has 
 the same  couplings as the massless  dark photon after choosing $ \sin \theta = 0$ (right-side of Fig.~\ref{fig:mix}); this case therefore represents the limit of vanishing mass of the massive dark photon. On the contrary, the massless dark photon proper---corresponding to the choice $ \tan \theta = \left[ \varepsilon/\sqrt{1 - \varepsilon^2} \right]$---is not related to any limiting case of the massive dark photon.
 
There are no electromagnetic milli-charged particles in the massive case; they are present only if both $U(1)$ gauge groups are spontaneously broken (or equivalently $M_b \neq 0$\ in the Stueckelberg Lagrangian in \eq{Stu})---which is not the case of our world where the photon is massless.

\subsection{Kinetic mixing: Electric or hyper-charge?}
There seems to be the choice in the kinetic mixing in \eq{kinetic} between the $U(1)_{e.m.}$ group of electric charge and the $U(1)_Y$ group of the hyper-charge, with mixing parameter $\varepsilon$ defined as in Eq.(\ref{kinetic}).
Concerning the massless dark photon, these two choices give rise to the same physics, since the dark photon remains decoupled from the SM fields at the tree-level. The only difference is that the photon and $Z$-boson are now both coupled to the dark-sector current, with $e_{\dd} \varepsilon/ \sqrt{1-\varepsilon^2} \cos \theta_W$  and $e_{\D} \varepsilon/ \sqrt{1-\varepsilon^2} \sin \theta_W$ strength, respectively.

Let now consider the massive dark-photon coupling to hyper-charge. In this case it is convenient to parametrize the coupling of the  dark photon to the hyper-charge as
\be
\tilde {\cal L} = - \frac{\varepsilon}{2 \cos \theta_W} \tilde F^\prime_{\mu\nu} B^{\mu\nu} \, . \label{new}
\ee
The usual diagonalization of the gauge bosons $W^3_\mu$ and $B_\mu$ now includes also the dark photon $\tilde{A}^{\prime}_\mu$ (in the non-diagonal basis) so that the physical gauge bosons $Z_\mu$ and $A_\mu$ also contain a dark-photon component $A^{\prime}_\mu$ in the mass eigenstate basis. In particular, at the $O (\varepsilon)$ in the expansion, we have
\be
\left(
\begin{array}{c}
W^3_\mu\\
B_\mu\\
\tilde{A}^{\prime}_\mu\\
\end{array}  \right)
= 
\left(
\begin{array}{ccc}
c_W & s_W & -s_W \varepsilon \\
-s_W & c_W & -c_W \varepsilon  \\
t_W \varepsilon &  0 & 1\\
\end{array}
\right)\,
\left(
\begin{array}{c}
Z_\mu\\
A_\mu\\
A^\prime_\mu\\
\end{array}
\right)\, ,
\ee
where $c_W$, $s_W$ and $t_W$ are the usual cosine, sine, and tangent of the Weinberg angle $\theta_W$, respectively. New couplings of the massive dark photon to the SM fermions appear for the photon and the $Z$ gauge boson up to $O (\varepsilon^2)$:
\be
{\cal L} \supset - e \, \varepsilon J^\mu A^\prime_\mu + e^{\prime} \, \varepsilon \, t_W J^{\prime \mu} Z_\mu \,  + e^{\prime} J^{\mu\prime}  A^{\prime}_{\mu}\, , \label{hyper}
\ee
where  $J_{\mu}$ is the EM current, while $J^{\prime}_{\mu}$ and $e^{\prime}$ are the matter current and coupling of the dark-photon in the dark sector, respectively.
After integrating out the $Z$ boson, we see that the coupling of the massive dark photon to the SM fermions is recovered as $-e\varepsilon$.

Which coupling is used depends then only on the energy of the processes considered, with the direct coupling to the photon for all processes below the electroweak scale breaking, and the hyper-charge above it. Since all limits are to be considered approximately within the order of magnitude, the presence of the factor $c_W$ in the definition in \eq{new} does not matter.
The Lagrangian in \eq{hyper} shows that, if the mixing is between the dark photon and the hyper-charge, the $Z$ gauge boson acquires  a milli-charged coupling strength  $e^{\prime}\, t_W \, \varepsilon$ to the dark sector current.

For completeness, let us  also recall  two other possibilities that have been discussed in the literature:
\begin{itemize}
\item[-]  There is no kinetic mixing as in \eq{kinetic} but the mass term between the dark photon and the  $Z$-boson is taken non-diagonal and therefore giving a mixing between 
these two states \cite{Appelquist:2002mw,Galison:1983pa,He:1991qd,Babu:1997st,Davoudiasl:2012ag}. The dark photon is named the dark $Z$ and there are characteristic experimental signatures in parity violating processes  and the coupling to neutrinos;
\item[-]  The  $B-L$ global symmetry (or other conserved flavor symmetries) are gauged and taken to be the $U(1)$ group of the dark photon, which  mixes with the hyper-charge~\cite{Heeck:2014zfa,Bauer:2018onh,Fayet:2016nyc}. There is direct coupling to the SM fermions in this case and the dark photon is no longer dark.
 \end{itemize}
 
Although their implementation is not discussed  in this review, other interesting generalizations---as, for instance, the dark photon to be considered  a Kaluza-Klein state in a model with large extra-dimensions~\cite{Rizzo:2018ntg} or the interplay between the neutrino see-saw mechanism and the dark photon~\cite{Bertuzzo:2018ftf}---should be borne in mind.

\subsection{Embedding  in a nonAbelian group}

In the massless case, the ordinary photon still couples to the dark sector with a milli-charge $\varepsilon e$. As reviewed in the next section, there are very stringent limits on the size of such a  milli-charge, at least for reasonably light dark states. To avoid the necessity of assuming a very small milli-charge, one can assume that the dark $U(1)$ group is a symmetry left over after the spontaneous breaking of a larger nonAbelian  group. 

The simplest realization of this symmetry breaking is provided by the group $SU(2)$ spontaneously broken to $U(1)$ by the vacuum expectation value of the neutral component of a scalar field in the adjoint representation.

In this scenario, the mixing term 
in \eq{kinetic} cannot be written because  the larger group  has traceless generators. The absence of mixing is in this case protected against radiative corrections and the dark and the ordinary photons see only their respective sectors (at least through renormalizable operators). 

This scenario is also suggested  by the extra Landau pole that otherwise would be present---assuming  that the Landau pole of the ordinary $U(1)$ is removed by the embedding of the SM in a scenario of grand unified theory. 

If we assume that the  dark  photon arises from a nonAbelian group, there is no milli-charged coupling of the dark sector to  ordinary photons. On the other hand, all states in the dark sector must come as multiplets of the nonAbelian group and the possible  experimental signatures of this additional structure can be searched for.

\section{UV models}
\label{sec:UV}

Because the massive dark photon couples directly to the SM electromagnetic current, its phenomenology is rather independent of the details of the underlaying UV completion. The two parameters $\varepsilon$ and $m_{A^\prime}$ suffice to fully describe the experimental searches.

The case of the massless dark photon is more complicated because the coupling to the SM particles only takes place through higher order operators whose structure heavily depends on the underlaying UV model. Even though it is possible to frame the experimental search in terms of the effective scale of these operators (as we do in section \ref{sec:massless}), the limits thus found begs to be framed in terms of the UV model parameters, namely the masses and the coupling of the dark sector states, in addition to the dark photon itself. For this reason, it is useful in this case to introduce a minimal UV model  (as we do in section~\ref{sec:model}) to  provide the relationships among the  parameters of the model and thus  possible to relate different limits that are instead  independent or not present under the portal interaction.
 
\subsection{Massive dark photon: Origin and size of the mixing parameter}

The size of the mixing parameter $\varepsilon$ is arbitrary. It is this feature that makes the charge not quantized. At the same time, it cannot be $O(1)$ because, if so,  the massive dark photon would have already been discovered. 

A natural suppression of $\varepsilon$  is achieved if the mixing only comes as a correction at one- or  two-loop level in some UV completion. This is achieved in a natural manner if the tree-level mixing is set to zero. One looks for the renormalization of the model and introduces the necessary counter-terms, of which the mixing in \eq{kinetic} is one.  If there are states in the UV completion carrying both ordinary and dark charges, 
 the loop of these states generates the mixing but it comes suppressed by the loop factor (neglecting logarithmic terms) and therefore of order, say, $1/(16 \pi^2)$ times the square of the coupling constant  and therefore approximately $O(10^{-3})$, for a perturbative value of such a coupling. One can further suppress such a term by assuming that the states carrying both charges come in doublets of opposite dark charges. In this case, the first contribution is at the two-loop level, and approximately of order $O(10^{-5})$.  If the mixing originates in the exchange of heavy messenger fields~\cite{Essig:2009nc} or in a multi-loop contribution~\cite{Koren:2019iuv,Gherghetta:2019coi}, its value can be  smaller.
 
 Even smaller  values of the parameter $\varepsilon$  are expected if the origin of the mixing  is non-perturbative; for example, values between $O(10^{-12})$ and $O(10^{-6})$  have been discussed---mostly within the broad heading of string compactification~\cite{Dienes:1996zr,Abel:2003ue,Abel:2006qt,Goodsell:2009pi,Goodsell:2009xc,Heckman:2010fh},
 or in scenarios of SUSY breaking~\cite{ArkaniHamed:2008qp} and hidden valley~\cite{Chan:2011aa}.
These arguments are often cited  to motivate experimental searches in the region of small mixing parameter $\varepsilon$ in the case of the 
massive dark photon---regardless of the large uncertainties in the predictions of the corresponding theoretical approaches.

\subsection{Massless dark photon: Higher-order operators}

The massless dark photon does not interact directly with the currents of the SM fermions. The higher-order operators through which  the interaction  with ordinary matter $\psi^i$ takes place start with the  dimension-five operators in the Lagrangian
\be
\mathcal{L} =   \frac{e_{\dd}}{2 \Lambda_5} \overline{\psi}^{\, i}\, \sigma_{\mu\nu} \left( \mathbb{D}_M^{ij} + i \gamma_5\, \mathbb{D}_E^{ij} \right)\psi^j 
\,F^{\prime \mu\nu}\, , \label{dipole}
\ee
where  $F^{\prime}_{\mu\nu}$ is  the  field strength associated to the dark photon field $A_{\mu}^{\prime}$,
and $\sigma_{\mu\nu}=i/2 \, [\gamma_{\mu},\gamma_{\nu}]$. The operator proportional to the coefficient $\mathbb{D}_M$ is the magnetic dipole moment and that  proportional to the coefficient $\mathbb{D}_E$ is the electric dipole moment. 
The indices $i$ and $j$ in the fermion fields keep track of the flavor and thus allow for flavor off-diagonal transitions.

The dimension-five operators in \eq{dipole} are best seen as operators of dimension six  with the gauge group $SU(2)_L$ taken as the unbroken symmetry of the Lagrangian and the SM fermion  grouped, like in the SM, into doublets $\psi_L$ and singlets $\psi_R$. In this case, the operators contain the Higgs boson field and can be written as
\be
\boxed{ \Biggr.
\mathcal{L} =  \frac{e_{\dd}}{ 2\Lambda^2} \overline{\psi}_L^{\,i}\, \sigma_{\mu\nu} \left( \mathbb{D}_M^{ij} + i \gamma_5\, \mathbb{D}_E^{ij}  \right) H \psi_R^j  \,F^{\prime \mu\nu} + \text{H.c.} \label{dipoleH} \Biggr.
 }
\ee
The effective scale is accordingly modulated by the vacuum expectation value (VEV) $v_h$ of the Higgs boson. This VEV  keeps track of the chirality breaking, with the whole operator vanishing as $v_h$ goes to zero.

In this review we shall only retain the magnetic dipole $\mathbb{D}_M$ term and set to zero the  electric dipole term proportional to $\mathbb{D}_E$. The inclusion of the latter would require the further assumption of CP-odd physics which is, we believe, premature at the moment.

Next, we have the dimension-six operators
\be
\mathcal{L}^\prime =  \frac{e_{\dd} }{2 \Lambda^2} \,  \overline{\psi}^{\, i}\, \gamma_{\mu} ( \mathbb{R}_r^{ij} \,+ i \gamma_5\,  \mathbb{R}_a^{ij} ) D_\nu \psi^j \, F^{\prime\mu\nu} \,,  \label{radius}
\ee
where the form factor $\mathbb{R}_r$ is related to the charge radius of the fermion; the term $\mathbb{R}_a$ is sometime referred to as the \textit{anapole}.

The operator in \eq{radius} contributes, via the equations of motion, to  four-fermion operators---which are accounted for in the effective field theory of the dimension-six operators~\cite{Grzadkowski:2010es} but are not relevant for the massless dark photon interaction to ordinary matter---and to the form factors of the interaction if the particles are off-shell. The latter provide a next-to-leading interaction between the massless dark photon and ordinary matter that has yet to be studied (and is not discussed in this review).

Higher-order operators give vanishingly small contributions and can be neglected.

The scale $\Lambda$ depends on the parameters of the underlaying UV model. Typically, it is the mass of a heavy state, or the ratio of masses of states of the dark sector, multiplied by the couplings of these states to the SM particles. In particular, the dipole operators in \eq{dipoleH}, as they require a chirality flip, can turn out to be enhanced, or suppressed, according to  the underlaying model  chirality mixing.

The fact that  the  interaction between the massless dark photon and the SM states  only takes place through higher-order operators  provide an appealing explanation for  its weakness. The structure of these operators leads directly to the possible underlaying UV models---a minimal example of which is discussed in section~\ref{sec:model}.

\section{Dark matter and the dark photon}
\label{sec:dark}

Dark matter  is part of the dark sector. The interplay between the dark photon and dark matter opens new windows on its physics and gives further constraints. 
Whereas in most scenarios dark matter is one of the fermion (or scalar) states in this sector, there also exists the possibility that dark matter could be a very light vector boson like the massive dark photon itself.

\subsection{Massless dark photon and galaxy dynamics }

Models of self-interacting dark matter  charged under Abelian or non-Abelian gauge groups and interacting through the exchange of massless as well as massive  particles have a long history.\footnote{The literature on the subject is already very extensive, see, for example,~\cite{Goldberg:1986nk,Holdom:1986eq,Gradwohl:1992ue,Carlson:1992fn,Foot:2004pa,Feng:2008mu,Ackerman:mha,Feng:2009mn,ArkaniHamed:2008qn,Kaplan:2009de,Buckley:2009in,Hooper:2012cw,Aarssen:2012fx,Cline:2012is,Tulin:2013teo,Gabrielli:2013jka,Baldi:2012ua,CyrRacine:2012fz,Cline:2013zca,Chu:2014lja,Boddy:2014yra,Buen-Abad:2015ova,Agrawal:2016quu}. 

Interacting dark matter can form bound states. The phenomenology of such  atomic dark matter \cite{Kaplan:2009de} has been discussed in the literature, see \cite{CyrRacine:2012fz}  and references therein.} 

The most obvious obstacle to having dark matter in the dark sector interacting via a long-range force as the one carried by the massless dark photon comes from the essentially collisionless dynamics of galaxies and  the ellipticity of their dark-matter halo.  

 The most severe observational limits come from  the present dark matter density distribution in collapsed dark matter structures, rather than effects in the early Universe or the early stages of structure formation~\cite{Ackerman:mha,Feng:2009mn,CyrRacine:2012fz}. 
 
 Bounds have been derived from the dynamics in merging clusters, such as the Bullet Cluster~\cite{Clowe:2006eq}, the tidal disruption of dwarf satellites along their orbits in the host halo, and kinetic energy exchanges among dark matter particles in virialized halos. The latter turns out to be the most constraining bound, noticing that self-interactions tend to isotropize dark matter velocity distributions, while there are galaxies whose gravitational potentials show a triaxial structure with significant velocity anisotropy; limits have been computed, with subsequent refinements, via estimating an isotropization timescale (through hard scattering and cumulative effects of many interactions, also taking into account Debye screening) and comparison to the estimated age of the object~\cite{Feng:2009mn}, or following more closely the evolution of the velocity anisotropy due to the energy transfer~\cite{Agrawal:2016quu}. The ellipticity profile inferred for the galaxy NGC720, according to \cite{Agrawal:2016quu} sets a limit of about
\be
m_\chi \left(\frac{0.01}{\alpha_{\D}}\right)^{2/3}  \agt 300 \; \text{GeV}\ ,
\label{galaxy}
\ee
where $m_\chi$  stands for the dark matter mass and the $\alpha_{\D}$ scaling quoted  is approximate and comes from the leading $m_\chi$ over $\alpha_{\D}$ scaling in the expression for the isotropization timescale. 

The limit in \eq{galaxy}  is subject to a number of uncertainties and assumptions; it is less stringent than earlier results, such as the original bound from soft scattering quoted in~\cite{Ackerman:mha}, 
\be
\frac{G_N m_\chi^4 N}{8 \alpha_{\dd}^2} \agt 50 \log \frac{G_N m_\chi^2 N}{2 \alpha_{\dd}} \, , \label{eq:limit_soft}
\ee
where $N$ is the number of dark-matter particles and $G_N$ is Newton's constant,
as well about a factor of 3.5 weaker than~\cite{Feng:2009mn} (see, also, \cite{Feng:2009hw,Lin:2011gj}). On the other hand, results on galaxies from $N$-body simulations in self-interacting dark matter cosmologies~\cite{Peter:2012jh}, which take into account predicted ellipticities and dark matter densities in the central regions, seem to go in the direction of milder constraints, about at the same level or slightly weaker than the value quoted in \eq{galaxy}---again subject to uncertainties, such as the role played by the central baryonic component of NGC720.

\subsection{Massless dark photon and  dark-matter relic density}

All the stable fields within the dark sector provide a multicomponent candidate for dark matter whose  relic density depends on the value of their couplings to the  dark photons and SM fermions (into which they  may annihilate, depending on the UV model) and masses.

 Not all of the dark fermions contribute to the relic density.
If these fermions are relatively light, their dominant annihilation is  into dark photons (see Fig.~\ref{fig:dark})
\be
\chi \chi \rightarrow A^\prime A^\prime 
\ee
 with a rate given by
\be
\langle \sigma_{\chi\chi \rightarrow A^\prime A^\prime} v \rangle = \frac{2 \pi \alpha_{\D}^2}{m_\chi^2}  \label{thermal-x-section1} \, .
\ee
For a  strength $\alpha_{\D} \simeq 0.01$, all fermions with masses  up to around 1 TeV have a large  cross section and their relic density(see \eq{omega} in the appendix~\ref{sec:cosmo})
\be
\Omega_\chi\, h^2 \approx \frac{2.5 \times 10^{-10} \; \mbox{GeV}^{-2}}{\langle \sigma_{\chi\chi \rightarrow A^\prime A^\prime}  v \rangle}
\ee
 is only a percent of the critical one; it is
roughly $10^{-4}$ the critical one for dark fermions in the 1 GeV range, even less for lighter states.  These dark fermions are not part of dark matter; they have (mostly) converted into dark photons by the time the universe reaches our age and  can only be produced in high-energy events. This is fortunate because, as we have seen, they are ruled out as possible dark matter candidates by the limit on galaxy dynamics.

Heavier dark fermions  can be dark matter. The dominant annihilation for these is not into dark photons but into SM fermions via the exchange of some messenger field $S$---the details depending  on the underlying UV model---and is proportional to the corresponding coupling which we denote $\alpha_L$ anticipating the discussion in section~\ref{sec:model}---with a thermally averaged cross section  approximately given by
\be
\langle \sigma_{\chi\chi \rightarrow f \bar f} v \rangle \simeq  \frac{2 \pi \alpha_{L}^2}{m_S^2} \label{thermal-x-section2}
\ee
instead of \eq{thermal-x-section1}.  The critical relic density can be reproduced if, assuming thermal production,
\be
2 \pi \alpha^2_{L}  \left( \frac{10 \, \mbox{TeV}}{m_S} \right)^2 \simeq 0.1 \, . \label{relic}
\ee

 These dark matter fermions belonging to the dark sector are in principle detectable through the long range exchange of the massless dark photon and its coupling to the magnetic (o electric) dipole moment of SM matter which is induced at the one loop level in the UV model of the dark sector. The somewhat  complementary problem of dark matter having dipole moment and interacting with nuclei through the exchange of a photon has been discussed in~\cite{Pospelov:2000bq,Sigurdson:2004zp,Banks:2010eh,Barger:2010gv,Fornengo:2011sz,DelNobile:2012tx,Chu:2020ysb}  This dipole interaction is now included within the basis of the operators in the effective field theory of dark matter detection~\cite{Fitzpatrick:2012ix,Liem:2016xpm,Brod:2017bsw}.

\subsection{Massive dark photon and light dark matter}

 \begin{figure}[t!]
\begin{center}
\includegraphics[width=3in]{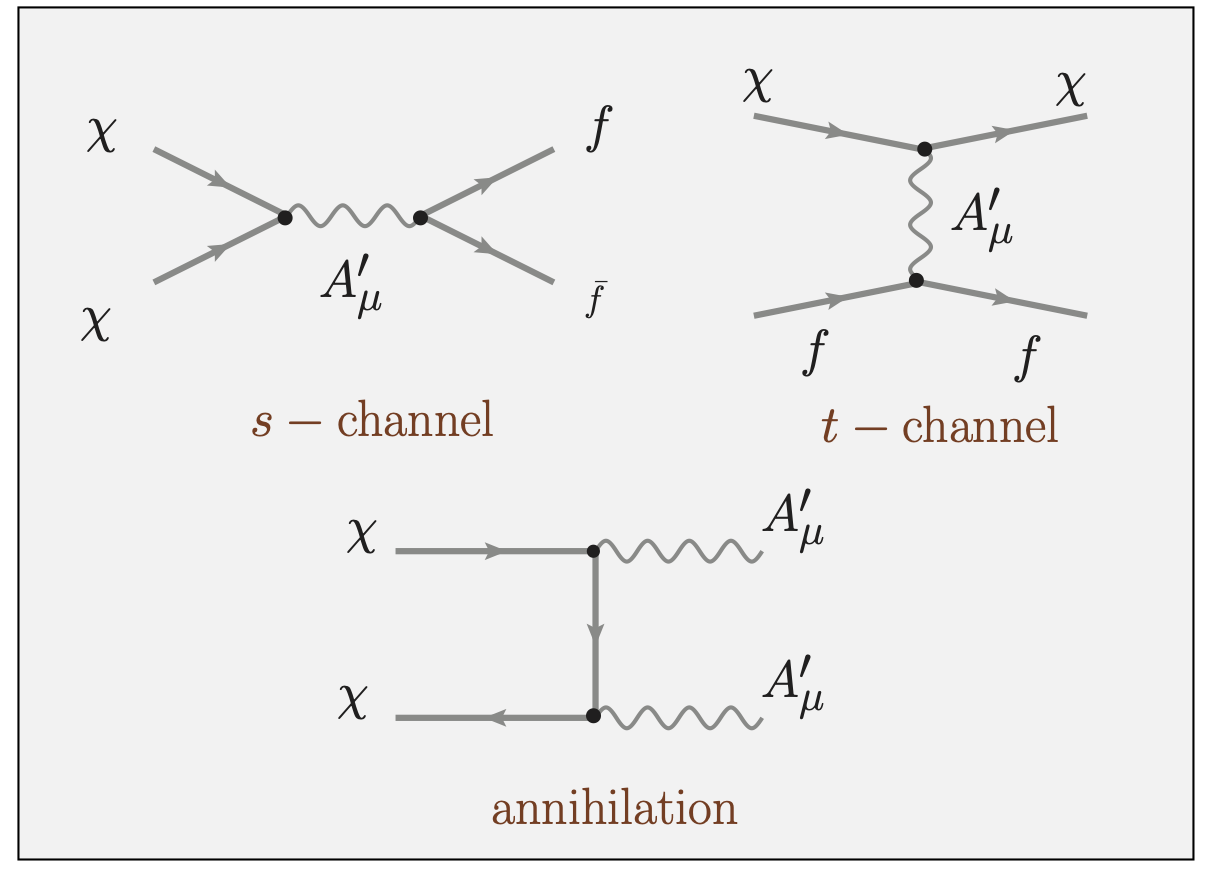}
\caption{\small Feynman diagrams for the three processes that are relevant for the discussion of the massive dark photon and dark matter.
\label{fig:dark} }
\end{center}
\end{figure}

 When dark matter is  lighter than the dark photon, and $m_{A^\prime} > 2 m_\chi$, the annihilation channel (see Fig.\ref{fig:dark})
\be
\chi \bar \chi \rightarrow A^\prime \rightarrow \bar f f
 \ee
  is open and we are in a scenario with light dark matter (LDM)~\cite{Boehm:2002yz,Knapen:2017xzo,Essig:2011nj}. The cross section is
 \bea
 \sigma_{\chi \chi \rightarrow f f}   &= & \frac{4 \pi}{3}  \varepsilon^2 \alpha \alpha_{\D} m_\chi^2
 \left( 1 + \frac{2\, m_e^2}{s} \right) \left( 1 + \frac{2\, m_\chi^2}{s} \right) \nn \\
 &\times & \frac{s} {(s-m_{A^\prime}^2 )^2 + m_{A^\prime}^2 \,  \Gamma_{A^\prime}^2} 
 \frac{\sqrt{1 -  \frac{4\, m_e^2}{m_{A^\prime}^2}}}{\sqrt{1 -  \frac{4\, m_\chi^2}{m_{A^\prime}^2}}}
 \label{cross}
\eea
The thermal average of $\sigma_{\chi \chi \rightarrow f f} v$ is defined in \eq{Taverage} of appendix~\ref{sec:cosmo}. In the non-relativistic limit where $s\simeq 4 m_\chi^2 $ we have that
\be
\langle \sigma_{\chi \chi \to f f} v \rangle \simeq  \varepsilon^2 \alpha \alpha_{\D} \frac{16 \pi m_\chi^2}{ (4m_\chi^2 - m_{A'}^2)^2} \label{xxx}
\ee
if we neglect $m_e$ with respect to $m_\chi$ and $\Gamma_{A^\prime}$ with respect to $m_{A^\prime}$.
The thermal average in \eq{xxx}  is related to the relic density $\rho_\chi$ (as reviewed in appendix~\ref{sec:cosmo}), or, in terms of the normalized quantity $\Omega_\chi = \rho_\chi/\rho_c$ as
\be
\Omega_\chi\, h^2 \approx \frac{2.5 \times 10^{-10} \; \mbox{GeV}^{-2}}{\langle \sigma_{\chi \chi \rightarrow f f} v \rangle}\, . 
\ee

The general  interplay between the massive dark photon and dark matter was originally discussed in \cite{Boehm:2003hm} and, more recently, in \cite{Hambye:2019dwd}.

It has been suggested~\cite{Izaguirre:2015yja} that the best variable to plot most effectively the constraints  in the case of LDM is by means of the \textit{yield} variable
\be
y\equiv \varepsilon^2 \alpha_{\D} \left( \frac{m_\chi}{m_{A^\prime}}\right)^4 \label{y}
\ee
because, from \eq{xxx} 
\be
\langle \sigma_{\chi \chi \rightarrow f f}  v \rangle  \simeq \frac{16 \pi \alpha  y}{m_\chi^2} 
\ee
and therefore the relic density is brought into the plot. Moreover, the scaling of these limits is made less dependent on the nature of the LDM. We add the limits in the plane \{$y$-$m_\chi$\} to those in the plane \{$\varepsilon$-$m_{A^\prime}$\} in section~\ref{sec:massive}. 

The  cross section in \eq{cross},  written in the $t$-channel (see Fig.~\ref{fig:dark}), controls the size of direct detection of dark matter in its scattering off the electrons of the detector thus producing ionization, in particular
\be
\sigma_e = \frac{16 \pi \mu^2_{\chi e} \alpha \alpha_{\D} \varepsilon^2}{(m_{A^{\prime}}^2+\alpha^2 m_e^2)^2} \left|F(q^2)\right|^2 \, ,
\ee 
where $\mu_{\chi e}$ is the reduced mass of the electron and $\chi$,
and $F(q^2)$ a form factor given by
\be
F(q^2)=\frac{m_{A^\prime}^2+\alpha^2 m_e^2}{m_{A^\prime}^2+q^2}\, ,
\ee
with $q^2$ the square of the exchanged momentum. This relationship translates into a differential event rate in a dark-matter detector with $N_T$ the number of target nuclei per unit mass
\be
\frac{dR}{d \ln E} = N_T \frac{\rho_\chi}{m_\chi} \frac{ d \langle \sigma_e v\rangle}{d \ln E}\, ,
\ee
where $E$ is the electron energy, $\langle \sigma_e v\rangle$ is the thermally averaged cross section with $v$  the $\chi$ velocity,  and $\rho_{\chi}$ the local density of $\chi$. This makes possible to utilize limits on LDM direct detection to constraint the dark photon parameter $\varepsilon$~\cite{Essig:2011nj}.

\subsection{Massive dark photon as dark matter}

 A very light massive dark photon could be a dark matter candidate\footnote{In addition,  the dark Higgs field breaking the $U(1)$ symmetry can provide yet another dark matter candidate~\cite{Mondino:2020lsc}.} if  produced non-thermally in the early Universe as a condensate, the same way as the axion is produced by the \textit{misalignement} mechanism~\cite{Preskill:1982cy,Abbott:1982af,Dine:1982ah}. In this mechanism, the value of the field is frozen by the fast expanding  Universe to whatever value it has at the initial moment. The rate of expansion is much larger than the mass and the field has no time to relax to the minimum of the potential. The unavoidable (and troublesome) Lorentz-invariance violation is estimated to be small and undetectable.
 
 In this scenario for the dark photon, as discussed in~\cite{Nelson:2011sf,Arias:2012az}, the mass arises via the Stueckelberg mechanism and there must be a non-minimal coupling to gravity. Once the Hubble constant value drops below the mass of the dark photon, its field starts to oscillate and  these oscillations behave like non-relativistic matter, that is, like cold dark matter.
  
There exist  two constraints on the parameters of this dark photon scenario. First of all, the initial value must be fine-tuned to reproduce the critical density. Second,  the decay into photons and SM leptons must not affect the cosmic microwave background. This latter requirement means that the mixing parameter $\varepsilon$ must not be too large (roughly, less than $10^{-9}$) and the mass $m_{A^\prime}$ must be less than 1 MeV. 

Production by fluctuations during inflation provides another possibility of having a massive dark photon as dark matter~\cite{Graham:2015rva,Nakai:2020cfw}.

The dark-photon dark matter is non-relativistic and interacts with ordinary matter mostly through the photo-electric process in which  a photon (with energy $m_{A^\prime}$)  is captured by an atom, with atomic number $Z$, with a cross section given, for ordinary photons, by
\be
\sigma_{p.e.}=4 \alpha^4 \sqrt{2} Z^5 \frac{8 \pi r_e^2}{3} \left( \frac{m_e}{\omega}\right)^{7/2} \, ,
\ee
where $\omega$ is the photon energy and $r_e$ the classical radius of the electron $r_e=\alpha/m_e$.
The cross section for the dark photons is that of ordinary photons rescaled by the mixing parameter $\varepsilon$:
\be
\sigma_{A^\prime}  = \varepsilon^2 \sigma_{p.e.} \, . \label{cross2}
\ee
This scenario is made accessible to the experiments by considering the  rate of absorption  of the dark photon by the detector~\cite{Pospelov:2008jk,Bloch:2016sjj}:
\be
\Gamma_{A^\prime} = \frac{\rho_{A^\prime}}{m_{A^\prime}} \, \sigma_{A^\prime} v_{A^\prime} \, ,
\ee
where the density $\rho_{A^\prime}$ is estimated  from the relic density (or the flux from the Sun).


\chapter{Phenomenology of the massless dark photon}
\label{sec:massless}

\vspace*{1cm}

{\versal The phenomenology of the massless dark photon} depends on the effect of the higher-order operator in \eq{dipoleH} which mediates its interaction with the SM particles. This operator enters the measured observables with an effective scale $\Lambda$ and the absolute value 
\be
d_M^{\, ij}\equiv |\mathbb{D}_M^{ij} |
\ee
of the magnetic dipole coefficient (neglecting the CP-odd  $\mathbb{D}_E$) which can eventually be related to the parameters of the underlying UV model like masses and coupling constants. 
The experimental searches can thus be framed in terms of the scale $\Lambda$, the dipole coefficient $d_M^{\, ij}$ and and the dark charge coupling $e_{\dd}$, which we rewrite as $\alpha_{\dd}=e_{\dd}^2/4\pi$. We do not assume this scale and coefficient to be universal.  Depending on the particular experimental set-up, the constraints are further sensitive to which particular lepton or quark is actually taking part in the interaction. The index, or indices, $i$ and $j$ keep track on the flavor dependence.

We discuss in section~\ref{sec:milli} the other side of the massless dark photon, namely the search for dark particles coupled to the ordinary photon by a milli-charge.

\section{Limits on the dark dipole scale $d_M/\Lambda^2$}
\label{sec:con1}

We collect in this section the known constraints on the size of the operator in \eq{dipoleH}. 

\begin{figure*}[t!]
\begin{center}
  \includegraphics[width=4.5in]{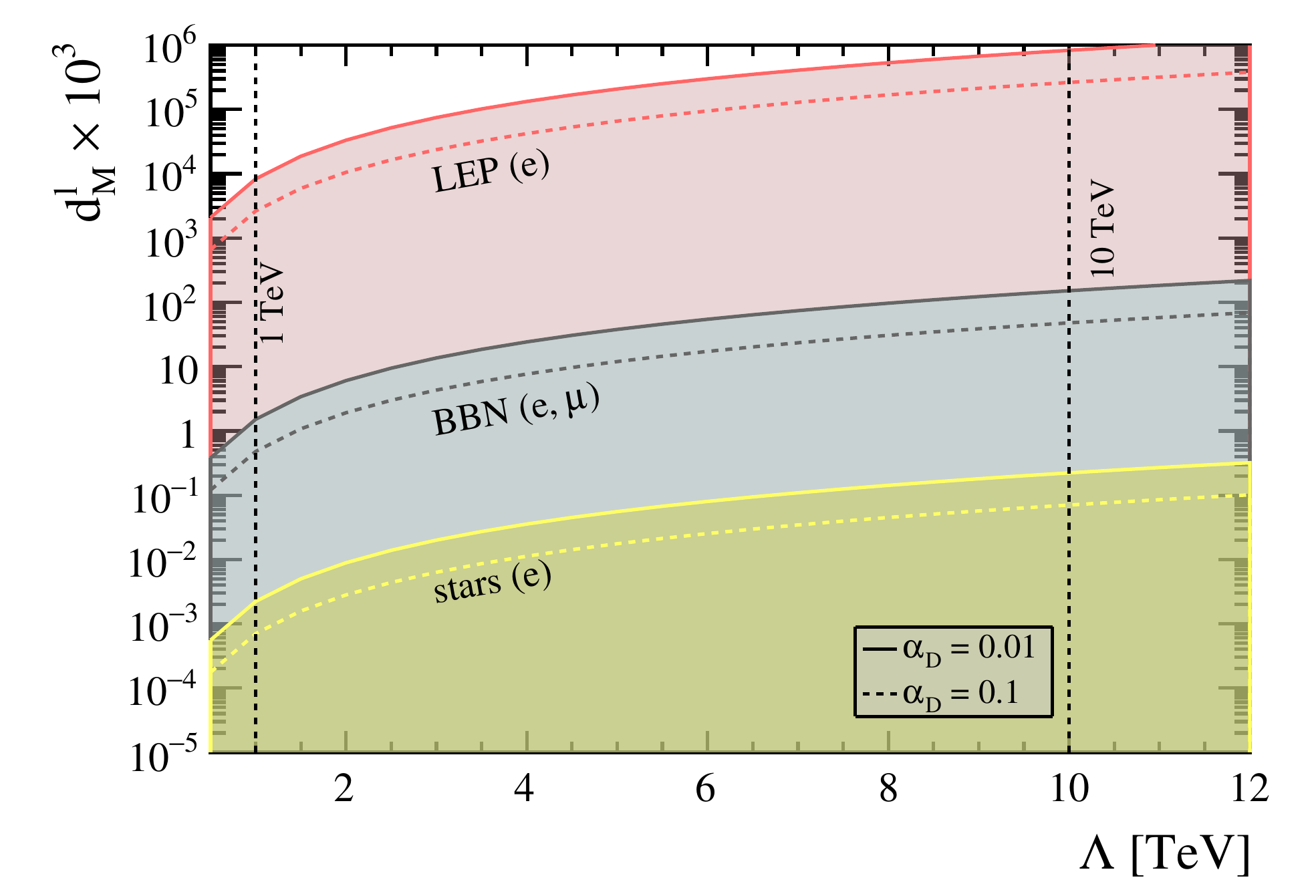}
\caption{\small Model-independent limits  for the interaction with \textbf{leptons}. The limits on the dark dipole operator  $d_M^{\, \ell}/\Lambda^2$ are shown by taking the coefficient $d_M^{\, \ell}$ as a function of the scale $\Lambda$ (for two representative values of $\alpha_{\dd}$).
Given an energy scale, the allowed values for  $d_M^{\, \ell}$  can  be read from the plot. The strongest bound on electrons comes  from stellar cooling (stars). Big bang nucleosynthesis (BBN) and collider physics (LEP) set the other depicted bounds. Solid lines  are for the representative value $\alpha_{\dd}=0.01$, dashed  lines for $\alpha_{\dd}=0.1$. \label{fig:massless1a} }
\end{center}
\end{figure*}

\begin{figure*}[t!]
\begin{center}
    \includegraphics[width=4.5in]{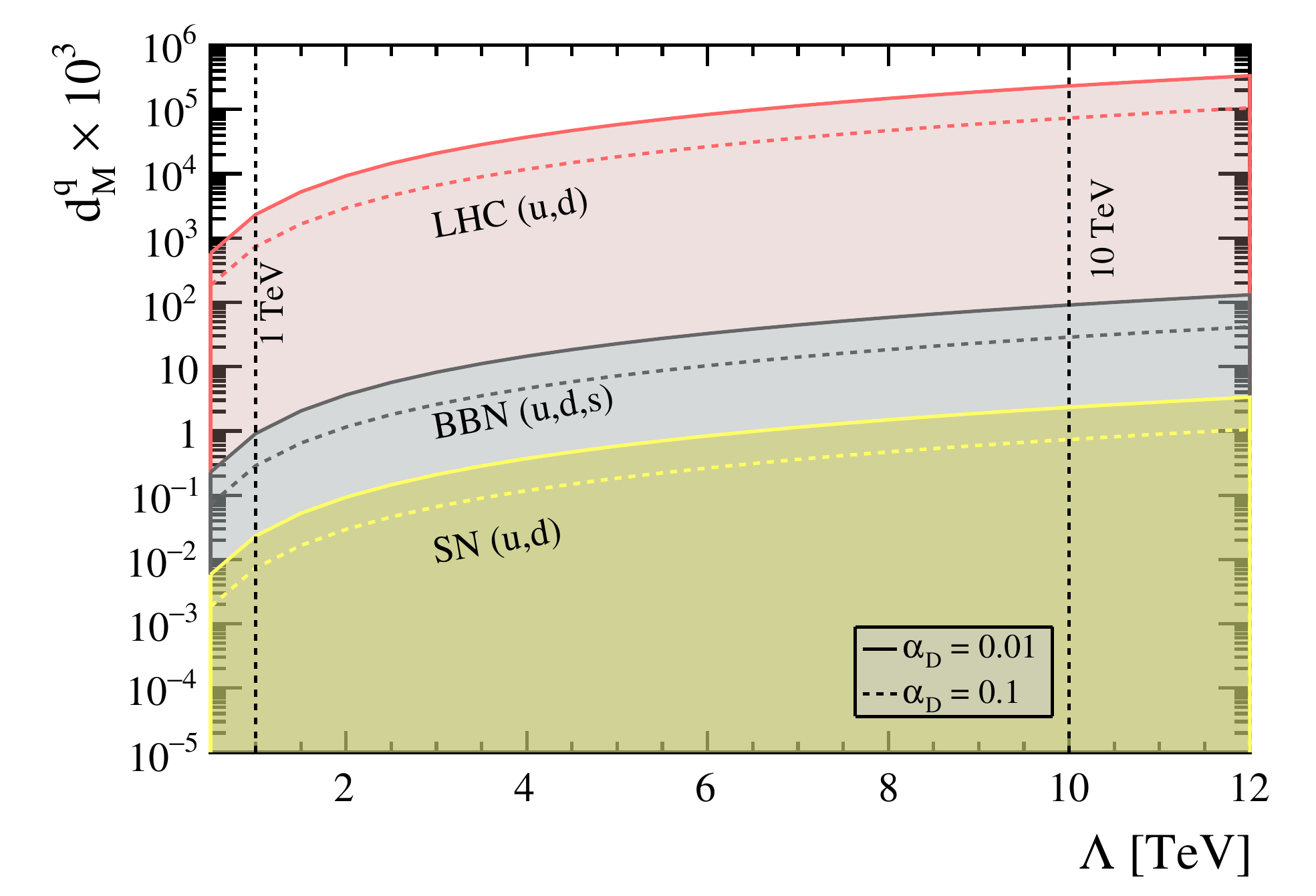}
\caption{\small Model-independent limits  for for the interaction with \textbf{quarks}. Same as in Fig.~\ref{fig:massless1a}.  The strongest bounds on light quarks comes from supernovae (SN). Primordial nucleosynthesis (BBN) and collider physics (LHC) set the other depicted bounds. Solid lines  are for the representative value $\alpha_{\dd}=0.01$, dashed  lines for $\alpha_{\dd}=0.1$. \label{fig:massless1b} }
\end{center}
\end{figure*}

 We show  in Fig.~\ref{fig:massless1a} and \ref{fig:massless1b} the more stringent  limits. Though these limits are on the combinations $d_M/\Lambda^2$, with  a factor depending on $\alpha_{\dd}$,   we find it convenient to plot them as $d_M$ as a function of $\Lambda$ so as to easily see what values of the dipole coefficient are allowed given a value for the scale $\Lambda$ (and two representative value of $\alpha_{\dd}$). 
 
\subsection{Astrophysics and cosmology}

 \begin{figure}[t!]
\begin{center}
\includegraphics[width=3.3in]{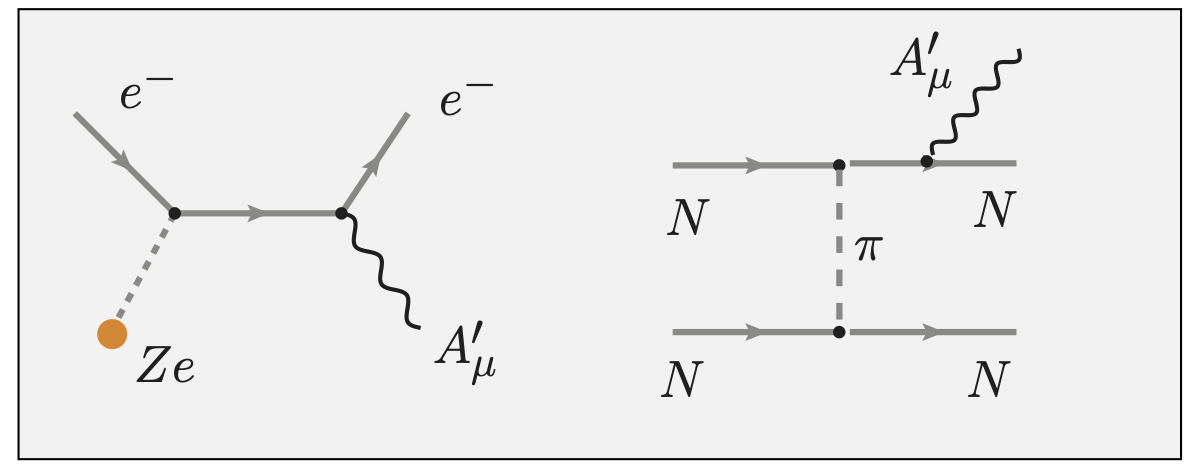}
\caption{\small  \textit{Bremsstrahlung} of dark photons from electrons in a star and from nucleons in a supernova. \label{fig:bremsstrahlung} }
\end{center}
\end{figure}

 Astrophysics and cosmology provide  very stringent limits on the  interaction of the dark photon with SM matter as given by the operator in \eq{dipoleH}.  
  It is understood that all the  limits are mostly  on the order of magnitude because of intrinsic uncertainties in the astrophysics of stellar medium, supernova dynamics and cosmological processes.
  
  Astrophysical constraints for models with a massless dark photon can be  derived from those obtained for axion-like particles  because the dipole operator in \eq{dipoleH} gives, in the non-relativistic limit, a derivative (and spin-dependent) coupling  of the dark photon with momentum $\mathbf{k}$ and polarization $\boldsymbol{\epsilon}$ to ordinary fermions $\psi$ given by
 \be
{\cal M}_{A^\prime_\mu} \approx \bar \psi \,  (\mathbf{k} \times \boldsymbol{\epsilon}) \cdot \boldsymbol{\sigma} \, \psi \, ,
 \ee
 which, after averaging over the polarizations,  gives the same contribution as that for a pseudo-scalar particle~\cite{Hoffmann:1987et,Raffelt:1996wa} like the axion, namely
 \be
{\cal M}_{a} \approx\bar \psi  \, \mathbf{k} \cdot \boldsymbol{\sigma} \, \psi  \, .
 \ee
 Only a factor of two must be included for the independent polarizations of the dark photon. 
 
 Because the massless dark photon does not mix with the ordinary photon, we can compute the limits in a kinetic theory in which the amplitude for the relevant process is computed in the vacuum and the effect of the medium---be it the stellar interior or the supernova nucleon gas---is included in the
abundances of the SM states at the given temperature.

\vskip0.3cm
$\bullet$ \underline{Stars}. The luminosity of stars is related to their energy balance. This balance is a sensitive probe of the stellar dynamics and the particle-physics processes on which is based. 
Three processes are important for energy loss in stars: Compton scattering, pair creation and \textit{Bremsstrahlung}. Of these three, it is the latter that provides the most stringent limit. The non-observation of anomalous energy transport, in various different types of stars, places strong constraints on the dipole coupling between SM states and the dark photon~\cite{Carlson:1986cu,Dobrescu:2004wz}.

The quantity  we need is the energy loss due to the emission of the extra particle. The energy loss per unit volume ${\cal Q}$ is given in the appendix~\ref{sec:TFT} in terms of the squared amplitude of the process of emitting, in our case, an axion.

For the \textit{Bremsstrahlung}  emission of axions, by electrons in the field of $n_j$ nuclei with charge $Z_j$, the squared amplitude is~\cite{Nakagawa:1987pga,Raffelt:1989zt}
\bea
\sum_\text{spin} |{\cal M}|^2  &= &\sum_j Z_j^2 n_j \frac{4 \alpha^2 \alpha^\prime_{ae}}{\pi} \frac{|\mathbf{p}_1| |\mathbf{p}_2| \omega^2}{ (\mathbf{q}^2 + \kappa_F^2)^2}
\left[ 2 \omega^2 \, \frac{p_1\cdot p_2 - m_e^2 + (p_2-p_1)\cdot k}{(p_1\cdot k)( p_2\cdot k)} \right. \nn \\
&+ &  2 - \left. 
\frac{p_1\cdot k}{p_2\cdot k}\ -\frac{p_2\cdot k}{p_1\cdot k}\right] \, 
\eea
where $p_1$ and $p_2$ and $k$ are the momenta of the initial electrons and $q=p_2-p_1$. $\omega$ and $k$ the energy and momentum of the axion and  
$\kappa_F = (4 \alpha p_F E_F /\pi)^{1/2}$ where   $p_F$ and  $E_F$ are the Fermi momentum and energy of the electrons in the plasma. The  coefficient $\alpha^\prime_{ae}$ is the coupling constant of the axion to the electrons.

In a degenerate medium (like the one for red giants and white dwarves) we have that the energy-loss rate per unit mass ${\cal Q}/\rho$ is given by~\cite{Raffelt:1996wa}
\bea
{\cal Q}/\rho&=& \frac{\pi^2\, \alpha^2\,\alpha^\prime_{ae} }{15}\frac{T^4}{m_e^2} \sum_j Z_j^2 n_j \; F (\kappa_F) \nn \\
& \simeq & \alpha^\prime_{ae}\,  1.08  \times 10^{27}   \left( \frac{T}{10^8 \text{K}}\right)^{4} \frac{Z^2}{A} \; F (\kappa_F) \, ,
\eea
the latter equation is written  in units of erg g$^{-1}$s$^{-1}$, and  the factor $F$ is approximately given in the relativistic limit as
\be
F(\kappa_F) \simeq \frac{2+\kappa_F^2}{2} \ln\frac{2+\kappa_F^2}{\kappa_F^2} -1 \, .
\ee

   The  most stringent limit for electrons comes from cooling in white dwarves~\cite{Bertolami:2014wua} and giant red stars~\cite{Viaux:2013lha} by axion \textit{Bremsstrahlung} in a degenerate medium. 
A combined fit of the data~\cite{Giannotti:2015kwo}
 finds (at $2 \sigma$) that the coupling must be
  \be
  \alpha^\prime_{ae}  \leq  3.0 \times 10^{-27} \, . \label{limit_alpha}
  \ee 
 
 The bound in \eq{limit_alpha} is translated into a bound for the dark photon by identifying the combination of parameters in the operator in \eq{dipoleH} that controls the same  process. This correspondence yields the equation
  \be
 \alpha^\prime_{ae} =  2 \frac{1}{4 \pi} \left( 2 \, e_{\dd} d_M^{\, e} \frac{  v_h \, m_e}{\Lambda^2}  \right)^2  \, , 
 \label{star_em}
 \ee
where the factor of 2 in front takes into account the two polarizations of the dark photon (with respect to the axion), $v_h=174$ GeV and  $m_e$ is the electron's mass.

To satisfy the limit in \eq{limit_alpha}, the dark photon parameters  in \eq{star_em}   must satisfy
\be
\frac{\Lambda^2 }{\sqrt{\alpha_{\dd}} \, d_M^{\, e}}   \agt 4.5 \times 10^6\; \mbox{TeV}^2\, , \label{stars}
\ee
after having included the numerical values of $m_e$ and $v_h$. 
The limit in \eq{stars} updates the one found in \cite{Dobrescu:2004wz}. 

\vskip0.3cm
$\bullet$ \underline{Supernovae}. An additional limit  is found from the neutrino signal of supernova 1987A, for which the length of the burst constrains anomalous energy losses in the explosion. 

As before, a bound can be derived from that for the coupling   between axions and nucleons.
The corresponding averaged square  amplitude is given in \cite{Brinkmann:1988vi,Raffelt:1993ix} as
 \bea
 \sum_\text{spin} |{\cal M}|^2  &= &  \frac{16 (4\pi)^3 \alpha_\pi^2 \, \alpha^\prime_{aN}}{3\, m_N^2}  
 \left[\left( \frac{\mathbf{k}^2}{\mathbf{k}^2 + m_\pi^2} \right)^2 + \left( \frac{\mathbf{l}^2}{\mathbf{l}^2 + m_\pi^2} \right)^2  \right.\nn \\
 &+& \left. \frac{\mathbf{k}^2\mathbf{l}^2 - 3 (\mathbf{k}\cdot \mathbf{l})^2}{(\mathbf{k}^2 + m_\pi^2)(\mathbf{l}^2 + m_\pi^2)} \right]
  \eea
 where $\alpha_\pi=(2m_N f/m_\pi)^2/4\pi \simeq 15$ is the pion-nucleon coupling and $\mathbf{k} =\mathbf{p}_2 -\mathbf{p}_4$ and $\mathbf{l} =\mathbf{p}_2 -\mathbf{p}_3$  and $\mathbf{p}_i$ the momenta of the nucleons (see Fig~\ref{fig:bremsstrahlung}). The  coefficient $\alpha^\prime_{aN}$ is the coupling constant of the axion to the nucleons.
 
 In the thermal medium $\mathbf{k}^2 \simeq 3 m_N T$ and we can neglect the pion mass to obtain
 \be
\sum_\text{spin} |{\cal M}|^2 =  \frac{32 (4\pi)^3 \alpha_\pi^2\, \alpha^\prime_{aN}}{ m_N^2}  
\ee
and the energy-loss rate per unit mass in the degenerate case is~\cite{Brinkmann:1988vi,Iwamoto:1984ir}
\be
{\cal Q}/\rho \simeq  \alpha^\prime_{aN} \; 1.74 \times 10^{33}  \frac{\rho}{10^{15}} \left( \frac{T}{\text{MeV}}\right)^{6} \, ,
\ee
in units of erg g$^{-1}$s$^{-1}$, which should not exceed the neutrino luminosity.
This limit yields, taking the most conservative estimate in~\cite{Keil:1996ju,Carenza:2019pxu},
\be
\alpha^\prime_{aN}  \leq   1.3 \times 10^{-18}  \, . \label{sn0}
\ee

The combination that controls energy transfer to dark photons in this  process from ordinary matter (the quarks in the nucleons)  is
 \be
 \alpha^\prime_{aN} = 2 \, \frac{1}{4 \pi} \left( 2 \, e_{\dd} d_M^{\, q} \frac{  v_h \, m_N}{\Lambda^2}  \right)^2  \, , \label{SN_em}
 \ee
where $m_N$ is the nucleon mass.  By taking the  limit in \eq{sn0}, we have
\be
\frac{\Lambda^2 }{\sqrt{\alpha_{\dd}} \, d_M^{\, q}}  \agt 4.3 \times 10^5\; \mbox{TeV}^2\, , \label{sn}
\ee
which applies to the  light $u$ and $d$ quarks---if we neglect small corrections due to the form factors in going from the nucleons to the quarks. The limit in \eq{sn} updates  the one found in \cite{Dobrescu:2004wz}.  

A caveat in the limit  in  \eq{sn0} is due to the fact that  if the coupling is too strong the emitted axions are re-absorbed by the expanding supernova and there is no energy loss; this happens for
\be
 \alpha^\prime_{aN} \geq 0.7 \times 10^{-14} \, , \label{sn1}
\ee
which yields 
\be
\frac{\Lambda^2}{\sqrt{\alpha_{\dd}} d_M^{\, q}}  \alt 5.9 \times 10^3  \; \mbox{TeV} ^2\, , \label{sn2}
\ee
There are however limits from laboratory physics, discussed in the next section, that almost close this window.

\vskip0.3cm
$\bullet$ \underline{Big bang nucleosynthesis}. A cosmological bound for the dark photon operator in \eq{dipoleH} comes from the determination of the effective number of relativistic species in addition to those of the SM partaking in the thermal bath---the same way the number of neutrinos is constrained.  This number is constrained  by  data on big bang nucleosynthesis (BBN) to be~\cite{Fields:2019pfx}:  
\be
N_\text{eff} =2.878\pm 0.278\,. \label{neff}
\ee 

We follow \cite{Dobrescu:2004wz} in deriving the corresponding limits.

The two degrees of freedom of the dark photon exceeds this limit at the big bang temperature   $T_{BBN}$  and must have decoupled before at temperature $T_d$ which is taken to be just above the QCD phase transition: $T_d=150$ MeV.
The request of decoupling before the BBN epoch can be translated in having the Hubble constant (see appendix~\ref{sec:cosmo})
 \be
 H (T_d) =\frac{T^2_d}{M_{Pl}} \left( \frac{ \pi^2}{90} g_* (T_d) \right)^{1/2}
 \ee
 be larger than the rate of interactions between SM states and the dark photon 
 \be
 \Gamma_{A^\prime}  = n_{A^\prime}  \langle \sigma v \rangle \, ,
 \ee
 where $\langle \sigma v \rangle$ is the thermally averaged cross section for the interaction of the dark photon with the SM particles present at the temperature $T_d$, $v=1$, and the number density of dark photon is given (see appendix~\ref{sec:cosmo}) by ($k_B=1$)
 \be
  n_{A^\prime} =\frac {2 \zeta (3)}{\pi^2} \, T^3 \, .
  \ee
  
  The cross section for SM fermions to Compton and annihilate into dark photon is approximately given by
  \be
 \langle \sigma v \rangle  \simeq \frac{\alpha_{\dd} d_M^2 v_h^2}{\Lambda^4}  \, .   
   \ee
   
 We thus find the condition  
\be
 \frac{2 \zeta(3)}{\pi^2} T^3_d \langle \sigma v \rangle  < \frac{T^2_d}{M_{Pl}} \left( \frac{2 \pi^2}{45} g_* (T_d) \right)^{1/2}\, ,
\ee
where the effective number of degrees of freedom $g_* (T_d)$ is bound from the limit on $N_\text{eff}$. This relationship is obtained from 
\be
 \left(  \frac{T_{BBN}}{T_d} \right)^{4}= \left( \frac{g_* (T_{BBN})}{g_* (T_d)} \right)^{4/3} < \frac{7}{4} \Delta N_\text{eff} \, ,
\ee
which, knowing that $g_* (T_{BBN})=43/4$, gives
\be
g_* (T_d) > (43/7)^{4/3} \Delta N_\text{eff}^{-3/4}\, ,
\ee
where $\Delta N_\text{eff} \equiv N_\text{eff}-3 \simeq 0.34$ by taking 2$\sigma$ of the result in \eq{neff}.

The limit applies to the interaction of leptons (electron and muon): 
\be
\frac{\Lambda^2 }{\sqrt{\alpha_{\dd}} \, d_M^{\,\ell}}   \geq 6.6 \times 10^3\; \mbox{TeV}^2\, , \label{bbn1l}
\ee
and quarks ($s, u, d$):
\be
\frac{\Lambda^2 }{\sqrt{\alpha_{\dd}} \, d_M^{\, q}}   \geq 4.3 \times 10^3\; \mbox{TeV}^2\, , \label{bbn1q}
\ee
which partake into the Compton and annihilation processes. The difference between \eq{bbn1l} and \eq{bbn1q} is due to the number of colors.

\subsection{Precision, laboratory and collider physics}

Laboratory physics can set  new constrains on  the dipole operator in \eq{dipoleH}. They are  less stringent than those from astrophysics and cosmology because the higher-order dipole operator always  yields a small number of events; these small numbers are amplified in the stars by the enormous density of particles in the medium but not in the laboratory experiments where the density is smaller. 

\vskip0.3cm
$\bullet$ \underline{Precision physics}.  The operator in \eq{dipoleH} gives rise to a macroscopic  spin-dependent (non-relativistic) potential \cite{Dobrescu:2006au}:
\be
V(r)=-\frac{\alpha_{\dd} v^2 d_M^a d_M^b}{4 \Lambda^4 r^3} \Big[ \boldsymbol{\sigma}_a  \cdot \boldsymbol{\sigma}_b - 3 \, (\boldsymbol{\sigma}_a \cdot \hat{\boldsymbol{r}})(\boldsymbol{\sigma}_b \cdot \hat{\boldsymbol{r}}) \Big] \, ,
\label{pot}
\ee
where ${\bf r}={\bf r_a}-{\bf r_b}$ is the vector distance and $r=|{\bf r}|$ and $\hat{\boldsymbol{r}}$ the corresponding unit vector. The potential in \eq{pot} is
between two fermions $f_a$ and $f_b$, with spin $\boldsymbol{\sigma}_a$ and $\boldsymbol{\sigma}_b$, and   magnetic dipole moments $d_M^{a,b}$,  as defined in \eq{dipoleH}---whose interaction  can affect atomic energy levels as well as macroscopic forces.

The potential in \eq{pot} can be used to explore atomic physics as well as macroscopic fifth-force like interactions. 

Many atomic energy levels are known with high precision. Unfortunately, the theoretical computation is lagging behind many of the experiments, mainly because of uncertainties in higher-order corrections like those due to the size of the nuclei. For this reason many results are given as energy differences where corrections proportional to $1/r^3$ are factorized out. This procedure  makes often impossible to use these results to test  the potential in \eq{pot}.  

The best limit is obtained in  the fine-structure spectroscopy of Helium. The  extra interaction between the two electrons   has been discussed in~\cite{Ficek:2016qwp} whose limits, obtained by the constraints from the $2^3 P_2$-$2^3 P_1$ transitions in $He$,  can be expressed as 
\be
 \frac{\Lambda^2}{\sqrt{\alpha_{\dd}} d_M^{\, e}}   \agt 872 \;\mbox{GeV}^2\, .
\ee

Bounds on long-range forces depending on spin set limits on the scale of the operator in \eq{dipole} based on the potential in \eq{pot} as discussed in \cite{Dobrescu:2006au}. The strongest bounds come from limits on macroscopic forces between electrons~\cite{Ni:1999di}
\be
 \frac{\Lambda^2 }{\sqrt{\alpha_{\dd} }d_M^{\,e}}   \agt 1.61 \; \mbox{TeV}^2\, , \label{pot_ee}
\ee
and electrons and nucleons~\cite{Wineland:1991zz} 
\be
\frac{\Lambda^2 }{\sqrt{\alpha_{\dd}} \sqrt{d_M^{\, e} d_M^{\, q}}}   \agt 1.94\; \mbox{TeV}^2\, . \label{pot_eN}
\ee
The limits among nucleons and electrons and protons are weaker.

Whereas the strong limits on the anomalous magnetic moments of the electron and the muon are traditionally used to set limits on new physics, they  cannot be used directly in our case because  they only apply to operators coupling to the visible photon.
The operator in \eq{dipole} enters in the computation of the magnet moments but only at higher order with two insertions in the loop computation. The limits are accordingly weak. The contribution of the dark photon to the anomalous magnetic moment is given by
\be
a_{\ell}^{A^\prime} =  -\frac{3}{2} \frac{\alpha_{\dd} }{\pi}\left(  \frac{m_{\ell} v_h d_M^{\ell}  }{\Lambda^2} \right)^2
\left[  \frac{5}{4}  + \log \frac{\mu^2}{m^2} \right] 
\ee
 in the $\overline{MS}$ scheme; contrary to the SM case, the result depends on the subtraction of a divergence. 
 
 We discuss below in section~\ref{sec:model} how in the UV model, where there are states coupled to both the dark and the visible photon,  the anomalous magnetic moment can be brought to bear directly on the limits. 

The  quantity $\Delta a_e$, the difference  between the experimental value of the electron anomalous magnetic moment~\cite{Hanneke:2008tm} and its SM prediction is very small. The  uncertainty on this difference  (at 1$\sigma$) is given by~\cite{Giudice:2012ms}
\be
\delta_{\Delta a_e} < 8.1 \times 10^{-13} \, . \label{limit-MME}
\ee
By requiring that the contribution of the dark photon does not exceed this value, and therefore does not contribute to the electron magnetic moment, we obtain
\be
 \frac{\Lambda^2}{\sqrt{\alpha_{\dd}}  d_M^{\, e}} \agt 0.075 \; \mbox{TeV}^2 \, ,
\ee
by taking the renormalization scale $\mu = m_e$. 

The  analogous quantity $\Delta a_\mu$, the difference  between the experimental value of the muon anomalous magnetic moment~\cite{Bennett:2006fi} and its SM prediction~\cite{Blum:2018mom}, is less   than 
\be
\Delta a_\mu  <  2.74  \times 10^{-9} \, , \label{limit-MMMU}
\ee
at $2 \sigma$ level, from which we derive
\be
 \frac{\Lambda^2}{\sqrt{\alpha_{\dd}}  d_M^{\, \mu}} \agt 0.5  \; \mbox{TeV}^2 \, ,
\ee
 for $\mu = m_\mu$. Notice that the current 3.2$\sigma$ discrepancy in $\Delta a_\mu$  could be explained by requiring
\be
 \frac{\Lambda^2}{\sqrt{\alpha_{\dd}}  d_M^{\, \mu}} \simeq  0.27  \; \mbox{TeV}^2 \, .
\ee 

Flavor changing processes can provide  constraints on possible dipole operator contributions to off-diagonal interactions. 

In the lepton sector the process  $\mu \rightarrow e X^0$, with $X^0$ a massless neutral boson, is bounded to~\cite{Bayes:2014lxz}   
\be
\BR (\mu \to e X^0) <  5.8 \times 10^{-5}  \, ,
\ee
which gives 
\be
\frac{\Lambda^2}{\sqrt{\alpha_{\dd}} d_M^{\, \mu e}}  \agt 5.1 \times 10^5 \;\mbox{TeV}^2\, .  
\ee

Similar limits in the hadron sector on, for example, the decays $K\to\pi X^0$ or $B\to K X^0$, cannot be used because they are forbidden when $X^0$ is a spin one boson like the dark photon. The decay $B\to K^* X^0$ is not forbidden but  gives a very weak bound.
Instead, the current limit on the rare decay $K^+\to \pi^+ \nu \bar \nu$ given by (at the 90\% CL, see, for example, \cite{Engelfried:2019pva})
\be
\BR(K^+\rightarrow \pi^+  \nu \bar \nu) < 1.85 \times 10^{-10} \label{Kpinunu}
\ee
 can be used, if we assume the dark photon to decay into light dark-sector fermions, and yields 
\be
\frac{\Lambda^2}{\sqrt{\alpha_{\dd}} d_M^{\, sd}}  \agt 9.5 \times 10^{6}  \;\mbox{TeV}^2\, ,
\ee
which is the strongest among all the limits on the dipole interaction we have discussed.

\vskip0.3cm
$\bullet$ \underline{Laboratory physics}. An interesting limit is derived by means, again, of the data from SN 1987A, this time indirectly from the counting of events in the Kamiokande detector. Axions from the star can, via inverse \textit{Bremsstrahlung}, excite the oxygen nuclei in the water tank as, in the process $a ^{16}\mbox{O} \rightarrow ^{16}\!\!\mbox{O}^*$, which subsequently decay producing $\gamma$ rays triggering the detector. The failure of observing these  extra events excludes the values for the coupling $\alpha^\prime_{aN}$~\cite{Engel:1990zd}
\be
6.5 \times 10^{-14} \leq \alpha^\prime_{a N} \leq 8.0 \times 10^{-8} \, , \label{kamio0}
\ee
which can be turned, taking the lower limit in \eq{kamio0}, in
\be
\frac{\Lambda^2}{\sqrt{\alpha_{\dd}} d_M^{\, q}}   \agt 1.9 \times 10^3  \; \mbox{TeV} ^2\, , \label{kamio}
\ee
for the massless dark photons. The limit in \eq{kamio} nicely closes the range left open by \eq{sn}.
A thin windows between \eq{sn1} and \eq{kamio0} is apparently left open for $\alpha^\prime_{a N} \simeq   10^{-14}$.

\vskip0.3cm
$\bullet$ \underline{Collider physics}. Limits from colliders are weaker but are worthwhile to be reported since they come from laboratory physics which is independent of all astrophysical assumptions. The process of pair annihilation into a dark and an ordinary photon provides a striking benchmark (mono-photon plus missing energy) for this search. It applies to electrons in searches at the LEP~\cite{Abbiendi:1998yu,Acciarri:1999kp,Heister:2002ut}:
\be
\frac{ \Lambda^2}{\sqrt{\alpha_{\dd}} d_M^{\, e}}   \agt 1.2 \; \mbox{TeV}^2\, , \label{lep}
 \ee
 and the first  generation of quarks at the LHC from CMS~\cite{Aaboud:2016uro} with luminosity of 35.9 fb$^{-1}$ (the ATLAS result~\cite{Sirunyan:2018dsf} is with smaller luminosity and less stringent):
 \be
\frac{ \Lambda^2}{\sqrt{\alpha_{\dd} }d_M^{\, q}}   \agt 4.3  \; \mbox{TeV}^2\, . \label{lhc}
 \ee
We computed the limits in \eq{lep} and \eq{lhc}  for this review by requiring that the  number of dark photon events be, bin by bin, less than the difference between the observed and the expected number of events.

\subsection{Can the massless dark photon be seen at all?}

The limits for the dark dipole of the massless dark photon, as summarized in Fig.~\ref{fig:massless1a} and  Fig.~\ref{fig:massless1b}, are  indeed very stringent. For an  effective scale $\Lambda$ around 1 TeV, for example, only values of dipole moments of  $O(10^{-6})$ for electrons and $O(10^{-5})$ for quarks are still allowed.  These are numbers making  detection  in an experiment very challenging. 

This does not mean that the massless dark photon cannot be searched for in the laboratory. We must look either to processes where SM particles heavier than the electron or the muon and the $u$ or $d$ quarks are involved---and the most severe astrophysical bounds do not apply---or  physics where the dipole operator   in \eq{dipoleH} is between fermions of different flavors or very high-energy processes where the large scale $\Lambda$ is partially compensated by the scaling of the  dipole and radius operators  in \eq{dipole} and  \eq{radius} and the overall contribution is less suppressed.

For example, for a first generation quark taken to be  a parton in a hadron collider, the limit at an energy scale of 10 TeV, is of $d_M^{\, u,d} \simeq 10^{-3}$ (see Fig.~\ref{fig:massless1b}) which would give a deviation in the cross section within the reach of future machines.  Similarly, for the electron, the limits in Fig.~\ref{fig:massless1a} show that a $d_M^{\, e} \simeq 10^{-6}$ is still allowed at the scale of 1 TeV and therefore accessible at future lepton colliders for the projected sensitivity. As much suppressed as these cross sections are, they are comparable with those of the case of the massive dark photon after the corresponding limits are taken into account (see section \ref{sec:con2}). 

These, and others possibilities, are discussed in section~\ref{sec:future1} where some of the proposed experiments to search for the massless dark photon are reviewed.

\section{Limits on milli-charged particles}
\label{sec:milli}

 \begin{figure}[t!]
\begin{center}
  \includegraphics[width=4.5in]{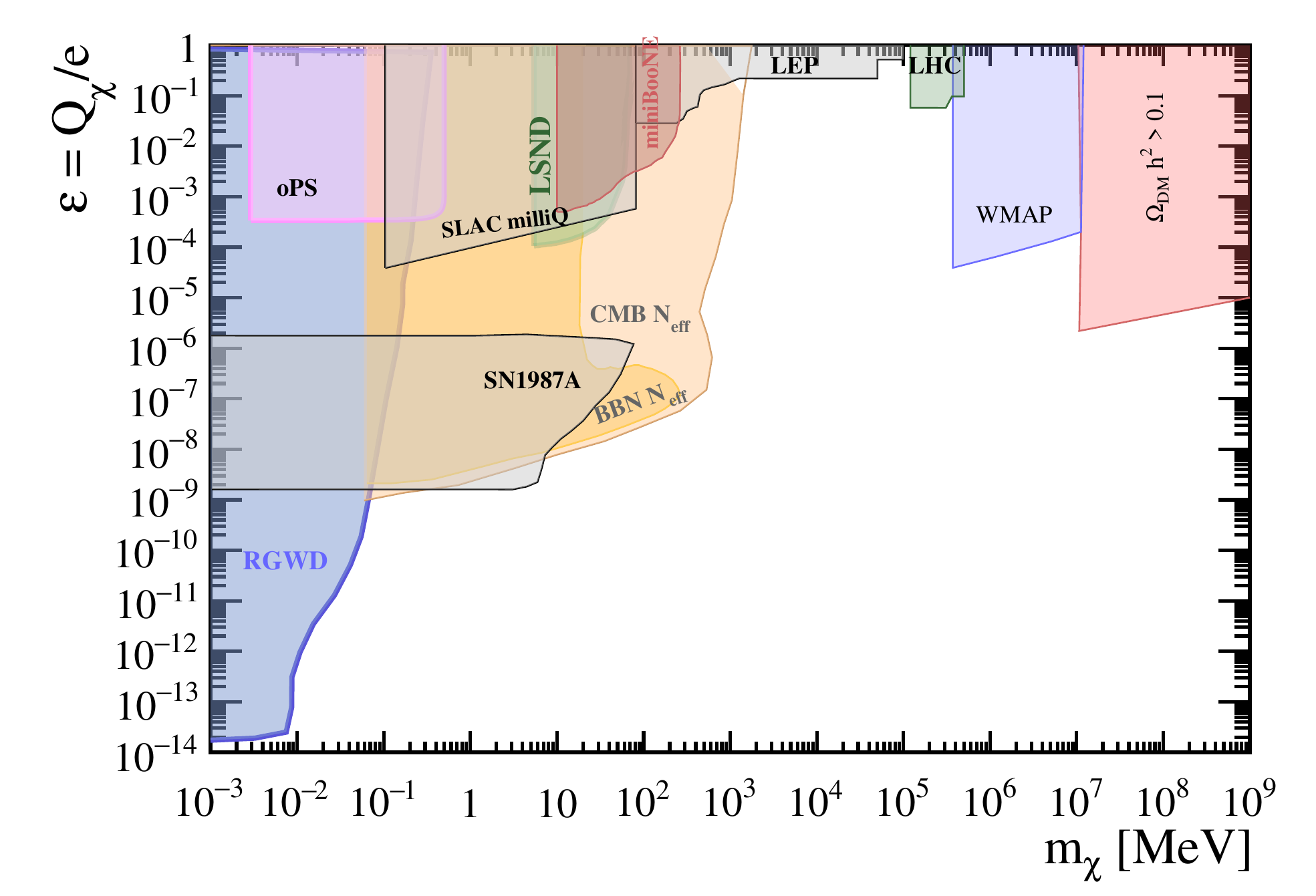}
  \caption{\small  \label{fig:milli1} Existing limits 
     for experiments 
    on milli-charged dark-sector matter.
    Limits from
    stellar evolution (RGWD~\cite{Vogel:2013raa} and SN1987A~\cite{Chang:2018rso});
    $N_{eff}$ during nucleosynthesis and in the cosmic microwave background~\cite{Vogel:2013raa};
    invisible decays of ortho-positronium (oPS)~\cite{Badertscher:2006fm};
    SLAC milliQ experiment~\cite{Prinz:1998ua};
    reinterpretation of data from LSND and miniBooNE~\cite{Magill:2018tbb};
    searches at LEP~\cite{Davidson:2000hf} and LHC~\cite{Jaeckel:2012yz};
    WMAP results and dark matter relic density abundance ~\cite{Jaeckel:2012yz}.    
     }
\end{center}
\end{figure}

 \begin{figure}[t!]
\begin{center}
  \includegraphics[width=4.5in]{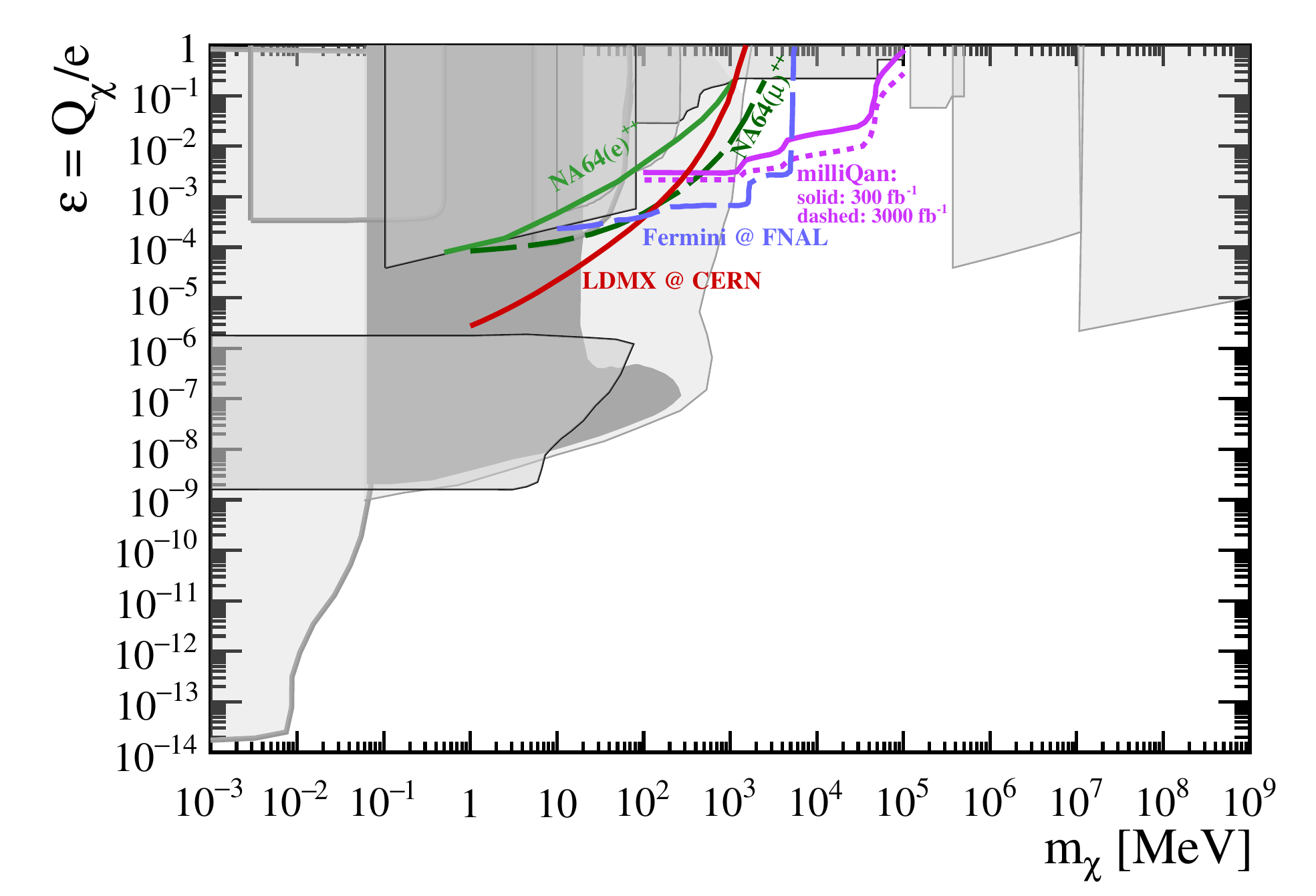}
  \caption{\small  \label{fig:milli2} Future sensitivities for proposed experiments 
    on milli-charged dark-sector matter.
 Future sensitivities of
    NA64(e)$^{++}$~\cite{NA64:eplus}; NA64($\mu$)~\cite{NA64:2018iqr}; FerMINI~\cite{Kelly:2018brz};
    milliQAN~\cite{Ball:2016zrp};LDMX~\cite{Akesson:2018vlm}. The sensitivity shown for LDMX @ CERN
    corresponds to $10^{16}$~electrons-on-target and a beam energy of 16~GeV. 
    The existing limits are shown as gray areas.
    The plot is revised from~\cite{Beacham:2019nyx}.
  }
\end{center}
\end{figure}

Milli-charged particles arise, as discussed in the section~\ref{sec:mass} of the Introduction, in the case of a massless dark photon because the rotation of the mixing term in \eq{kinetic} leaves the photon coupled to the dark sector particles $\chi$  with strength $\varepsilon e^\prime$. Searches are accordingly parameterized in terms of the mass $m_\chi$  and the electromagnetic coupling  (modulated by  $\varepsilon$) of the supposedly milli-charged dark-sector particle.

The physics of stellar evolution for
horizontal branches, red giants, and white dwarves (RGWD~\cite{Vogel:2013raa}), 
together with supernovae (SN1987~\cite{Chang:2018rso})
provide bounds in the region of small masses ($m_\chi \ltap 1$ MeV).
In this region constraints on $N_{eff}$ during nucleosynthesis  and in the
cosmic microwave background ($N_{eff}$ BBN and CMB~\cite{Vogel:2013raa})
limits the possibility of having  milli-charged particles.
These limits are derived along the same lines discussed in the case of the massless dark photon.

Further limits can be derived  from precision measurements in QED, notably
from the Lamb shift in the transition $2S_{1/2}$-$2P_{3/2}$ in the Hydrogen atom~\cite{Hagley:1994zz}
and the non-observation of the invisible decay of ortho-positronium (oPS~\cite{Badertscher:2006fm}).
Limits in the intermediate mass range $1-100$~MeV come from a SLAC dedicated
experiment (SLAC milliQ~\cite{Prinz:1998ua}) and from the reinterpretation of data from the neutrino
experiments LSND and miniBooNE~\cite{Magill:2018tbb}.

Searches at LEP~\cite{Davidson:2000hf} and LHC~\cite{Jaeckel:2012yz}
cover larger values of the mass (100 MeV $\ltap m_\chi \ltap 1$~TeV). 

Finally, for very large masses ($m_\chi \gtap 10$ TeV) the impact on the cosmological parameters severely
restricts the possible values of milli-charges
(WMAP and dark matter relic density constraint, ~\cite{Jaeckel:2012yz} and references therein).

All these limits are shown as filled area in the plot of Fig.~\ref{fig:milli1}.

Milli-charged particles as dark matter have been proposed
(see for example~\cite{Kovetz:2018zan} and~\cite{Liu:2019knx}) to
explain the anomalous 21~cm hydrogen absorption signal reported by the
EDGES experiment~\cite{Monsalve:2018fno}. Given the preliminary nature of the results, we have not included them in 
Fig.~\ref{fig:milli1}.

The projected limits of future experiments are depicted in Fig.~\ref{fig:milli2} 
together with the current limits in gray background to show the expected advances.
Of these, the most significative for masses around 1~GeV comes from the proposed milliQAN
experiment~\cite{Ball:2016zrp} proposed to be installed on the surface above one of the LHC
interaction points. MilliQAN could improve the collider limits by two orders of magnitude.
The range in mass between 10-100~MeV can be optimally covered by the FerMINI experiment~\cite{Kelly:2018brz}
proposed in the DUNE near detector hall at Fermilab. Finally the search for milli-charged particles 
below 10~MeV mass may be improved by almost two orders of magnitude by the LDMX experiment~\cite{Akesson:2018vlm}
proposed both at CERN~\cite{LDMX:CERN} and at SLAC~\cite{Raubenheimer:2018mwt}.

\section{ A  minimal model of the dark sector}
\label{sec:model}

As discussed in section~\ref{sec:intro},  it is useful to underpin the phenomenology of the massless dark photon to a UV model. We  consider a \textit{minimal} model   consisting of   dark  fermions that are, by definition, singlets under the SM gauge interactions. These dark fermions interact with the visible sector through a portal provided by scalar messengers which carry both SM and dark-sector charges.  These scalars  are phenomenologically akin to the sfermions  of supersymmetric models.

In general, we can have  as many  dark fermions as there are in the SM;  they can  be classified conveniently  according  to whether they couple (via the corresponding messengers) to quarks ($q_L$, $u_R$, $d_R$)  or leptons ($l_L$, $e_R$): We denote  the first (hadron-like) $Q$ and the latter (lepton-like) $\chi$. 
  
The Yukawa-like interaction Lagrangian for flavor-diagonal interactions can  be written as~\cite{Gabrielli:2013jka,Gabrielli:2016vbb}:
\bea
 \mathcal{L} &\supset & -g_L \left(\phi^\dag_L \bar{\chi}_R l_L + S_L^{U\dag} \bar{Q}^{\uu}_R q_L + S_L^{D\dag} \bar{Q}^{\dd}_R q_L\right) \nn \\
& -&  g_R \left(\phi^\dag_R \bar{\chi}_L e_R + S_R^{\uu\dag} \bar{Q}^{\dd}_L u_R + S_R^{\uu\dag} \bar{Q}^{\dd}_L d_R\right) + \text{H.c.} \label{LLRR} \, 
\eea
 where $q_L$ ($q_R$)  and $e_L$ ($e_R$) are $SU(2)_L$ doublets (singlets) for quarks and leptons respectively. Sum over flavor and color indices, that we omitted for simplicity, is understood.
The $L$-type scalars are doublets under SU(2)$_L$, while the $R$-type scalars are singlets under SU(2)$_L$. The $S_{L,R}$ messengers carry color indices (unmarked in (\ref{LLRR})), while the messengers $\phi_{L,R}$ are color singlets.  The Yukawa coupling strengths  are parameterized by $\alpha_{L,R}\equiv g_{L,R}^2/(4 \pi)$; they can be different for different fermions and as many as the  SM fermions. 
For simplicity, we take them to be equal and, in addition, $\alpha_L=\alpha_R$.

In order to generate chirality-changing processes, we must  have the mixing terms
 \be
\mathcal{L} \supset -\lambda_s S_0\left(H^\dag \phi_R^\dag \phi_L + \tilde{H}^\dag S_R^{\uu\dag} S_L^{\uu} + H^\dag S_R^{\dd\dag} S^D_L\right) + \text{h.c.} 
 \, , \label{mix} 
\ee
where $H$ is the SM Higgs boson, $\tilde{H}=i\sigma_2 H^\star$, and $S_0$ a scalar singlet of the dark sector. 
After both   $S_0$ and $H$   take a vacuum expectation value  ($\mu_S$  and $v_h$---the electroweak vacuum expectation value---respectively), the Lagrangian in \eq{mix} gives rise to  the mixing between right- and left-handed states.

 Dark sector and messenger states are both charged under an unbroken $U(1)_{\dd}$ gauge symmetry which is the same of  the corresponding massless  dark photon, with coupling strength $\alpha_{\dd}$. 
We  assign different dark $U(1)_{\dd}$ charges to the various dark sector fermions to ensures,  by charge conservation, their stability.  Since SM fields are neutral under  $U(1)_{\dd}$ interactions, messengers and associated dark-fermions field in Eq.(\ref{LLRR}) must carry the same  $U(1)_{\dd}$ quantum charge.

When the dark sector scalar $S_0$ and the Higgs boson acquire their vacuum expectation values,  the scalar messengers must be rotated to identify the physical states.  Before this rotation,
 $\phi_{L\nu}, S^U_{Ld}, $and $S^D_{Lu}$ are degenerate mass eigenstates with mass $m_S$.   After the rotation, 
the  mass eigenstates  (labeled by $\pm$)  are given by 
$
\phi_{\pm} \equiv \frac{1}{\sqrt{2}}\left(\phi_{L} \pm \phi_R\right) \,, ~~~
S^{U,D}_{\pm} \equiv \frac{1}{\sqrt{2}}\left(S^{U,D}_{L} \pm S^{U,D}_R\right) \, \, ,
$
corresponding to  masses 
  \be
  m_\pm = m_{\phi, S} \sqrt{1 \pm \eta_s}
  \ee
where we defined the mixing parameters for the $S$ and $\phi$ messengers
\be
\eta_{\phi, S} \equiv \frac{\lambda_s \mu_{\phi, S} v_h}{m_S^2}\, . \label{mixing}
\ee

In the new basis, the  interaction terms in \eq{LLRR} in the lepton sector  is given  by
\bea
\label{intdiag}\mathcal{L}^{(lep)} &\supset & -g_L \phi_{L\nu}^\dag \left(\bar{\chi}_R \nu_L\right) - \frac{g_L}{\sqrt{2}}\left(\phi_+^\dag + \phi_-^\dag\right)\left(\bar{\chi}_R e_L\right)\nn \\
&-& \frac{g_R}{\sqrt{2}}\left(\phi_+^\dag - \phi_-^\dag\right)\left(\bar{\chi}_L e_R\right) + \text{h.c.}  \, .
\eea
The corresponding interaction terms in the hadronic sector have the same form. 

Looking at (\ref{intdiag}), we can see that if $\chi$ is a stable dark-sector species, then its mass must be at most $m_- + m_e$. Similarly, for a dark-sector species $Q$, the mass must be no heavier than $m_- + m_q$, where $m_q$ is the mass of the SM species corresponding to $Q$. This sets an upper bound for the mixing $\eta_{\phi, S}$:
\be
\label{mixinglimit}\eta_{\phi, S} < 1 - \left(\frac{M}{m_{\phi, S}}\right)^2  \, . 
\ee
In \eq{mixinglimit}, $M$ is the mass of the heaviest stable dark-sector species. We  assume that $M$ is heavier than any SM species. The upper bound in \eq{mixinglimit} also guarantees that the scalar messengers are heavier than the dark fermion into which they can thus decay.

This model
can be considered as a template for many models of the dark sector with the scalar messenger as stand-in for more complicated portals.  It is 
a simplified version of the model in \cite{Gabrielli:2013jka}, which might  provide a natural solution to the SM flavor-hierarchy problem.

 The discussion above is restricted to the flavor-diagonal interactions. A more general flavor structure in the portal interaction, including the off-diagonal terms, arising as a consequence of the simultaneous diagonalization of the dark-fermion mass and quark interaction basis, can be simply obtained by generalizing the above terms as follows \cite{Gabrielli:2016cut}
\bea
S^{\uu_i\dag}_L\bar{Q}^{\uu_i}_Rq^i_L &\to& S^{\uu_i\dag}_L\bar{Q}^{\uu_i}_R(\rho_L^{\uu})_{ij}q^j_L\nonumber\\
S^{\uu_i\dag}_R\bar{Q}^{\uu_i}_Lq^i_R &\to& S^{\uu_i\dag}_R\bar{Q}^{\uu_i}_L(\rho_L^{\uu})_{ij}q^j_R\, ,
\label{subst}
\eea
and analogously for the down and lepton sectors, where $i,j$ are explicit flavor indices and sum over $i,j$ is understood. 

\begin{figure}[t!]
\begin{center}
\includegraphics[width=3.4in]{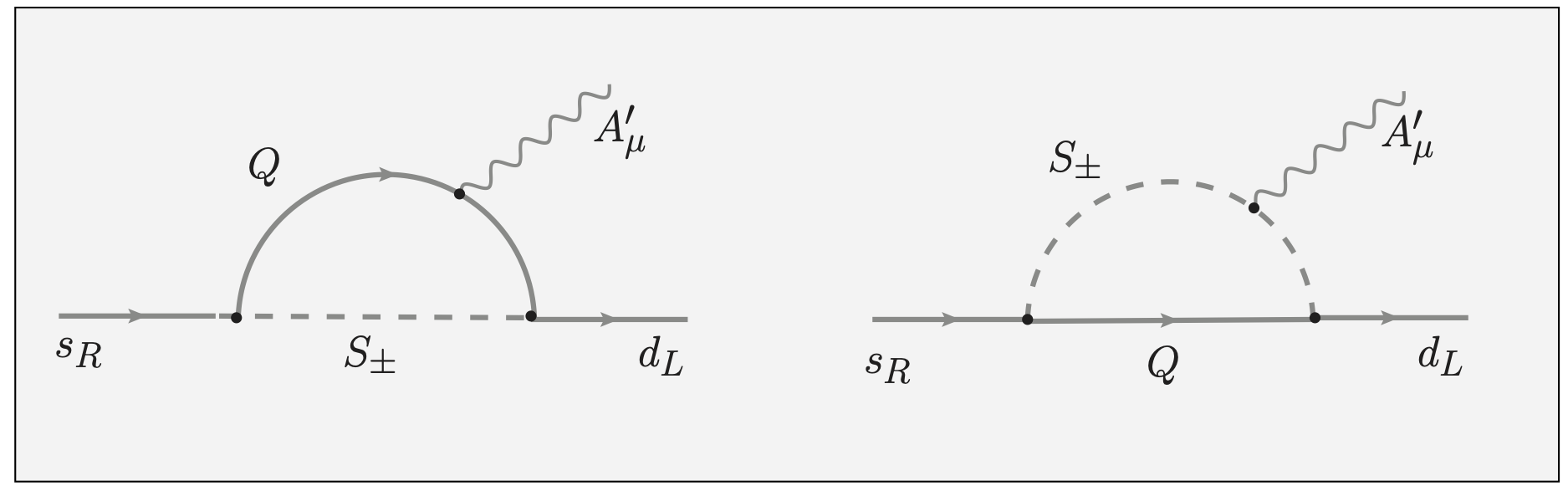}
\caption{\small Vertex diagrams for the generation of the dipole operators in the  model of the dark sector. 
\label{vertex} }
\end{center} 
\end{figure}

To keep the contribution to the dipole coefficient simple, lest the generality obfuscates the estimate, we follow the guidelines of the model in \cite{Gabrielli:2013jka}.
We assume that the masses of the messengers $\phi^i$, $S^{\uu, i}_{L,R}$ and $S^{\dd, i}_{L,R}$ are the same and  the mixing matrices $\rho_{ij}$ have a hierarchical structure (like in the SM) with the off-diagonal  smaller than the diagonal terms. The former hypothesis is a consequence of the $SU(N_F)$ flavor symmetry in the free lagrangian of messenger sector (with $N_F=6$)) 
\cite{Gabrielli:2013jka}, while the latter follows from the requirement of minimal flavor violation hypothesis \cite{Gabrielli:2016vbb}.

We also take $\rho_{ij}\equiv \rho^{\dd}_{ij} = \rho^{\uu}_{ij}$. This way, the loop of dark sector particles is dominated by the contribution with  the heaviest dark fermion coupled to the SM fermions of flavor $i$ and $j$  with one coefficient off-diagonal $\rho_{ij}$ and one diagonal $\rho_{ii}$. In the following, in order to distinguish the contribution from the up and down sector couplings we will use the notation $\rho_{uu}\equiv \rho^{\uu}_{11}$, $\rho_{dd}\equiv \rho^{\dd}_{11}$,  $\rho_{sd}\equiv \rho^{\dd}_{21}$, and similarly for the other coefficients.

Matching the model to the effective Lagrangian   given in \eq{dipole} after  integrating the loop, and identifying the scale $\Lambda$ as
\be
\frac{v_h}{ \Lambda^2} \simeq \frac{m_{Q^i} }{m^2_S} \, ,
 \ee
  with $m_{Q^i}$ the heaviest dark-fermion running in the loop, we can
re-express
 the magnetic dipole  explicitly in terms of the parameters of the model. For example, in  the case of the generic (quark) flavor transition from $i\to j$, with $i$- and $j$  mixing,  neglecting the SM masses, according to the Lagrangian in (\ref{LLRR}) and substitutions (\ref{subst}), we have  \cite{Gabrielli:2016cut}
\be
\mathbb{D}_M^{ij} = \rho_{jj} \rho_{ij}^* \, \Re \left[\frac{g_L g_R}{(4 \pi)^2}\right]  F_M(x,\eta_s)    \, . \label{ds}
\ee
where $x=(m_{Q^i})^2/m_S^2$ and $\eta_s$ the mixing parameter defined in (\ref{mixing}). In the following, we will introduce the notation of $m_{S^{\uu}}$ and
$m_{S^{\dd}}$ to distinguish the common messenger mass in the up and down $SU(2)_L$ sectors respectively, and  $\eta^{\uu,\dd}_s$ for the corresponding mixing parameters. The function $F_M(x,y)$ is given by~\cite{Gabrielli:2016cut}
\be
F_M(x,y)=\frac{1}{2} \Big[ f(x,y)-f(x,-y) \Big] \, , \label{F}
\ee
where
\be
f(x,y)=\frac{1-x+y+(1+y)\log{\left(\frac{x}{1+y}\right)}}{
  \left(1-x+y\right)^2}\, .
\ee
CP-violating phases, relevant for flavor  changing processes,  can arise from the mixing parameters. For instance, in the $n\to m $ flavor transition, we can have CP-violating phase $\delta_{\mbox{\tiny CP}}$ from the relation 
\be
 \rho_{nm} \rho_{mm}^* - \rho_{nm}^* \rho_{mm} = 2\,  i \sin \delta_{\mbox{\tiny CP}} \, . \label{phase}
 \ee

\subsection{Constraints on the UV model parameters}

The introduction of the UV model  makes possible to re-discuss the bounds of section~\ref{sec:con1} on the massless dark photon in terms of the parameters of the model. 

 There are no laboratory limits for the masses of the dark fermions from events in which they are produced because they are SM singlets and do not interact directly with the detector. 
  Cosmological bounds have been considered in~\cite{Acuna:2020ccz} where, in particular,  avoiding distortions of the cosmic microwave background is shown to require the masses of the dark fermions to be larger than 1 GeV or, if lighter, that the coupling $\alpha_L$ and $\alpha_R$  be less than $10^{-3}$.

The messenger states have the same quantum numbers and spin as the supersymmetric squarks.
At the LHC they are copiously produced in pairs through QCD interactions and decay at tree level into a quark and a dark fermion. The final state arising from their decay is thus the same as the one obtained from the $\tilde q \to q \chi^0_1$ process.
Therefore limits on the messenger masses can be obtained by reinterpreting supersymmetric searches on first and second generation squarks decaying into a light jet and a massless neutralino~\cite{Aaboud:2017vwy}, assuming that the gluino is decoupled. 
A lower bound on their masses is thus obtained~\cite{Barducci:2018rlx} to give
\be
m_S^i \gtap 940 \; \mbox{GeV}\, ,
\ee
for the messenger mass related to the dark fermions $Q^U$ and $Q^D$. This limit
 increases up to 1.5 TeV by assuming that messengers of both chiralities associated to the first and second generation of SM quarks are degenerate in mass.

For the masses of the lepton-like scalar messengers,  constraints on the mass of sleptons~\cite{Sirunyan:2018nwe}  give the following lower bound on the messenger mass in the lepton sector:
\begin{eqnarray}
m_\phi \gtap \; 290 \; \mbox{GeV}.
\end{eqnarray}

All the limits discussed in section~\ref{sec:con1} can be re-expressed in terms of the UV model parameters.

For example, the limit from  stellar cooling in \eq{stars} becomes
\be
\frac{m_\phi^2/m_{\chi^e} }{\sqrt{\alpha_{\dd} \alpha_L \alpha_R} \, |\rho_{ee}|^2| \, F_M (x_{e},\eta_{\phi})}\gtap 2.1 \times 10^6\; \mbox{TeV}\, ,
\ee
where $x_{e}=m_\phi^2/m_{\chi^e}^2$, with $m_{\chi^e}$ the dark fermion mass
associated to the electron, and  $\eta_{\phi}$ the corresponding mixing parameter in the colorless messengers sector, and the loop function $F_M(x,y)$ is given in \eq{F}. This limit, which is obtained by rescaling the right-hand side of \eq{stars} for $1/(4\pi v_h)$, applies specifically to the Yukawa coupling of electrons and the corresponding messenger state. 

For a quick estimate of the bound above and those that follow, the loop function $F_M(x,y)$ can be considered a coefficient  of order $O(10^{-1})$  as long as  $\eta_\phi$ is not too small. For instance, for $x \simeq 1$ and $y \simeq 0.5$, the loop function $F_M \simeq 0.09$.

Similarly, by using the same rescaling factor, the neutrino signal of supernova 1987A and the limit in \eq{sn} yields now
\be
\frac{m_{S}^2/m_{Q^u}}{\sqrt{\alpha_{\dd} \alpha_L \alpha_R} \, |\rho_{uu}|^2\, F_M (x_{u},\eta_{S})}\gtap 2.0 \times 10^5\; \mbox{TeV}\, ,
\ee
where now $x_u=m_{S}^2/m_{Q^u}^2$, with $m_{Q^u}$ the dark-fermion associated to the light $u$ quark. A similar limit holds for the case of the $d$ quark sector. 

The others bounds in section~\ref{sec:con1} can be written in terms of the parameters of the model in the same way.

Instead, new bounds can be set  now that we have un underlying UV model because  the scalar messengers carry also the electromagnetic charge. Processes with the visible photon can thus be used; these processes  were not available for the model-independent case in section~\ref{sec:con1} for which only the coupling to the dark photon was  taken into account. 

The magnetic moment of the SM fermions arises from the one-loop diagram of the states of the UV model. 

From \eq{limit-MME} in section~\ref{sec:con1}, we find
 \be
 \frac{m_\phi^2/m_{\chi^{e}}}{\sqrt{\alpha_L \alpha_R}\, |\rho_{e e}|^2\,  G_M (x_{e},\eta_{\phi})} \gtap 9.8 \times 10^4  \; \mbox{TeV}\, ,
 \label{mu1}
 \ee 
 where $x_{e}=m_{\chi^{e}}^2/m_{\phi}^2$, with $m_{\chi^{e}}$ the dark-fermion mass associated to the muon. The loop function is in this case given by
\cite{Gabrielli:2016cut}
\be
G_M (x,y)=\frac{1}{2}\Big[ g (x,y)- g (x,-y) \Big]\, ,\ee
where
 \be
 g  (x,y)=\frac{(1+y)^2-x^2+2x\left(1+y\right)
   \log{\left(\frac{x}{1+y}\right)}}{2\left(x-1-y\right)^3}\, .
 \ee

Also  interesting is the anomalous magnetic moment of the muon because of the lingering discrepancy between theory and experiments.  From \eq{limit-MMMU} in section~\ref{sec:con1}, we find
 \be
 \frac{m_\phi^2/m_{\chi^{\mu}}}{\sqrt{\alpha_L \alpha_R}\, |\rho_{\mu\mu}|^2\,  G_M (x_{\mu},\eta_{\phi})} \gtap 6.3 \times 10^3\; \mbox{TeV}\, ,
 \label{mu2}
 \ee 
 where $x_{\mu}=m_{\chi^{\mu}}^2/m_{\phi}^2$, with $m_{\chi^{\mu}}$ the dark-fermion mass associated to the muon.  
Again, for a quick estimate of the bounds above and those that follow, the loop function $G_M(x,y)$ can be considered a coefficient  of order $O( 10^{-1})$  as long as  $\eta_\phi$ is not too small. For instance, for $x \simeq 1$ and $y \simeq 0.5$, the loop function $G_M \simeq 0.05$.

The various Yukawa couplings and messenger and fermion masses are  probed in a selective manner in flavor physics where  we must distinguish among the various couplings and states.  Mixing (proportional to a coefficient $\rho_{ij}$ in the equations below) between different flavor states must be included. 

The strongest bound comes from the limit on the $\BR(\mu \rightarrow e \gamma) < 4.2\times 10^{-13}$ (CL 90\%)~\cite{TheMEG:2016wtm} of the MEG experiment. From this result, we find that
 \be
 \frac{m_\phi^2/m_{\chi^{\mu}}}{\sqrt{\alpha_L \alpha_R} \, |\rho_{\mu\mu}\rho^{\star}_{\mu e}|\,
  G_M (x_{\mu},\eta_{\phi})}\gtap 4.9\times 10^8\; \mbox{TeV}\, .
 \label{muegamma}
 \ee  

A weaker bound can be extracted, in the hadronic sector, from the difference between the experimental limit on the  
$\BR(B \to X_s \gamma) < (3.21 \pm 0.33) \times 10^{-4}$~\cite{Lees:2012ym} of the BaBar collaboration and its SM estimate~\cite{Misiak:2006zs}. It yields
 \be 
 \frac{m_{S}^2/m_{Q^b}}{\sqrt{\alpha_L \alpha_R} \, |\rho_{bb}\rho^{\star}_{bs}|\,
   G_M (x_{b},\eta_{S})}\gtap 1.3\times 10^4\; \mbox{TeV}\, ,
 \label{bsgamma}
 \ee
where $x_b=m^2_{Q^{b}}/m_{S}^2$,
 with $m_{Q^{b}}$ the mass of dark fermion associated to the $b$-quark,

The limits in  \eq{muegamma} and \eq{bsgamma} apply specifically to the off-diagonal terms in the  Yukawa couplings $\rho_{ij}$ of the muon-electron and $b$-$s$ quark mixing respectively, and to the corresponding mass of messenger states. 

The mass mixing in the Kaon system~\cite{Gabrielli:2016cut,Fabbrichesi:2017vma}  gives a further limit
\be
\frac{m^2_{S}}{\left(\alpha_L^2+\alpha_R^2\right)\left|\rho_{ss}\rho_{sd}^*\right|^2}\gtap 3\times 10^5\; \mbox{TeV}^2 \, , \label{sd}
 \ee
 which is not related to the dark photon and its coupling $\alpha_{\dd}$  because it comes from the box-diagram insertion of the dark scalars and fermions.
 
The  limit in \eq{sd} is obtained by requiring that the messenger contribution to the box diagram for the $K^0$-$\bar{K}^0$ mixing does not exceed the experimental value of the mixing parameter $\Delta m_K= 3.48\times 10^{-12}{\rm MeV}$~\cite{Tanabashi:2018oca}. 
 Due to chirality arguments, the leading contribution to the box diagram in \eq{sd}  does not depend on the dark fermion mass, which is assumed to be much smaller than the corresponding messenger mass in the down sector and therefore very weakly on the loop function. 
 
  The limit in \eq{sd} 
  applies specifically to the off-diagonal term in the Yukawa coupling of $d$-$s$ quark mixing and the corresponding messenger state. A similar but weaker bound can be found from $B$-meson mixing.

 As displayed in the equations above, all these limit  can be made weaker by taking $m_\chi$ (or $m_Q$) sufficiently light or by varying the corresponding mixing parameters $\eta_{s}$, $\eta_{\phi}$.  In the UV model is thus possible to play with the parameters to make room for larger values of the dipole coefficient by absorbing part of the suppression in the connection between the scale $\Lambda$ and the mass ratios $m_\chi/m_\phi^2$ and $m_Q/m_S$. For instance a scale $\Lambda=1$ TeV for the new physics of the dark sector is still allowed by the stringent bound in \eq{limit-MMMU} if we take $m_\chi$ sufficiently small. This way, there is some additional freedom in comparing limits from different processes as compared to the model-independent  case  where the scale $\Lambda$ is taken to be the same for all bounds.

\section{Future  experiments}
\label{sec:future1}
The massless dark photon has been neglected so far from the experimental point of view as compared to the massive one.
It is one of the aims of the present review to boost the community scrutiny in this direction.
In the past few year several proposals have been put forward and new experiments are in the planning:
\begin{itemize}

\item \underline{Flavor physics}: This is one of the most promising areas for searching for the dark photon and the dark sector in general because none of the stringent astrophysical constrains discussed in section~\ref{sec:con1} applies given the flavor off-diagonal nature of the dipole operator in these cases.

Proposals exist for processes in Kaon physics
  at NA62~\cite{NA62:2017rwk}. The Kaon decay $K\to \pi A^\prime$ is forbidden by the conservation of angular momentum but the decay $K^+ \to \pi^0 \pi^+ A^\prime$ is allowed and the estimated branching ratio~\cite{Fabbrichesi:2017vma} is within reach of the current sensitivity.  The rare decays $K^+\to \pi^+ \nu \bar \nu$~\cite{CortinaGil:2018fkc} and $K_L \to \pi^0 \nu \bar \nu$~\cite{Ahn:2018mvc} are  other two processes where the physics of the dark photon can play a crucial role~\cite{Fabbrichesi:2019bmo}. Also Hyperion decays can be used for detecting the production of $A^\prime$~\cite{Su:2019ipw} and in the decay of charmed hadrons~\cite{1795128}  and BESIII. 
  
 In addition, decays into invisible states of $B$-mesons at BaBar~\cite{Lees:2012wv} and Belle~\cite{Hsu:2012uh} and $K_{L,S}$ and other neutral mesons at NA64~\cite{Gninenko:2015mea,Gninenko:2014sxa} can be used to study the dark sector (assuming the invisible states belong to it). These decays are greatly enhanced by the Fermi-Sommerfeld~\cite{sommerfeld,fermi} effect due to their interaction with the dark photon---the same way as ordinary decays, like the $\beta$-decay, are enhanced by the same effect---making this another exciting area for searching the dark sector~\cite{Barducci:2018rlx}.

\item \underline{Higgs and $Z$ physics}: The striking signature of a mono-photon plus missing energy can be used to search Higgs~\cite{Gabrielli:2014oya,Biswas:2015sha,Biswas:2017lyg} and $Z$-boson~\cite{Fabbrichesi:2017zsc,Cobal:2020hmk} decay into a visible and a dark photon. Again,  the stringent astrophysical constrains discussed in section~\ref{sec:con1} do not apply because the size of the dipole operator is dominated (in the loop diagram) by the heavy-quark contribution's giving raise to the coupling to the dark photon, as discussed in section~\ref{sec:model}.

\item \underline{Pair annihilation}: Collider experiment at higher energies and luminosities can use the same  striking signature of a mono-photon plus missing energy  to  search for the dark photon. Even though the dipole interaction is suppressed and severely constrained in this case by the astrophysical and cosmological bounds discussed in section~\ref{sec:con1}, it is no more suppressed than the equivalent cross sections for the massive case. Moreover, the dipole operator scales as the center-of-mass energy in the process and higher  energies make it more and more relevant;  

\item \underline{Magnons}: An interesting possibility is the use of magnons in ferromagnetic materials and their interaction with dark photons (QUAX proposal)~\cite{Barbieri:2016vwg,Chigusa:2020gfs}. The estimated sensitivity is again done for axions but can be translated for massless dark photons as in the discussion about stars above.

\item \underline{Astrophysics}: Gravitation waves emitted during the inspiral phase of neutron star collapse can test the presence of other forces beside gravitation. Dipole radiation by even small amount of charges on the stars modifies the energy emitted;  the dark photon is a prime candidate for this kind of correction~\cite{Croon:2017zcu,Alexander:2018qzg,Kopp:2018jom,Fabbrichesi:2019ema}.
\end{itemize}


\chapter{Phenomenology of the massive dark photon}
\label{sec:massive}

\vspace*{1cm}

{\versal The phenomenology of the massive dark photon}  is  discussed in terms of its interaction with the SM particles, as given by \eq{Lmassive}:
\be
\boxed{
{\cal L} = -\varepsilon e J^\mu A^\prime_\mu \, , \label{current}
}
\ee
where $J^\mu$ is the electromagnetic current. The strength of this interaction is modulated by the parameter $\varepsilon$. The parameter space for the experimental searches is  given by the mass of the dark photon  $m_{A^\prime}$ and the mixing parameter $\varepsilon$.

\subsection{Production, decays and detection}

Because the current in \eq{current} is the same as the usual electromagnetic current, dark photons $A^\prime$ can be produced like ordinary photons. The main production mechanisms are:
\begin{itemize}
\item[-] \underline{\textit{Bremsstrahlung}}. The incoming electron scatters off the target nuclei ($Z$), goes off-shell and can thus emit the dark photon: $e^- Z \rightarrow e^- Z A^\prime$;
\item[-] \underline{Annihilation}: An  electron-positron pair annihilates into an ordinary and a dark photon:  $e^-\, e^+ \rightarrow \gamma A^\prime$
\item[-] \underline{Meson decays}: A meson $M$ (it being a $\pi^0$  $\eta$, or a $K$ or a $D$) decays as $M \rightarrow \gamma A^\prime$;
\item[-] \underline{Drell-Yan}: A quark-antiquark pair annihilates into the dark photon, which then decays into a lepton pair (or hadrons): $q \bar q \rightarrow A^\prime (\rightarrow \ell^+ \ell^-$ or $h^+ h^-)$. 
\end{itemize}
Different experiments use different production mechanisms and, sometime, more than one simultaneously.

 \begin{figure}[t!]
\begin{center}
\includegraphics[width=3in]{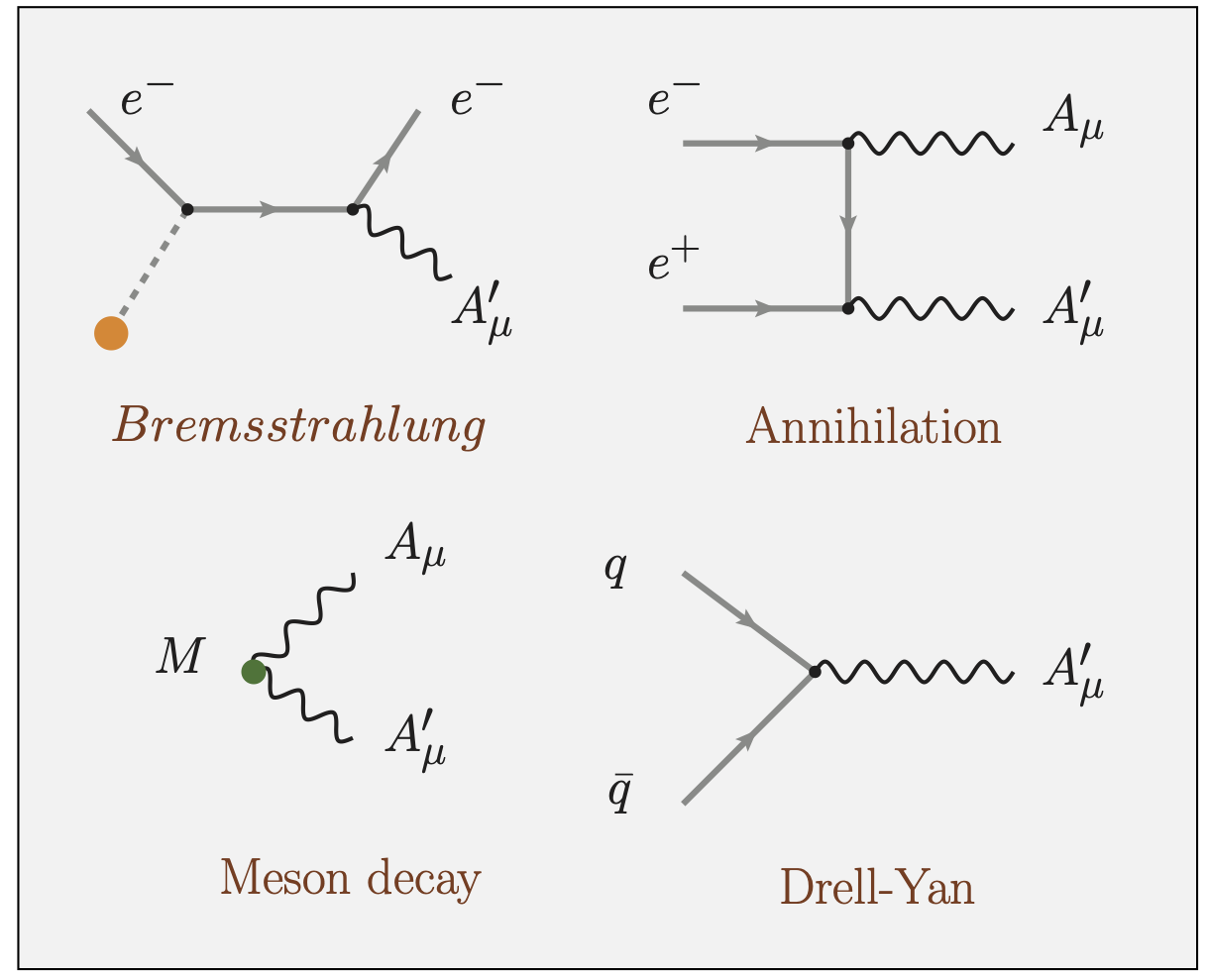}
\caption{\small 
\label{fig:production} Production of dark photons:  \textit{Bremsstrahlung}, Annihilation, Meson decay and Drell-Yan.}
\end{center}
\end{figure}

Detection of $A^\prime$ is based on its decays modes.
The  decay width of the massive dark photon $A^\prime$ into SM leptons $\ell$ is 
\be
\Gamma ( A^\prime \rightarrow \ell^+ \ell^-) = \frac{1}{3} \alpha \,\varepsilon^2 m_{A^\prime} \sqrt{ 1 - \frac{4 m_\ell^2}{m_{A^\prime}^2}} \left(1 + \frac{2 m_\ell^2}{m_{A^\prime}^2 } \right) \, ,
\ee
 which is only open for $m_{A^\prime} > 2 m_e$. Similarly, the width into hadrons is
\be
\Gamma ( A^\prime \rightarrow \mbox{hadrons}) =  \frac{1}{3} \alpha \varepsilon^2  m_{A^\prime} \sqrt{ 1 - \frac{4 m_\mu^2}{m_{A^\prime}^2}} \left(1 + \frac{2 m_\mu^2}{m_{A^\prime}^2 } \right)  R \, ,
\ee
where $R\equiv \sigma_{e^+e^- \rightarrow \mbox{\tiny had}}/\sigma_{e^+e^- \rightarrow \mu^+ \mu^-}$.

Since all visible widths are proportional to $\varepsilon$, the branching ratios are independent of it.

At accelerator-based experiments, several  approaches can be pursued to search
for dark photons depending on the characteristics of the available beam line and the detector.
These can be summarized as follows:

\begin{itemize}
\item[-] \underline{Detection of visible final states:}
  dark photons with masses above $\sim 1~$MeV can decay to visible final states.
  The detection of visible final state is a technique mostly used in beam-dump and collider experiments,
  where typical signatures are expected to show up as narrow resonances over an irreducible background.
  Collider experiments are typically sensitive to larger values of  $\varepsilon$ ($\varepsilon > 10^{-3}$) than
  beam dump experiments which typically cover couplings below $10^{-3}$.
  The use of this technique requires  high  luminosity  colliders  or  large  fluxes  of  protons/electrons  on  a  dump because the dark photon detectable rate is proportional to the fourth power of the coupling involved, $\varepsilon^4$, and so very suppressed for very feeble couplings.

  The  smallness  of  the  couplings  implies  that  the  dark photons  are  also very  long-lived (up  to   0.1  sec)  compared to  the  bulk of  the  SM particles.   Hence: The decays to SM particles can be optimally detected using experiments with long decay volumes followed by spectrometers with excellent tracking systems and particle identification capabilities.
  
\item[] \underline{Missing momentum/energy techniques:}
  invisible decay of dark photons can be detected in fixed-target reactions as, for example, 
  $e^- Z \to e^-Z A'$ ($Z$ being  the nuclei atomic number) with $A' \to  \chi \overline{\chi}$ and $\chi$ being a putative dark matter particle,
  by measuring the missing momentum or missing energy carried away from the escaping invisible particle or particles.
  The main challenge for this approach is the very high background rejection that must be achieved,
  which relies heavily on the detector being hermetically closed and, in some cases, on the exact knowledge of the
  initial and final state kinematics.
  
  These techniques guarantee an intrinsic better sensitivity for the same luminosity than the technique based on the detection of dark photons decaying to visible final states,  as it is independent of the probability of decays and therefore scales only as the SM-dark photon
  coupling squared, $\varepsilon^2$.

\item[-]  \underline{Missing mass technique:}

This technique is mostly used to detect invisible particles (as DM candidates or particles with very long lifetimes) in reactions with a well-known initial state, as for example, at $e^+ e^-$ collider experiments using the process $e^{+} e^{-} \to A' \gamma$, where $A'$ is on shell, using the single photon trigger.

Characteristic signature is the presence of a narrow resonances  emerging over a smooth background in the distribution of the missing mass. 
  
   It requires detectors with very good hermeticity that allow to detect all the other particles
   in the final state. Characteristic signature of this reaction is the presence of a narrow
   resonance emerging over a smooth background in the distribution of the missing mass.
   The main limitation of this technique is the required  knowledge of the background arising from processes
   in which particles in the final state escape the apparatus without being detected.

\end{itemize}

\subsection{Visible and invisible massive dark photon}

 \begin{figure}[t!]
\begin{center}
\includegraphics[width=3in]{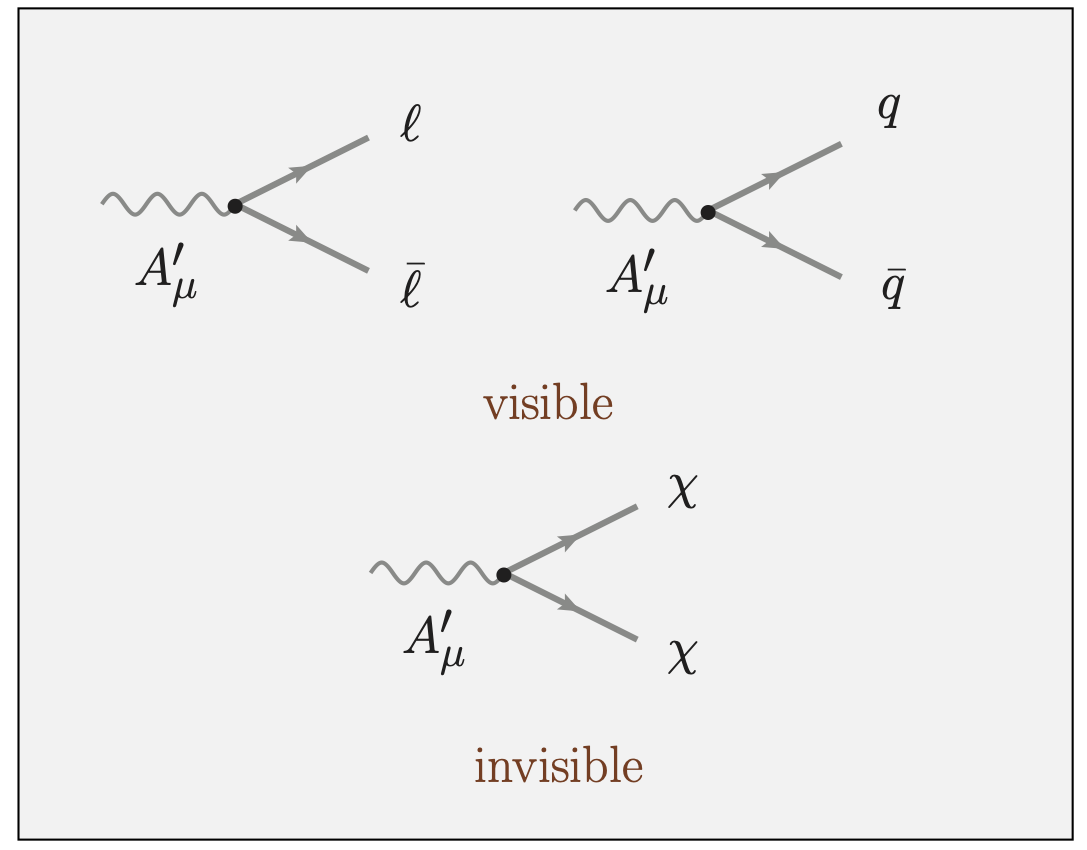}
\caption{\small 
\label{fig:decay} Decay of the massive dark photon into visible (SM leptons or hadrons) and invisible (DM) modes}
\end{center}
\end{figure}

In collecting the limits on the parameters of massive dark photon  is important to distinguish two cases accordingly on whether its mass is smaller or larger than twice the mass of the electron, the lightest charged SM fermion.

The dark photon is \textit{visible} if its mass is $M_{A^\prime} > 2 m_e \simeq 1$ MeV  because it can decay into SM charged states which leave a signature in the detectors. We discuss the  limits on the visible dark photon in section~\ref{sec:con2}. 

In the same regime for which  $M_{A^\prime} > 1$ MeV, however, the massive dark photon  could also    decay into dark sector states if their masses are light enough.  In this case we have a non-vanishing branching ratio into invisible final states. 
The invisible decay into these states of the dark sector $\chi$ in  given by
\be
\Gamma ( A^\prime \rightarrow \chi \bar \chi) = \frac{1}{3} \alpha_{\dd} \, m_{A^\prime} \sqrt{ 1 - \frac{4 m_\chi^2}{m_{A^\prime}^2}} \left(1 + \frac{2 m_\chi^2}{m_{A^\prime}^2 } \right) \, .
\ee
Dark photons decays into this invisible channel if $m_{A^\prime} > 2 m_\chi$; this channel dominates if $\alpha_{\dd}  \gg \alpha \varepsilon^2$.

Most of the experimental searches with dark photon in visible decays assume that the dark-sector states are not kinematically accessible and the dark photon is \textit{visible} only through its decay into SM states.
The limits need to be re-modulated if the branching ratio into invisible states is numerically significant or even dominant.
We discuss this case in section~\ref{sec:con3} below.

If the mass of the dark photon is less than 1 MeV, it cannot decay in any known SM charged fermion and its decay is therefore completely \textit{invisible}.  The experimental searches for dark photon into invisible final states are based on the energy losses that the production of dark photons, independently of his being stable or decaying into dark fermions,  implies on astrophysical objects like stars or in signals released in direct detection dark matter experiments.
The experimental limits in the case of the invisible dark photon are discussed in section \ref{sec:con4} below.

\section{Limits  on the parameters $\varepsilon$ and $m_{A^\prime}$}

As discussed, the space of the parameters  (the mixing $\varepsilon$ and the mass $m_{A^\prime}$  of the dark photon) is best spanned in two regions according on whether the mass $m_{A^\prime}$ is larger or smaller than twice the mass of the electron: Roughly 1 MeV.

\subsection{Constraints for $m_{A^\prime} > 1$ MeV with $A'$ decays to visible final states}
\label{sec:con2}

 \begin{figure*}[t!]
\begin{center}
  \includegraphics[width=4.7in]{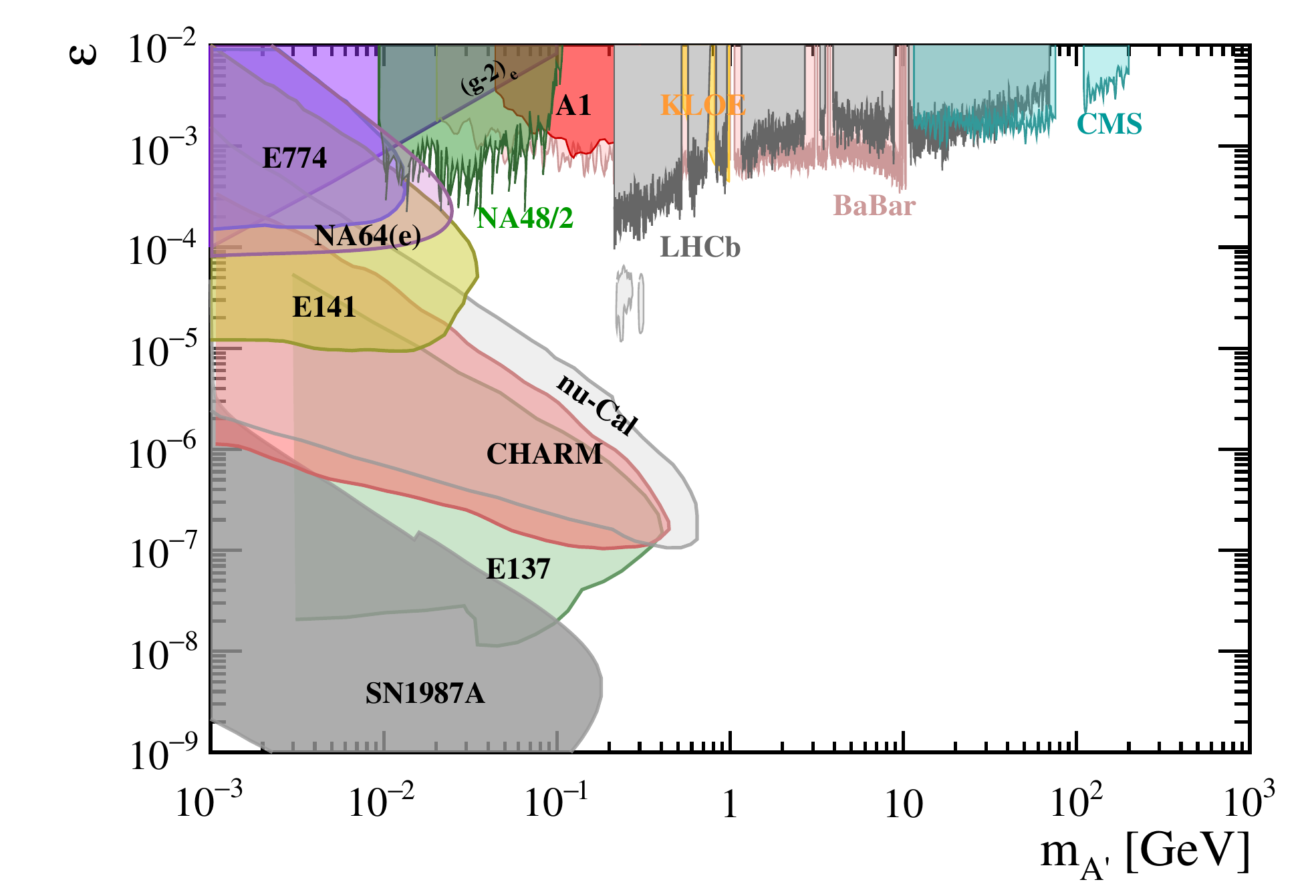}
\caption{\small 
  \label{fig:massive11}
  Existing limits on the massive dark photon for $m_{A^{\prime}} >1$ MeV from di-lepton
  searches at experiments at collider/fixed target (A1~\cite{Merkel:2014avp}, LHCb~\cite{Aaij:2019bvg},
  CMS~\cite{CMS:2019kiy},
  BaBar~\cite{Lees:2014xha}, KLOE~\cite{Archilli:2011zc,Babusci:2012cr,Babusci:2014sta,Anastasi:2016ktq},
  and NA48/2~\cite{Batley:2015lha}) and old  beam dump:
  E774~\cite{Bross:1989mp}, E141~\cite{Riordan:1987aw}, E137~\cite{Bjorken:1988as,Batell:2014mga,Marsicano:2018krp}),
  $\nu$-Cal~\cite{Blumlein:2011mv,Blumlein:2013cua},
  and CHARM (from~\cite{Gninenko:2012eq}.
  Bounds from supernovae~\cite{Chang:2016ntp} and $(g-2)_e$~\cite{Pospelov:2008zw}
  are also included.
}
\end{center}
\end{figure*}

 \begin{figure*}[t!]
\begin{center}
  \includegraphics[width=4.7in]{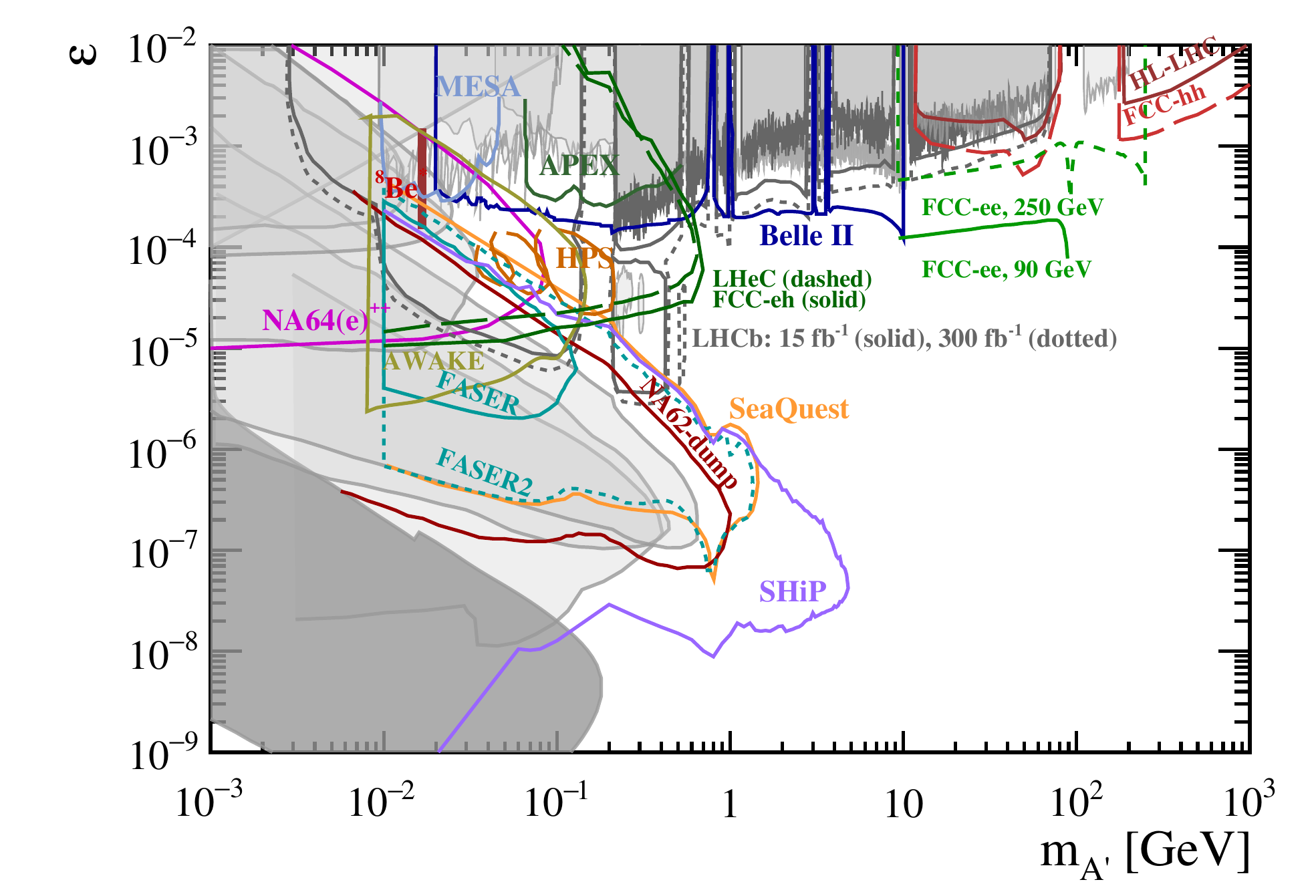}
\caption{\small 
  \label{fig:massive12}
   Colored curves are projections for existing and proposed experiments  on the massive dark photon for $m_{A^{\prime}} >1$ MeV:
  Belle-II~\cite{Kou:2018nap} at SuperKEKb;
  LHCb upgrade~\cite{Ilten:2016tkc,Ilten:2015hya} at the LHC; NA62 in dump mode~\cite{NA62:dump} and
  NA64(e)$^{++}$~\cite{NA64:eplus} at the SPS; FASER and FASER2~\cite{Feng:2017uoz} at the LHC;
  SeaQuest~\cite{Berlin:2018pwi} at Fermilab; HPS~\cite{Adrian:2018scb} at JLAB;
  an NA64-like experiment at AWAKE~\cite{Caldwell:2018atq},
  and an experiment dedicated to dark photon searches at MESA~\cite{Doria:2019sux,Doria:2018sfx}.
  For masses above 10 GeV projections obtained for ATLAS/CMS 
  during the high luminosity phase of the LHC
  (HL-LHC~\cite{Curtin:2014cca}) and for experiments running at a future FCC-ee~\cite{Karliner:2015tga},
  LHeC/FCC-eh~\cite{DOnofrio:2019dcp}, and
  FCC-hh~\cite{Curtin:2014cca} are also shown. The vertical red line
  shows the allowed range of couplings of a new gauge boson $X$  to electrons
  that could explain the $^8$Be anomaly~\cite{Feng:2016jff,Feng:2016ysn}.
  The existing limits are shown as gray areas. The bottom plot is revised from~\cite{Beacham:2019nyx}.
}
\end{center}
\end{figure*}

 Two kinds of experiments provide the existing limits on the visible massive dark photon in the region
 of $m_{A^\prime} > 1$  MeV: experiments at colliders and at fixed-target or beam dumps.
 In both cases the experiments search for resonances over a smooth background,
 with a vertex prompt or slightly displaced with respect to the beam interaction point in case of collider,
 or highly displaced in case of beam dump based experiments.
 The two categories are highly complementary, being the first category mostly
 sensitive to relatively large values of the mixing parameter $\varepsilon$, ($\varepsilon > 10^{-3}$)
 and the dark photon mass (up to several tens of GeV for $pp$ collider experiments),
 while the second is sensitive to relatively small values ($10^{-7} \ltap \varepsilon \ltap 10^{-3}$)
 in the low mass range, $m_{{A^{\prime}}}$ less than few GeV.
 
\begin{itemize}
\item \underline{Experiments at colliders.} These experiments search for resonances in the
  invariant mass distribution of $e^+ e^-, \mu^+ \mu^-$ pairs.
  Different dark-photon production mechanisms are
  used in the different experiments: 
  meson decays ($\pi^0 \to \gamma A'$, NA48/2~\cite{Batley:2015lha}),
  \textit{Bremsstrahlung} ($e^- Z \to e^- Z A'$, A1~\cite{Merkel:2014avp}),  
    annihilation ($e^+ e^- \to \gamma A'$, BaBar~\cite{Lees:2014xha}), and
  all these processes in different searches at
  KLOE~\cite{Archilli:2011zc,Babusci:2012cr,Babusci:2014sta,Anastasi:2016ktq}.
  In a proton-proton ($pp$) collider the dark photon is produced via the $\gamma-A'$ mixing in all the processes
  where an off-shell photon $\gamma^*$ with mass $m(\gamma^*)$ is produced:
  meson decays, \textit{Bremsstrahlung}, and Drell-Yan production. LHCb~\cite{Aaij:2017rft,Aaij:2019bvg} has performed
  a search for dark photon decaying in $\mu^+ \mu^-$ final states using 1.6~fb$^{-1}$ of data collected
  at the LHC $pp$ collisions at 13~TeV centre-of-mass energy. CMS~\cite{CMS:2019kiy} has performed
  the same search using 137~fb$^{-1}$ of fully reconstructed data and 96.6~fb$^{-1}$ of data
  collected with a reduced trigger information.

  Fig.~\ref{fig:massive11}  shows the existing limits for NA48/2, A1, LHCb, and BaBar;
  only one set of limits from
  KLOE is shown since the others have been superseded by the limits from BaBar.

\item \underline{Beam-dump experiments}.
  These experiments use the collisions of an electron or proton beam with a fixed-target or a dump to generate the
  dark photon via \textit{Bremsstrahlung} (electron and proton beams), meson production and QCD processes (proton beams only).
  The products of the collisions are mostly absorbed in the dump and the dark photon is searched
  for as a displaced vertex with two opposite charged tracks in the decay volume of the experiment.

  Fig.~\ref{fig:massive11} shows the limits from experiments at extracted electron beams (E141~\cite{Riordan:1987aw}
  and E137~\cite{Bjorken:1988as,Batell:2014mga, Marsicano:2018krp} at SLAC, E774~\cite{Bross:1989mp} at Fermilab)
  and at extracted proton beams from CHARM at CERN
  (\cite{Gninenko:2012eq} based on CHARM data~\cite{Bergsma:1985qz}).
\end{itemize}

In addition, bounds on energy losses in supernovae
provide further limits in the region of small masses. These limits where discussed in~\cite{Dent:2012mx,Dreiner:2013mua} and updated in~\cite{Chang:2016ntp,Hardy:2016kme} by including the effect of finite temperature and plasma density.

Also the electron magnetic moment, with its very precise experimental determination,
can be used to set an indirect limit~\cite{Pospelov:2008zw}.
These limits are included in Fig.~\ref{fig:massive11}. 

Recent constraints from ATLAS~\cite{Aad:2014yea,Aad:2015sms} and CMS~\cite{Khachatryan:2015wka}
would nominally cover the interesting
region around 1~GeV for $\varepsilon$ between $10^{-6}$ and $10^{-2}$ but unfortunately they have been
framed within a restrictive model 
and are not on the same footing that the limits included in Fig.~\ref{fig:massive11}.

Additional limits (not included in Fig.~\ref{fig:massive11}) from cosmology (in the cosmic microwave background and nucleosynthesis) exist
in the very dark region of very small $\varepsilon < 10^{-10}$~\cite{Fradette:2014sza}.

Looking at Fig.~\ref{fig:massive11}, it is clear that it would be desirable
to first close the gap between the beam-dump and the collider based experiments
in the region between tens of MeV up to 1~GeV in the dark photon mass,
and then extend the limits for larger masses. Both of these goals could be achieved through
a series of experiments summarized here below whose sensitivity is shown in Figure~\ref{fig:massive12} 
as colored curves.

\begin{itemize}
\item[-] {\it Belle-II at SuperKEKB} will search for visible dark photon decays $A^{\prime} \to e^+ e^-, \mu^+ \mu^-$ where
  $A^{\prime}$ is produced in the process $e^+ e^- \to A^{\prime} \gamma$.
  The projections shown in Fig.~\ref{fig:massive1} is based on
  50~ab$^{-1}$ of integrated luminosity~\cite{Kou:2018nap}.
  
\item[-] {\it LHCb upgrade (phase I and phase II) at the LHC}:
  LHCb phase I will search for dark photon in visible final states both using
  the inclusive di-muon production~\cite{Ilten:2016tkc}
  and the $D^{*0} \to  D^0 e^+ e^-$ decays~\cite{Ilten:2015hya}.
  The projections are based on 15~fb$^{-1}$, 3 years data taking with 5 fb$^{-1}$/year
  with an upgraded detector after the LHC Long Shutdown 2. This can be further improved with a
  possible Phase II upgraded detector~\cite{Bediaga:2018lhg} collecting up to 300~fb$^{-1}$ of integrated luminosity after Long Shutdown 4.
  
\item[-] {\it NA62$^{++}$ or NA62 in dump mode at the SPS, CERN,} will search for a multitude of
  feebly-interacting particles, including dark photon, decaying into visible final states and possibly emerging
  from the interactions of 400~GeV proton beam with a dump.
  NA62 aims to collect approximately $10^{18}$ protons-on-target in 2021-2024~\cite{NA62:dump}.
  
\item[-] {\it NA64(e)$^{++}$ at the SPS, CERN,} is the upgrade of the
  existing NA64(e) experiment. It aims to collect about $5 \times 10^{12}$ electrons-on-target 
  after the CERN Long Shutdown 2~\cite{NA64:eplus}.
  
\item[-] {\it NA64-like experiment at AWAKE, CERN:}
  progress in the coming years in proton-driven plasma wake-field acceleration
  of electrons at the AWAKE facility at CERN could allow an NA64-like experiment be served by a
  high-intensity high energy primary electron beam for search for dark photons in visible
  final states~\cite{Caldwell:2018atq}.
  The sensitivity plot has been obtained assuming $\sim 10^{16}$ electrons-on-target with an energy of 50~GeV.
  
\item[-] {\it FASER  and FASER2 at the LHC, CERN:} FASER~\cite{Feng:2017uoz}
  is being installed in a service tunnel of the LHC
  located along the beam collision axis, 480~m downstream from the ATLAS interaction point.
  At this location, FASER (and possibly its larger successor FASER2) will enhance the LHC discovery potential
  by providing sensitivity to dark photons, dark Higgs bosons, heavy neutral leptons,
  axion-like particles, and many other proposed feebly-interacting particles~\cite{Ariga:2018uku}.
  FASER and FASER2 aim to collect 150~fb$^{-1}$ and 3000~fb$^{-1}$ of integrated luminosity, respectively.
  
\item[-] {\it HPS at Jefferson Laboratory (JLab)}
  The HPS experiment~\cite{Adrian:2018scb}, proposed at an electron beam-dump at CEBAF electron beam
  (2.2-6.6~GeV, up to 500~nA), search for visible ($A^{\prime} \to  e^+ e^-$)
  dark photon (prompt and displaced) decays produced via \textit{Bremsstrahlung} production in a thin $W$ target.
  The experiment makes use of the 200 nA electron beam available in Hall-B at Jefferson Lab.

\item[-] {\it SeaQuest at Fermilab (FNAL) } will search for visible dark photon decays $A^{\prime} \to e^+ e^-$ at the
  120~GeV  main injector proton beamline at FNAL~\cite{Berlin:2018pwi}.
  It plans to accumulate approximately $10^{18}$ protons-on-target by 2024.

\item[-] {\it MAGIX or Beam Dump Experiment at MESA, Mainz:}
  The MESA accelerator is a {\it continuous wave} linac that will be able to provide an electron beam
  of  $E_{max} =~155$~MeV energy and up to 1~mA current~\cite{Doria:2019sux}.
  The MAGIX detector is a twin arm dipole spectrometer placed around
  a gas target and will search for search for visible ($A^{\prime} \to e^+ e^-$)
  dark photon (prompt and displaced) decays produced via \textit{Bremsstrahlung} production~\cite{Doria:2018sfx}.
  The possibility of a beam dump setup experiment is also under
  study. Timeline: targeted operations in 2021-2022 and 2 years of data taking.

\item[-] {\it Experiments at a future $e^+ e^-$ circular collider, FCC-ee:}
  a powerful technique to be exploited at experiments running at a
  future $e^+ e^-$ circular collider is the radiative return,
  $e^+ e^- \to A^{\prime} \gamma, A^{\prime} \to \mu^+ \mu^-$. The results obtained in~\cite{Karliner:2015tga}
  have been rescaled to the integrated luminosities of 150~fb$^{-1}$ at $\sqrt{s} = 90$~GeV and
  5~ab$^{-1}$ at $\sqrt{s} = 250$~GeV, as in~\cite{Strategy:2019vxc}.

\item[-] {\it ATLAS/CMS at the high-luminosity phase at the LHC  and
  at a future $pp$ circular collider:}
  at $pp$ colliders the dark photon can be produced via a Drell-Yan process,
  $pp \to A^{\prime} \to e^+ e^-, \mu^+ \mu^-$. The physics reach of ATLAS/CMS like experiments
  have been computed for $\sqrt{s}=14$~TeV and 3~ab$^{-1}$ and $\sqrt{s} = 100$~TeV,
  3~ab$^{-1}$~\cite{Curtin:2014cca}.

\item[-] {\it ATLAS/CMS at a possible LHeC collider in the LHC tunnel
  and a future FCC-eh circular collider:}
  At the LHeC (FCC-eh) a 7~TeV (50~TeV) a proton beam collides with a 60~GeV electron
  beam achieving a center-of-mass energy of 1.3~TeV (3.5~TeV) and
  a total integrate luminosity of 1~ab$^{-1}$ (3~ab$^{-1}$). 
  At $eh$ colliders the main production process for the dark photon is
  the deep inelastic scattering $e^- +$ parton $\to e^-$ parton $A^{\prime}$,
  with $A^{\prime} \to$ charged fermions ~\cite{DOnofrio:2019dcp}.
\end{itemize}

\subsection{Constraints for $m_{A^{\prime}} > 1$ MeV with $A'$ decays to invisible final states}
\label{sec:con3}

 \begin{figure*}[t!]
\begin{center}
  \includegraphics[width=4.7in]{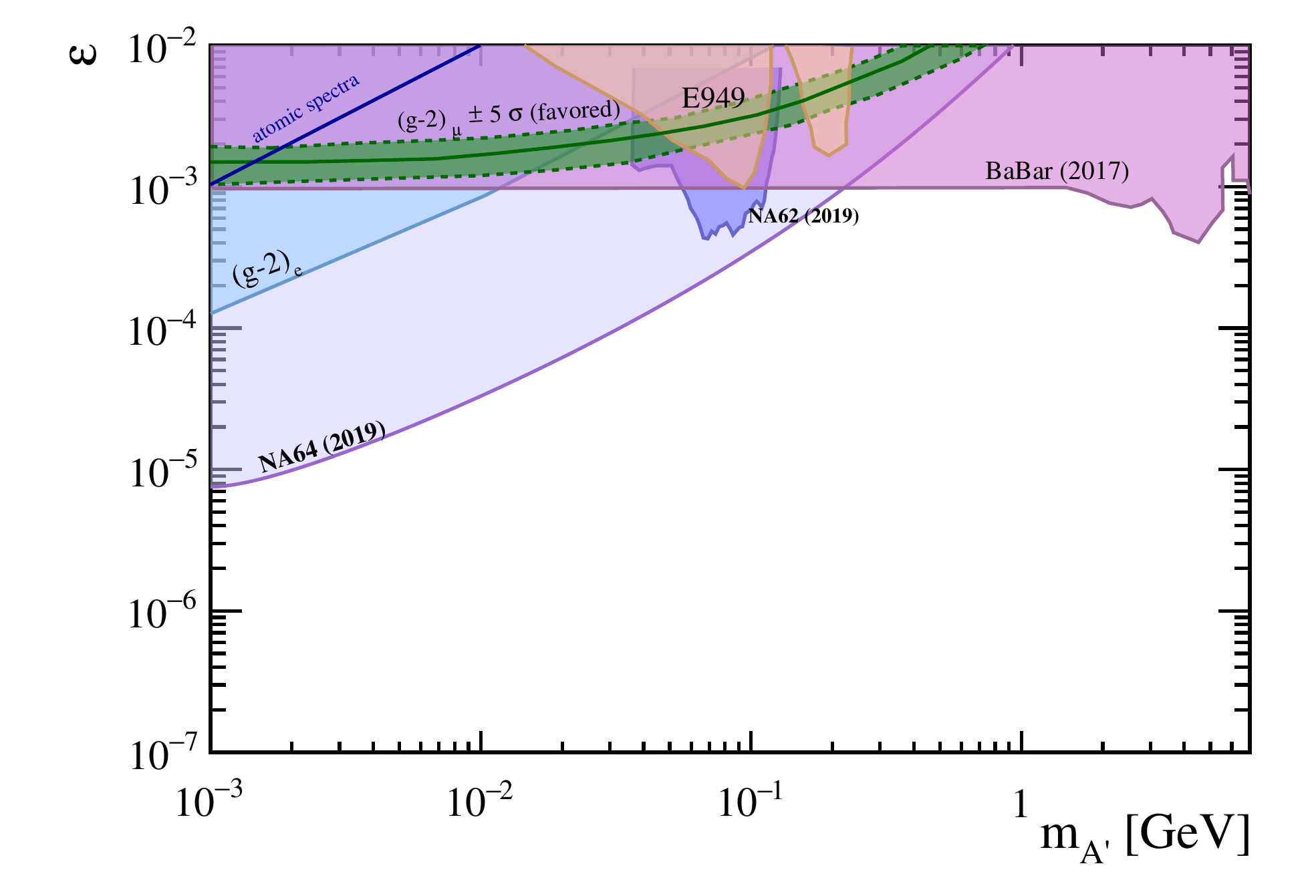}
\caption{\small 
  \label{fig:massive31}
  Existing limits  for  a massive dark photon going to invisible final states ($\alpha_{\dd}>> \alpha \varepsilon^2 $).
 Existing limits from Kaon decay experiments (E787~\cite{Adler:2001xv},
  E949~\cite{Artamonov:2009sz}, NA62~\cite{CortinaGil:2019nuo}), BaBar~\cite{Lees:2017lec},
  and NA64(e)~\cite{NA64:2019imj}. The constraints from $(g-2)_{\mu}$~\cite{Bennett:2006fi}
  and $(g-2)_e$ are also shown.
}
\end{center}
\end{figure*}

 \begin{figure*}[t!]
\begin{center}
  \includegraphics[width=4.7in]{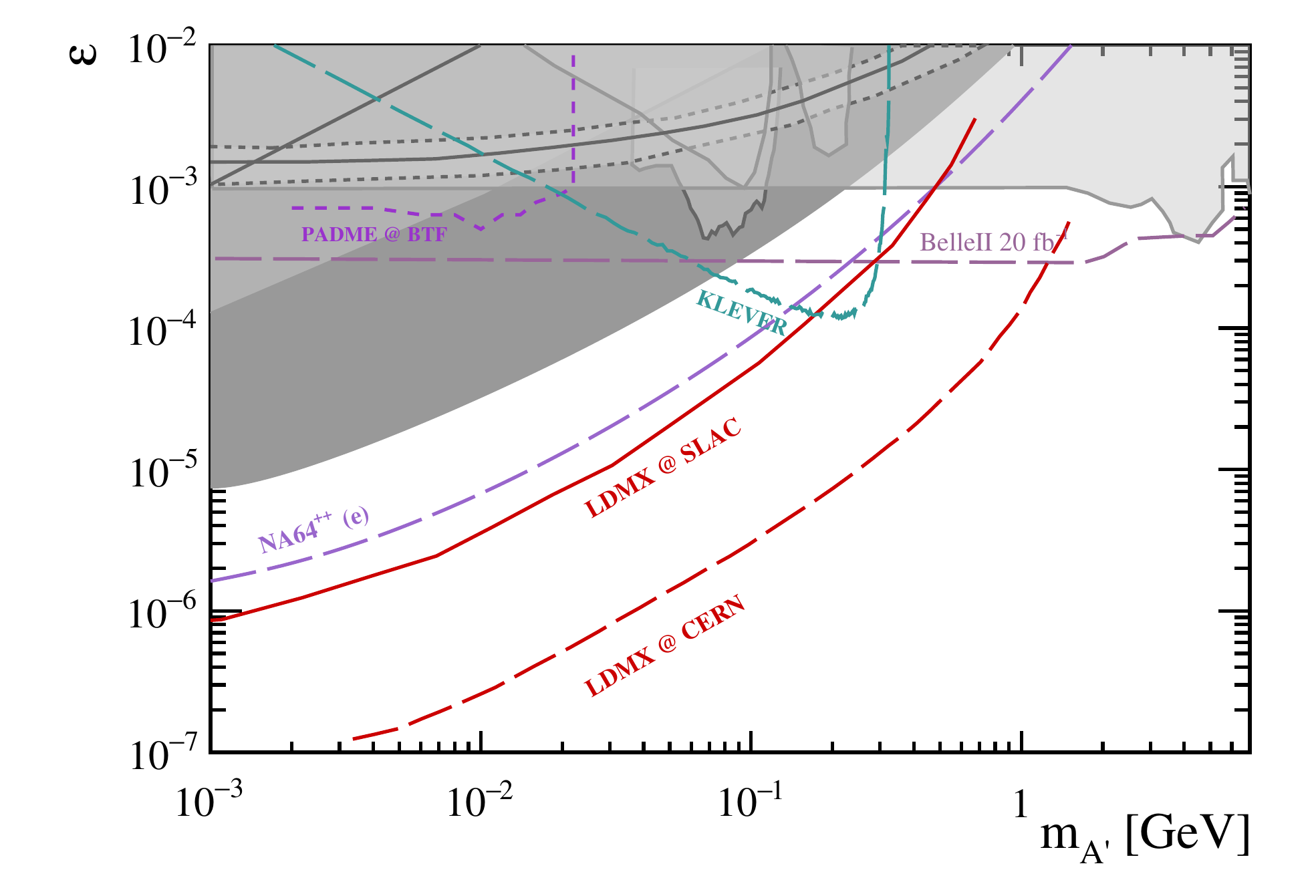}
\caption{\small 
  \label{fig:massive32}
  Future sensitivities for proposed experiments
   for a massive dark photon going to invisible final states ($\alpha_{\dd}>> \alpha \varepsilon^2 $).
   Future sensitivities for NA64(e)$^{++}$~\cite{NA64:eplus},
  Belle II~\cite{Kou:2018nap}, KLEVER~\cite{Ambrosino:2019qvz}, PADME~\cite{Raggi:2015gza}, LDMX@SLAC~\cite{Akesson:2018vlm,LDMX:CERN}, and LDMX@CERN~\cite{Akesson:2018vlm,Raubenheimer:2018mwt}.
  The sensitivity curves for LDMX@SLAC and LDMX@CERN assume $10^{14}$~electrons-on-target and $E_{\rm beam} = 4$~GeV and $10^{16}$~electrons-on-target and $E_{\rm beam} = 16$~GeV, respectively.  The bottom plot is revised from~\cite{Beacham:2019nyx}. See text for details.
}
\end{center}
\end{figure*}

Different constraints apply in the case of massive dark photon going into
invisible final states in the mass region $m_{A^\prime} >1$~MeV.
In this case techniques like missing momentum, missing energy, and missing mass are used in order to
identify a possible massive dark photon decaying into invisible final states.

The most stringent bounds come from BaBar~\cite{Lees:2017lec} and the electron beam dump NA64(e) experiment at CERN~\cite{NA64:2019imj} which recently superseded
the results from Kaon experiments (E787~\cite{Adler:2001xv} and E949~\cite{Artamonov:2009sz} at
BNL, NA62~\cite{CortinaGil:2019nuo} at CERN).
The existing bounds are depicted in the top plot of Fig.~\ref{fig:massive31} as colored areas.
These limits overlap with the exclusion regions defined by the dark photon decays into visible final states for masses
$m_{A^{\prime}}>1$~GeV and
complement them in the range  of masses 10~MeV$\ltap m_{A^{\prime}} \ltap$~1~GeV and kinetic mixing strength
$10^{-5} \ltap \varepsilon \ltap 10^{-3}$, where the searches of dark photon into visible decays are typically weaker.

Sensitivities of existing or proposed experiments are shown in  Fig.~\ref{fig:massive32} as colored
lines. These include:

\begin{itemize}
\item[-] {\it NA64(e)$^{++}$} with $5\times 10^{12}$~electrons-on-target will search $A^{\prime} \to$ invisible final states
  with a missing energy technique using a secondary electron beam at $\sim 100$ GeV at the CERN SPS~\cite{NA64:eplus}.
  \item[-] {\it Belle II} will search for dark photons in the process $e^+ e^- \to A^{\prime}$ and $ A^{\prime} \to$
    invisible~\cite{Kou:2018nap}. Projections are based on 20~fb$^{-1}$ of integrated luminosity.
  \item[-] {\it KLEVER}, proposed at the SPS, could search for dark photons in invisible final states
    as a by-product of the analysis of the $K_L \to \pi^0 \nu \overline{\nu}$ rare decay, 
    pushing further the investigation performed by traditional Kaon experiments
    in the mass region between 100-200 MeV~\cite{Ambrosino:2019qvz}.
  \item[-] {\it PADME}~\cite{Raggi:2015gza}
    will search for $A^{\prime} \to$ invisible final states using the
    missing momentum technique at the Beam Test Facility (BTF)
    at Laboratori Nazionali di Frascati (INFN). It will use a 550~MeV positron beam on a diamond target.
    A first commissioning run was performed in late 2018 and early 2019 to assess the detector
    performance and beam line quality. A physics data taking to collect $5 \times 10^{12}$ positrons
    on target is expected in the second part of 2020.
\end{itemize}

\subsection{Constraints for $m_{A^\prime} < 1$ MeV}
\label{sec:con4}

 \begin{figure*}[th!]
\begin{center}
  \includegraphics[width=4.7in]{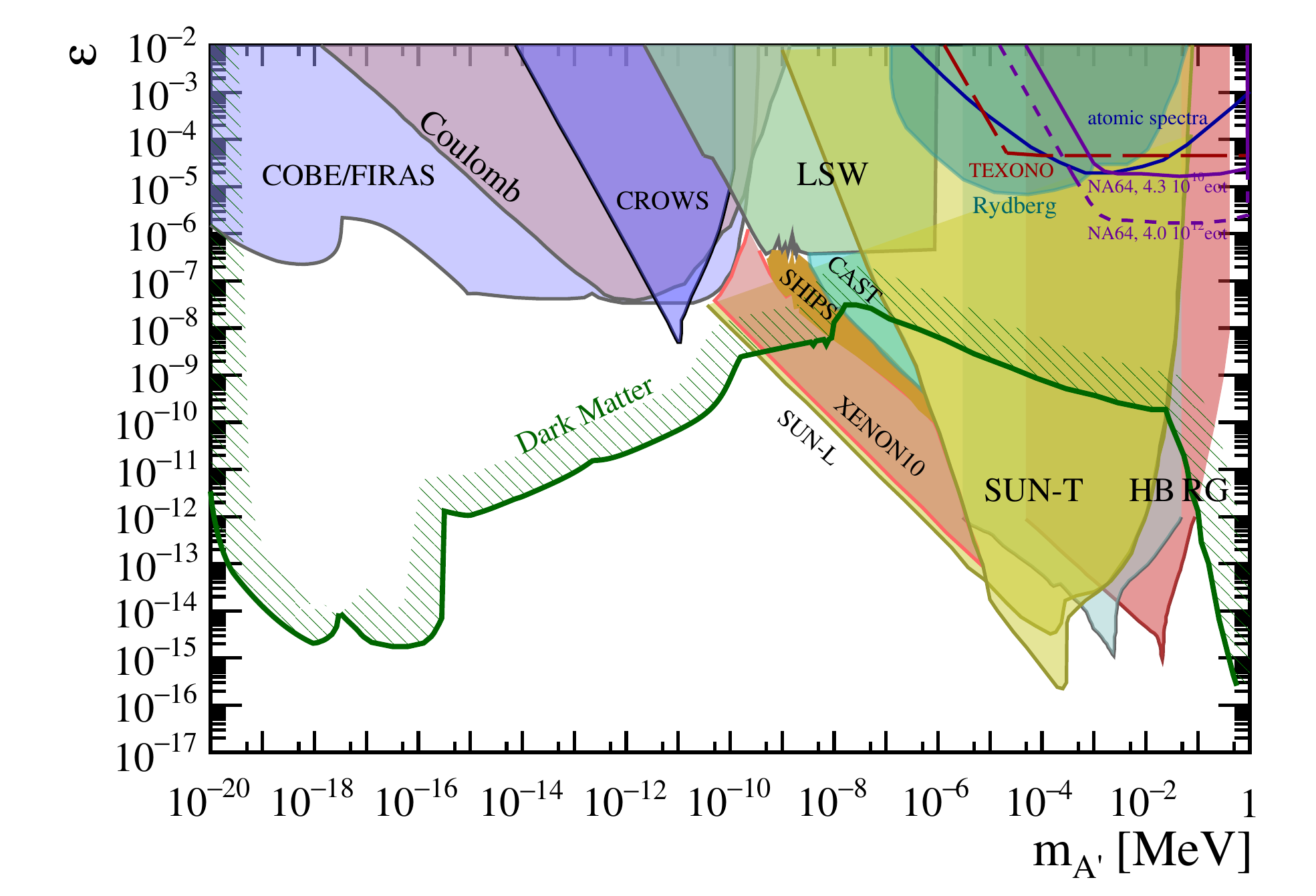}
\caption{\small 
  \label{fig:massive21} Current limits on massive dark photon for $m_{A^\prime} < 1$ MeV.
  Bounds from cosmology (COBE/FIRES~\cite{Caputo:2020bdy,Garcia:2020qrp,Witte:2020rvb,McDermott:2019lch}), 
  light through a wall (LSW)~\cite{Ehret:2010mh}, CROWS~\cite{Betz:2013dza}, CAST~\cite{Redondo:2008aa}, XENON10~\cite{An:2013yua}, SHIPS~\cite{Schwarz:2015lqa}, TEXONO~\cite{Danilov:2018bks},
  atomic experiments (Coulomb, Rydberg and atomic spectra~\cite{Jaeckel:2010xx}) and astrophysics: Solar lifetime (SUN-T and SUN-L), red giants (RG),
  horizontal branches (HB)~\cite{Redondo:2013lna,An:2013yfc,Hardy:2016kme}.  
  Additional limits under the assumption that the dark photon  is the dark matter: The curve ``Dark Matter" includes the combination of  the constraints from the references discussed in the main text.
  }
\end{center}
\end{figure*}

 \begin{figure*}[th!]
\begin{center}
  \includegraphics[width=4.7in]{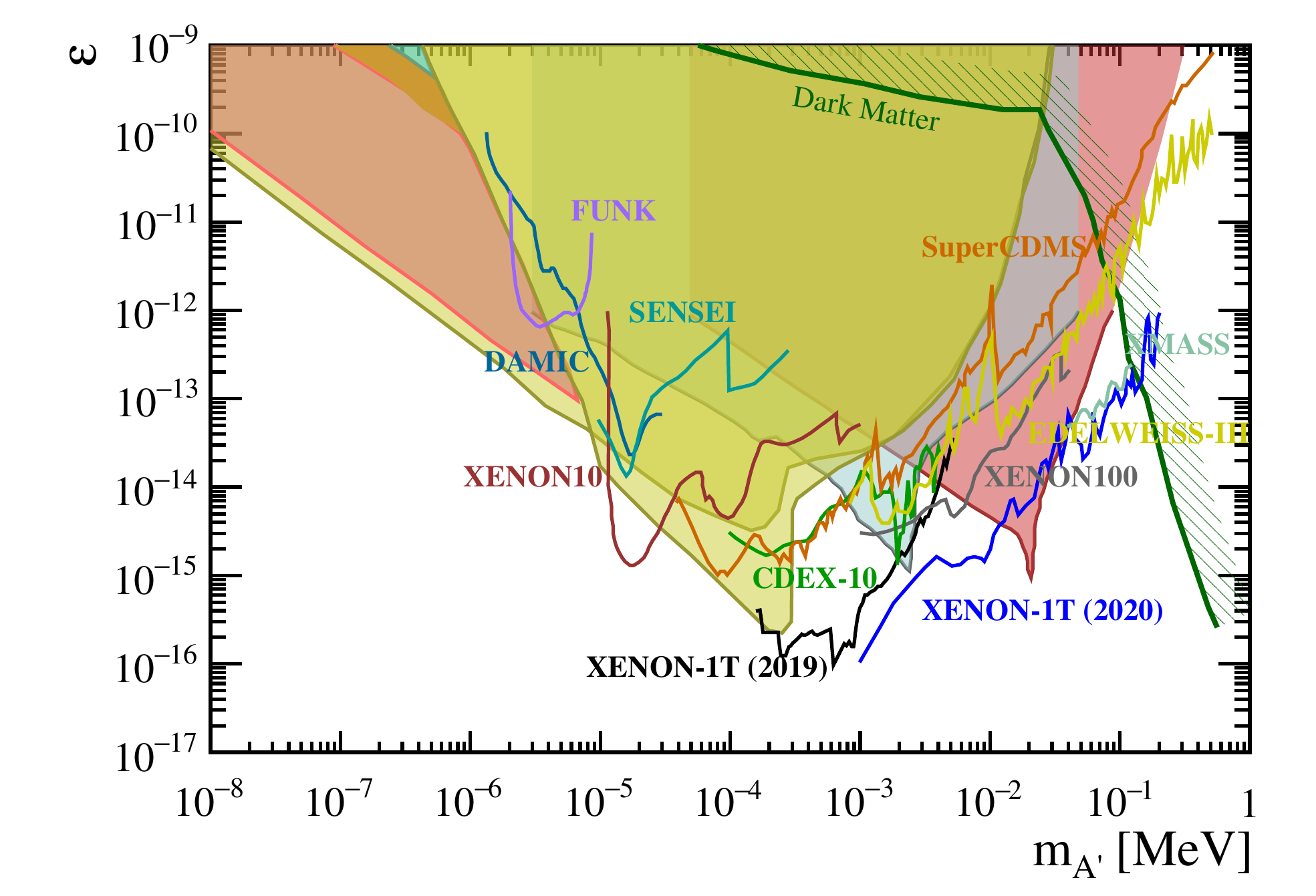}
\caption{\small 
  \label{fig:massive22}   Zoom in the range $10^{-8}$~MeV$\ltap M_{A^{\prime}} \ltap$~1~MeV and $10^{-17} \ltap \varepsilon \ltap 10^{-9}$ for the current limits on massive dark photon for $m_{A^\prime} < 1$ MeV.
  Results from dark matter direct detection experiments
  (XENON10 and XENON100 (\cite{An:2014twa} based on XENON10~\cite{Angle:2011th} and XENON100~\cite{Aprile:2014eoa} data),
  XENON1T (2019)~\cite{Aprile:2019xxb}; XENON1T (2020)~\cite{Aprile:2020tmw}; DAMIC~\cite{Aguilar-Arevalo:2016zop}; SuperCDMS~\cite{Aralis:2019nfa};
  CDEX-10~\cite{She:2019skm}; EDELWEISS-III~\cite{Armengaud:2018cuy};
  SENSEI~\cite{Abramoff:2019dfb}; XMASS~\cite{Abe:2018owy}; FUNK~\cite{Andrianavalomahefa:2020ucg}). 
  }
\end{center}
\end{figure*}

 Strong constraints exist for the invisible massive dark photon in the region $m_{A^\prime} < 1$  MeV.
 They come from different sources:  
\begin{itemize}
\item \underline{Atomic and nuclear experiments}: These experiments aim to detect modifications of the
 Coulomb force (as discussed in~\cite{Bartlett:1970js}) due to the dark photon.  Corrections in Rydberg atoms, Lamb shift  and hyperfine splitting in atomic hydrogen have been translated into bounds on the massive dark photon mixing parameter~\cite{Jaeckel:2010xx}. The  results of the TEXONO neutrino experiment~\cite{Deniz:2009mu} have been interpreted in terms of dark photon parameters in \cite{Danilov:2018bks};
\item \underline{Axion-like particles and helioscopes}: Experiments  of  light shining through a wall (LSW) for axions and axion-like particles can be adapted to the dark photon and limits can accordingly be estimated~\cite{Ehret:2010mh}. The same phenomenon has  been used  in the experiment CROWS~\cite{Betz:2013dza} at CERN. The CAST result, on the flux of axion-like particles from the Sun (Helioscope), can be translated~\cite{Redondo:2008aa} into a bound on the massive dark photon parameters.  The same is true for XENON10, whose  data set provides further limits~\cite{An:2013yua} and the results from the  experiment SHIPS~\cite{Schwarz:2015lqa}; 
\item \underline{Astrophysics}:  The non-observation of anomalous energy transport (by the mechanism discussed in section~\ref{sec:con1}) in stars on the horizontal branch (HB),
  red giants (RG) and the Sun (SUN-T and SUN-L)
   imposes severe constraints on the mixing parameter of the massive dark photon. Mixing effects are important in these processes for both the longitudinal (L) and transverse (T) modes and one must use thermal field theory~\cite{Redondo:2013lna,An:2013yfc,Hardy:2016kme}. The dark photon partakes of the plasmon modes (see appendix~\ref{sec:TFT}) in an effective mixing with the ordinary photon
  proportional to its mass (and vanishing as it goes to zero). 
\item \underline{Cosmology}:  The oscillation between the ordinary and the massive dark photon $\gamma \to A^\prime$ induces deviations on the black body spectrum (as measured by COBE/FIRAS~\cite{Fixsen:1996nj}) in the   cosmic microwave background. This effect depends on the effective plasma mass of the dark photon and it is enhanced when this mass is equal to $m_{A^\prime}$. The bound depicted in  Fig.~\ref{fig:massive2} follows the most recent evaluation~\cite{Caputo:2020bdy,Garcia:2020qrp,Witte:2020rvb}---which  includes  inhomogeneities in
 the plasma mass---for  values $m_{A^\prime} <  10^{-15}$ MeV,   
and  \cite{Mirizzi:2009iz,McDermott:2019lch} for  larger values.
  \end{itemize}
  
 Even stronger constraints can be derived under the assumption  that the dark photon is itself the dark matter.
 The combination (in order of increasing values of $m_{A^\prime}$) of 
 \begin{itemize}
 \item[-] astrophysical bounds on dwarf galaxies~\cite{Wadekar:2019xnf}. These limits apply for values of $m_{A^\prime}< 10^{-19}$ MeV and are not shown in Fig.~\ref{fig:massive2}, 
 measurements of the temperature of the intergalactic medium at the epoch of He$^{++}$ re-ionization in the presence of inhomogeneities~\cite{Caputo:2020bdy,Garcia:2020qrp,Witte:2020rvb},
  \item[-] cold Galactic Center gas clouds  heating rates~\cite{Bhoonah:2018gjb};
  \item[-] cosmic microwave background spectral distortions ($\mu$ and $y$-type)~\cite{Witte:2020rvb};
 \item[-] energy deposition during the dark ages~\cite{McDermott:2019lch},
\item[-] the number of relativistic species $\Delta N^{eff}$ during big-bang nucleosynthesis and recombination~\cite{Arias:2012az} and 
 \item[-] the diffuse X-ray background~\cite{Redondo:2008ec} 
 \end{itemize}
 yield a series of limits on the upper value of $\varepsilon$  for different values of $M_{A^\prime}$. These limits are  depicted together by the curve labelled ``Dark Matter'' in Fig.~\ref{fig:massive21}.
 
  In addition, there are limits from:
 \begin{itemize}
  \item \underline{Dark matter direct detection experiments}: These experiments are part of the on-going search
  for dark matter through its direct detection. Data from XENON10/XENON100 (\cite{An:2014twa} based on XENON10~\cite{Angle:2011th} and XENON100~\cite{Aprile:2014eoa}),
  XENON1T~\cite{Aprile:2019xxb}, DAMIC~\cite{Aguilar-Arevalo:2016zop}, SuperCDMS~\cite{Aralis:2019nfa},
  CDEX-10~\cite{She:2019skm}, EDELWEISS-III~\cite{Armengaud:2018cuy},
  SENSEI~\cite{Abramoff:2019dfb}, XMASS~\cite{Abe:2018owy} and  FUNK~\cite{Andrianavalomahefa:2020ucg} can be used to constrain the massive dark photon parameters;
 \item \underline{Haloscopes}: Searches with microwave cavities for relic axion converting to photons~\cite{Sikivie:1983ip} can be translated into limits (not shown in Fig.~\ref{fig:massive2}) on the dark photon parameter $\varepsilon$ to be less than  $10^{-13}$-$10^{-15}$  in the range around $10^{-11}$-$10^{-12}$ MeV~\cite{Arias:2012az}.
 \end{itemize}

The limits from dark-matter direct detection are shown in  Fig.~\ref{fig:massive22}.

Some of the limits on the right side  of  Fig.~\ref{fig:massive21} are the continuation of the corresponding left side of the limits in Fig.~\ref{fig:massive31}. The two figures are   back-to-back at $m_{A^\prime}=1$ MeV thus covering the full range of the dark-photon masses.

\section{Limits on the parameters $y$ and $m_\chi$}
\label{sec:con5}

 \begin{figure*}[t!]
\begin{center}
  \includegraphics[width=4.7in]{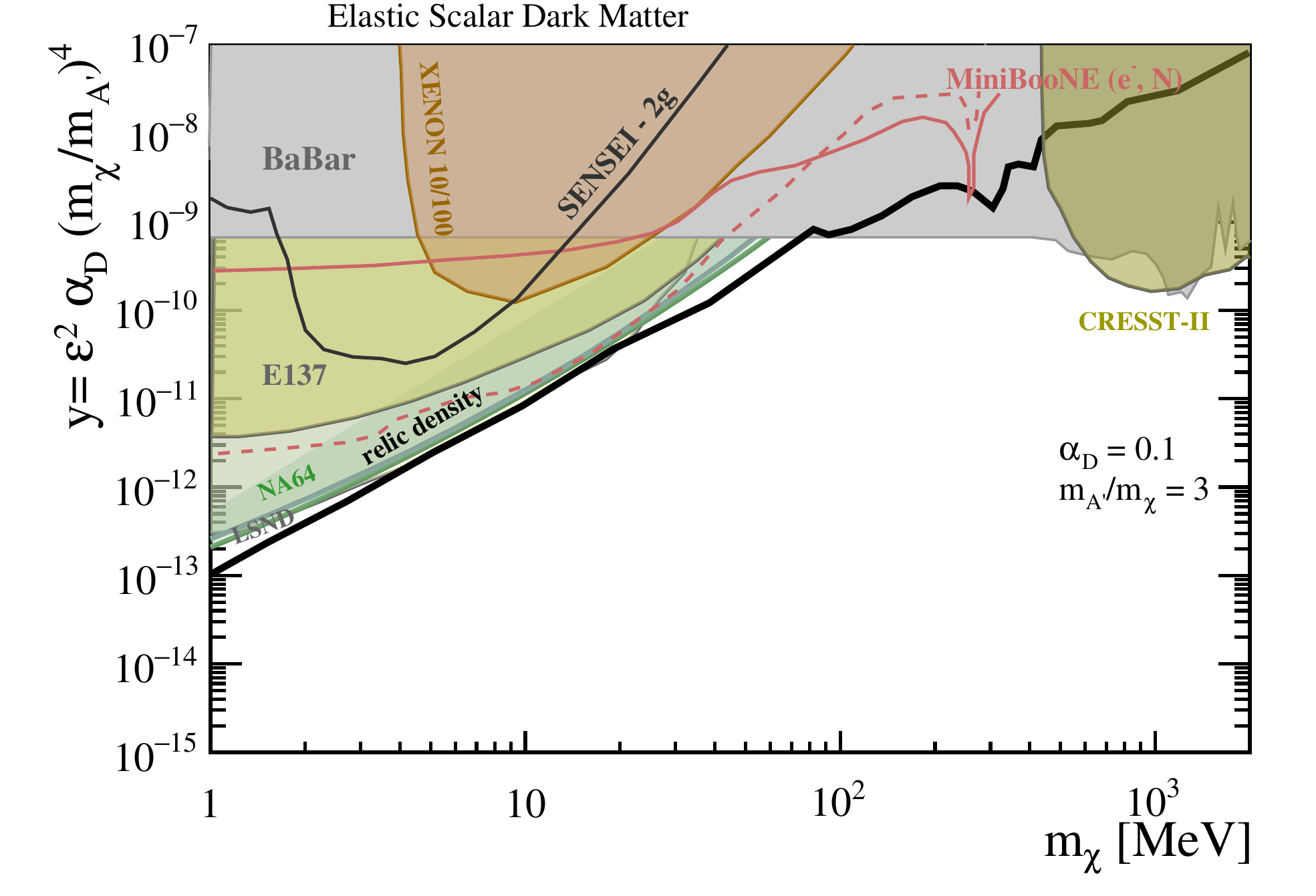}
\caption{\small 
  \label{fig:massive41}  Existing limits for existing experiments
   for massive dark photon for $m_{A^\prime} > 1$ MeV in the plane
  of the yield variable $y$ as a function of dark matter mass $m_{\chi}$ for an elastic scalar dark matter particle.
 Limits from BaBar~\cite{Lees:2017lec}, NA64(e)~\cite{NA64:2019imj}, reinterpretation of the data from
  E137~\cite{Batell:2014mga} and LSND~\cite{deNiverville:2011it}; result from MiniBooNE~\cite{Aguilar-Arevalo:2018wea};
  interpretation in the dark photon framework of data from
  CRESST-II~\cite{Angloher:2015ewa}.
  }
\end{center}
\end{figure*}

 \begin{figure*}[t!]
\begin{center}
  \includegraphics[width=4.7in]{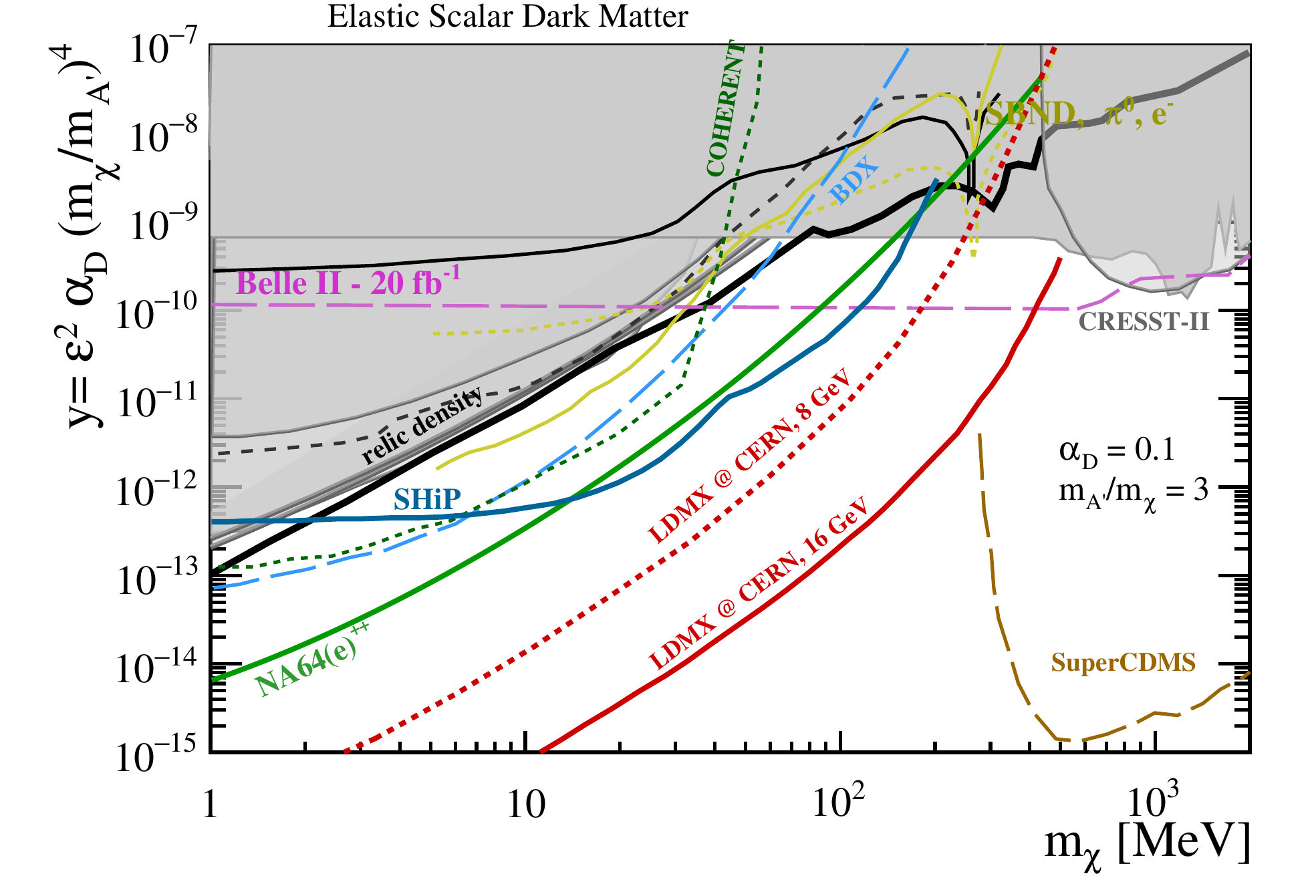}
\caption{\small 
  \label{fig:massive42}  Future sensitivities for proposed experiments
  for massive dark photon for $m_{A^\prime} > 1$ MeV in the plane
  of the yield variable $y$ as a function of dark matter mass $m_{\chi}$ for an elastic scalar dark matter particle.
  Projections for SHiP~\cite{Anelli:2015pba}, BDX~\cite{Battaglieri:2016ggd},
  SBND~\cite{Antonello:2015lea}, LDMX@CERN~\cite{Akesson:2018vlm,Raubenheimer:2018mwt},
  SENSEI with a proposed 100~g detector operating at SNOLAB~\cite{Battaglieri:2017aum},
  and SuperCDMS at SNOLAB~\cite{Agnese:2016cpb}.
  The plot is revised from~\cite{Beacham:2019nyx}.}
\end{center}
\end{figure*}

If the dark sector states into which the invisible dark photon decays are taken to be  dark matter,
there are new limits involving also the coupling strength $\alpha_{\D}$ and the
connection to the direct-detection searches for dark matter.
As discussed in section~\ref{sec:dark}, the best way to plot the experimental limits
in this case is in terms of the yield variable $y$, defined in \eq{y}, and the dark matter mass $m_\chi$.

 The corresponding limits strongly depend on the nature of the dark-matter state $\chi$ because the velocity
 dependence of the averaged cross sections. In the case of Dirac fermions,  Planck data~\cite{Ade:2015xua} rule out  sub-GeV dark matter because of their too large annihilation rate at the cosmic microwave background epoch.
 For this reason, pseudo-Dirac fermions and scalars, which have velocity suppressed annihilation cross sections,
 are usually studied. 

The current bounds and future perspectives in the plane $y$ versus dark matter mass are shown
in Fig.~\ref{fig:massive41} and \ref{fig:massive42} under the hypothesis that the dark matter is a scalar particle and for a specific choice
of $\alpha_{\D}$ ($\alpha_{\D} = 0.1$) and the ratio between the mediator and the dark matter masses ($m_{A^{\prime}}/m_{\chi} = 3$).
In these plots, the lower limit for the thermal relic density is also shown, under that
hypothesis that a single dark-matter candidate is responsible for the whole dark-matter abundance.
It is worth noting that results from accelerator-based experiments are largely independent
of the assumptions on a specific dark matter nature as dark matter at accelerators is produced in relativistic
regime and the strength of the interactions with light mediators and SM particles is only
fixed by thermal freeze-out.

Current bounds come from the same experiments using missing energy/missing momentum techniques
contributing to the \{$\varepsilon, m_{A^{\prime}}$\} sensitivity plot (BaBar and NA64(e))
with the addition of the re-interpretation of data from old neutrino experiments (E137~\cite{Batell:2014mga}
and LSND~\cite{deNiverville:2011it}) and results from current neutrino
experiments (MiniBooNE~\cite{Aguilar-Arevalo:2018wea})
exploiting dark matter scattering on nucleons and/or electrons. 
Bounds can also be derived by using a superfluid He-4 detector, as shown in ~\cite{Caputo:2019xum}, but they lie at the margin of the range included in Fig.~\ref{fig:massive41}.

In all the bounds shown for electron beam-dump or missing energy experiments, we are neglecting the contribution of secondary positrons in dark photon production through annihilation on atomic electrons, as for example studied in ~\cite{Marsicano:2018krp}.

Future initiatives that could explore a still uncovered parameter space in the plane
\{$y, m_{\chi}$\} for dark matter masses below 1~GeV are all those that have sensitivity in the
plane \{$\varepsilon, m_{A^{\prime}}$\} and, in addition, accelerator-based and dark matter direct detection
experiments exploiting dark matter scattering against the nucleons and/or electrons.
Accelerator-based experiments are SHiP at CERN~\cite{Anelli:2015pba}, and BDX at JLab~\cite{Battaglieri:2016ggd}
and SBND~\cite{Antonello:2015lea} at FNAL as explained below.

\begin{itemize}
\item[-] {\it BDX at JLAB}: The Beam Dump eXperiment (BDX)~\cite{Battaglieri:2016ggd} is aiming
to detect light dark matter $\chi$ produced in the interaction of an intense (100~$\mu$A) 10~GeV electron beam
with a dump. The experiment is sensitive to elastic dark matter scattering $e^- \chi \to  e^- \chi$ in
the detector after production in $e^-Z \to e^-Z A^{\prime} (A^{\prime} \to \chi \chi)$.

\item[-] {\it SBND} is planned to be installed at the 8 GeV proton Booster Neutrino Beamline at
  FNAL about 470~m downstream  of the beam dump~\cite{Antonello:2015lea}.
  The dark matter beam is primarily produced via pion decays out of collisions from the
  primary proton beam, and identified via dark-matter-nucleon or dark-matter-electron elastic scattering in a LAr-based detector.
  SBND is expected to improve upon MiniBooNE by more than an order of magnitude with $6\times 10^{20}$ protons-on-target.

\end{itemize}

Also dark matter direct-detection experiments with sensitivity below 1 GeV mass
contribute to this plot. These are:

\begin{itemize}
  
\item[-] {\it SENSEI} is a direct detection experiment~\cite{Tiffenberg:2017aac}
  that will be able to explore dark matter candidates with masses in the 1~eV and few~GeV range,
  by detecting the signal released in dark-matter-electron scattering interactions in a fully depleted silicon CCD.
  A 2-gram detector is already operating in the NUMI access tunnel~\cite{Barak:2020fql}. A larger project
  (100~grams) can be deployed at SNOLAB if funding is obtained~\cite{Battaglieri:2017aum}. 

\item[-] {\it CRESST-II}~\cite{Angloher:2015ewa}
  uses cryogenic detectors to search for nuclear recoil events induced by
elastic scattering of dark-matter particles in CaWO$_4$ crystals. Because of its low-energy
threshold, the sensitivity to dark matter was extended in the sub-GeV region.
Current bounds are derived from a dataset corresponding to 52 kg live days. 

\item[-] {\it super-CDMS}~\cite{Agnese:2016cpb}
  at SNOLAB (Canada)Start mid-2021, and uses 30~kg of Germanium and Silicium detectors.
\end{itemize}


\chapter{Outlook}
\label{sec:conclusiones}

{\versal  In the past 50 years} it has been assumed that physics  beyond the SM interacted through (at least) some of the same gauge interactions of the SM. The minimal supersymmetric SM  and weakly interacting massive dark matter are the two preeminent and most influential models based on this paradigm. 

This program is now running out of some of the initial momentum because of the lack of  the discovery of new particles. In the absence of new states,  the many parameters, for instance, of the minimal supersymmetric SM  are  working against its usefulness as a foil for the SM in mapping possible experimental discrepancies.

In more recent times---mostly under the influence of this lack of any real signal of  the breaking up of the SM---a more general scenario has been attracting increasing interest. Matter beyond the SM is  part of a new sector  which is dark because it does not interact through the SM gauge interactions. The dark sector may contain a wealth of physics with many particles (some of which are dark matter) and interactions. 

From our side, in the visible world, we may  glimpse this dark sector through a portal. If it exists,  this portal can take various forms depending on the spin of the mediator. We have reviewed in this primer  the vector case in which the portal arises  from the kinetic mixing between the SM electric (or hyper) charge gauge group and an $U(1)$ gauge symmetry of the dark sector.

The discovery of the dark photon associated to this new Abelian gauge symmetry is by far more interesting than finding just a new particle because, if found, this new gauge boson would be the harbinger of  a new interaction  and of the existence of a whole new sector of elementary particles.

Past and current experiments have already restricted an important part of the space of the parameters of the vector portal, both for the massless and the massive dark photon. Compared to other searches for models beyond the SM, the parameters are fewer and the signatures more easily interpreted.

We are now on the verge of a new wave of experiments aiming at further closing the windows left still open in the interaction between ordinary matter and the dark photon. The long years of searching in vane for physics beyond the SM has thought us to be very patient and persevering. 

The constraints in the massless case seem to relegate the possible detection  of the dark photon  to very large values of the effective scale $\Lambda$ in the dark dipole interaction, as we discuss in  section~\ref{sec:con1}. Exploring physics at such a large energy scale requires the high sensitivity that can only be achieved either in future lepton colliders (where the scaling with the energy of the dark dipole operator will also enhance its contribution) or in 
searches for rare flavor-changing decays like those of  the Kaon and $B$-meson systems. 

The constraints in the case of the massive dark photon have left open two important regions in the parameter space. The first one is for the visible dark photon  with masses around 100 MeV or larger and mixing parameter between  $10^{-6}$ and $10^{-4}$. Many future experiments aim at looking into this range, as we review in section~\ref{sec:con2}. If also this window will be closed, it means that the already feeble  interaction of the vector portal is very weak indeed. Which leaves us with the second window still left unexplored: an invisible dark photon with a  very light  mass and a  mixing parameter of order $O(10^{-8})$ or even lighter and with smaller mixing parameter, as discussed  in section~\ref{sec:con3} and \ref{sec:con4}. These two latter regions are of great interest for astrophysics and cosmology and a very active area of speculations.  

No  single  experiment  or  experimental  approach  is  sufficient  alone  to  cover  the large parameter space in terms of masses and couplings that dark photon models  suggest:  
Synergy and complementarity among a great variety of experimental facilities are paramount, calling for a broad collaboration across different communities.

\begin{flushleft}
{\Large \bfseries Acknowledgements}
\end{flushleft}	
\vspace*{0.2cm}

The digital inclusion of some of the experimental limits in the figures was done by means of 
{\sc WebPlotDigitizer}.\footnote{Website: \url{https://automeris.io/WebPlotDigitizer}}
Numerical results for some of the dark-photon limits were provided by the {\sc Darkcast} package.\footnote{Website: \url{https://gitlab.com/philten/darkcast}}

We would like to thank  Tram Acu\~{n}a, Andrea Celentano,  Monica D'Onofrio, Angelo Esposito, Oliver Fischer, Phil Ilten, Sam McDermott, Maxim Pospelov, Diego Redigolo, Josh Ruderman,  Piero Ullio, Alfredo Urbano  and Mike Williams for useful discussions and suggestions.

MF is affiliated to the Physics Department of the University of Trieste and the \textit{Scuola Internazionale Superiore di Studi Avanzati} (SISSA), Trieste, Italy. 
MF and EG are affiliated to the Institute for Fundamental Physics of the Universe (IFPU), Trieste, Italy. The support of all these institutions is gratefully acknowledged. EG thanks the Department of Theoretical Physics of CERN  for its kind hospitality during the preparation of this work.

\chapter{Appendix}

\section{Dark sector portals}
\label{sec:appA}

\vspace*{1cm}

 In this appendix we give the Lagrangian for the four portals mentioned in the introduction as well as a minimal bibliography to provide their context within the search of phenomenological models beyond  the Standard Model.
 
 The dark sector is assumed to interact with the visible, SM sector through relevant operators of dimension four and five (and possibly sub-leading higher-order operators).

These portals are classified according to the spin of the mediator field. We can have
\begin{itemize}
\item \underline{dark photon} (spin $1$): The portal operator arises from the kinetic is mixing between the SM photon field strength $F_{ \mu\nu} $ and a dark photon $F^{\mu\nu \prime}$: 
$$ \frac{\varepsilon}{2} \,F_{ \mu\nu} F^{\mu\nu \prime}\, ,$$
it is an operator of dimension four. It is assumed that the dark photon is the main carrier of the interaction among the dark sector states. 

The existence of an independent $U(1)$ group symmetry was originally proposed, in the context of supersymmetric theories, in \cite{Fayet:1980ad,Fayet:1980rr} and, more in general, in \cite{Okun:1982xi,Georgi:1983sy}; 
 \item \underline{axion} (spin 0): The operator comes from the  interaction between a pseudo-scalar, the axion $a$, and  the SM photon and fermions $\psi$: 
 $$ \frac{a}{f_a} \, F_{ \mu\nu} \tilde F^{\mu\nu} +\frac{1}{f_a} \partial_\mu a\, \overline{\psi} \,\gamma^\mu \gamma_5 \psi \, ,$$
 with operators of dimension  five; the physics of this portal is based on that of the axion and related to the strong CP problem as well as axion dark matter. In many cases, the portal is generalized to an axion-like particle (ALP) with similar couplings but without the constraints of the QCD axion. The parameters of the portal are two: the mass $m_a$ of the axion, or the ALP and the scale $f_a$. Often the ALP is the only member of the dark sector of these models. 
 
 The original axion emerged from addressing~\cite{Peccei:1977hh}  the strong CP problem  induced by instantons in QCD. Light pseudo-scalar bosons are found in many models of physics beyond the Standard Model;
\item \underline{scalar} (spin 0): Interaction between  a scalar  $S$ and the SM Higgs boson $H$: 
$$ (\mu S + \lambda S^2 ) H^\dag H\, ,$$
in this case the operators are of dimension three and four. The experimental limits are often expressed in terms of the two parameters $\nu$ and the mass $m_S$ of the scalar singlet, and neglecting the quartic coupling $\lambda$. In most models, the dark sector states have Yukawa-like interactions with the scalar $S$. 

The idea of a scalar singlet interacting with the Higgs boson originated within the framework of the next-to-minimal supersymmetric Standard Model~\cite{Ellis:1988er,Derendinger:1983bz,Frere:1983ag} and developed independently  in~\cite{Silveira:1985rk,Binoth:1996au,Patt:2006fw};
\item \underline{sterile neutrino} (spin $1/2$): Interaction between  a heavy fermion   $N$, which is a SM singlet, the SM Higgs boson and the SM fermions $L$:
$$ y_N \overline{L} H N\, , $$
with, again, an operator of dimension four.  The existence of heavy lepton-like fermions is suggested by neutrino see-saw models and the possible origin of baryon-number asymmetry in the leptonic sector. The experimental searches are framed in terms of the parameter $y_N$ and the mass of the  heavy fermion $N$. The sterile neutrino can be the only member of the dark sector or be one among many other dark fermions. 

The structure of the neutrino portal closely follows  that of the see-saw mechanism~\cite{Minkowski:1977sc,Yanagida:1980xy,GellMann:1980vs,Mohapatra:1980yp}---which was introduced to generate  small masses for the neutrinos---and, more in general, left-right symmetric models~\cite{Pati:1974yy,Mohapatra:1974gc,Mohapatra:1974hk,Senjanovic:1975rk}.
\end{itemize}

More details on the various portals can be found in the same references cited in the introduction: \cite{Hewett:2012ns,Essig:2013lka,Raggi:2015yfk,Deliyergiyev:2015oxa,Alekhin:2015byh,Curciarello:2016jbz,Alexander:2016aln,Beacham:2019nyx}.


\section{Boltzmann equation and relic density}
\label{sec:cosmo}

This appendix includes a short summary of some results necessary to follow the discussion in the main text about the relic density of dark matter and limits based on cosmology. We  follow the excellent review~\cite{Plehn:2017fdg}.

The rate $\Gamma$ for a the interaction between two particles is given as
\be
\Gamma = n\, \sigma\, v \, ,
\ee
the product of the corresponding  cross section $\sigma$ times the number density of the particles partaking $n$, times the their relative velocity $v$.

This  process proceeds as long as the rate is larger than the Hubble constant 
\be
H(T) =  \frac{\pi \sqrt{g_{*}(T)}}{\sqrt{90}}  \frac{T^2}{m_{Pl}} \, ,
\ee
where  $m_{Pl}$ the Planck mass and $g_*(T)$ is the number of effective degrees of freedom at the given temperature 
 is given by
\be
g_{*}(T) = \sum_\text{bosons} g_b \left( \frac{T_b}{T} \right)^4 + \frac{7}{8} \sum_\text{fermions} g_f \left( \frac{T_f}{T} \right)^4  \, , 
\ee
where $g_{b,f}$ is the number of degrees of freedom of the corresponding particle.
The value of the function  $g_{*}(T)$  goes from 106.5 above the EW phase transition to  3.38 at temperature around 0.1 MeV.

After  $\Gamma< H$, the particles are decoupled and their number density frozen.

The number density at the equilibrium at a given temperature $T$ (for $k_B=1$) is given by
\bea
n_{eq}(T) &=& g_{*} \int \frac{d^3p}{(2\pi)^3} \frac{1}{e^{E/T} \pm 1} \nn \\
&=& \left\{ \begin{array}{ll}{\displaystyle
g_{*} \left( \frac{mT}{2 \pi}\right)^{3/2} e^{-m/T} }& \mbox{non-relativistic} \quad (T \ll m)\\[0.3cm]
{\displaystyle \frac{\zeta(3)}{\pi^2} g_{*} T^3 }&  \mbox{relativistic bosons}\quad (T\gg m) \\[0.3cm]
{\displaystyle \frac{3}{4}\frac{\zeta(3)}{\pi^2} g_{*} T^3} &\mbox{relativistic fermions}\quad (T\gg m) \, ,
\end{array} \right.
\eea
where $\zeta(3)\simeq 1.2$ 

The number density $n(t)$ of a weakly interacting, massive particle $\chi$ at a certain time $t$ in the evolution of the Universe is computed by means of the Boltzmann equation 
\be
\dot{n}(t) + 3 H(t) n(t) = - \langle \sigma_{\chi \chi \rightarrow f f} v \rangle \left( n^2(t) - n^2_{eq}(t) \right) \, , \label{boltzmann}
\ee
where $H(t)$ is the Hubble constant and $\langle \sigma_{\chi \chi \rightarrow f f} v \rangle$ is
the thermal average of the cross section for a pair of the particles $\chi$, with relative velocity $v=(s-4 m_\chi^2)/m_\chi^2$, to annihilate into SM fermions $f$; this term depletes the density  as the particles $\chi$ turns into SM fermions. The thermal average is defined as
\be
\langle \sigma_{\chi \chi \rightarrow f f} v \rangle =  \frac{ \int_{\infty}^{4 m_\chi^2} d s \sqrt{s} (s - 4 m_\chi^2) K_1 \left( \frac{\sqrt{s}}{T} \right) \sigma_{\chi \chi \rightarrow f f} }{8 m_\chi^4 T \left[K_2  \left( \frac{m_\chi}{T} \right) \right]^2} \, ,\label{Taverage}
\ee 
where $K_1$ and $K_2$  are the  Bessel function of second kind. It is usually computed after expanding
\be
\langle \sigma_{\chi \chi \to f f} v \rangle = \langle s_0 + s_1 v^2 + O(v^4) \rangle
\ee
with $s_0$ the cross section in the $s$-wave and $s_1$ the first correction in the $p$-wave. The leading term is  $s_0$ for the dark sector Dirac fermions  interacting through the dark photon, in both the $s$- and $t$-channel.

 \eq{boltzmann} is usually re-written in terms of the function $Y(t) = n(t)/T^3$ and the variable $x=m_\chi/T=\sqrt{2 t H(T=m_\chi)}$ as
\be
\frac{d Y}{d x} = - \frac{\lambda(x)}{x^2} \left[ Y^2(x) - Y^2_{eq} \right] \label{b2}
\ee
with
\be
\lambda(x) = \frac{m_\chi^3 \langle \sigma_{\chi \chi \rightarrow f f} v \rangle}{H(T=m_\chi)}\, ,
\ee
and in this form numerically solved. 

\eq{b2} can be solved analytically by dropping the second term $Y^2_{eq}(x)$---which is small because decreasing like $e^{-x}$---approximating
\be
\langle \sigma_{\chi \chi \rightarrow f f} v \rangle = \sigma_{\chi \chi \rightarrow f f} v + O(v^2) \, ,
\ee
where $v=\sqrt{2/x}$ and writing
\be
\lambda (x)  =  \frac{\sqrt{180}\, m_{Pl} \, m_\chi}{\pi \sqrt{g_* x}} \sigma_{\chi \chi \rightarrow f f} 
\ee
by means of 
\be
H(T=m_\chi) = \frac{\pi \sqrt{g_*(T=m_\chi)}}{90} \frac{m_\chi^2}{m_{Pl}} \, .
\ee

The solution for $x^\prime$ larger than decoupling temperature $x_d$ is
\be
Y(x^\prime) =\frac{x_d}{\lambda} \, .
\ee
This quantity is related  to the relic density 
\be
\rho_\chi = m_\chi n(x^\prime) = m_\chi^4 \frac{Y(x^\prime )}{28 x_d}
\ee
or, in terms of the normalized quantity $\Omega_\chi = \rho_\chi/\rho_c$ as
\bea
\Omega_\chi h^2 &\simeq& 0.12 \frac{x_d}{23} \frac{\sqrt{g_*}}{10} \frac{1.7 \times 10^{-9} \mbox{GeV}^{-2}}{\langle \sigma_{\chi \chi \rightarrow f f} v \rangle}\nn \\
&\simeq & \frac{2.5 \times 10^{-10} \, \mbox{GeV}^{-2}}{\langle \sigma_{\chi \chi \rightarrow f f} v \rangle}  \, ,
\label{omega}
\eea
which provides the relationship between relic density and the annihilation cross section. 


\section{Thermal field theory }
\label{sec:TFT}
The energy loss rate ${\cal Q}$ (energy per volume and unit time)  for the emission of a pseudoscalar particle (the axion)   in a process with matrix element ${\cal}$,which is computed in the vacuum, is given by
\bea
{\cal Q} &=&\prod_{i=1} \int \frac{d^3 \mathbf{p}_i}{2 E_i (2\pi)^3} f_i (E_i) \prod_{f=1} \int \frac{d^3 \mathbf{p}_f}{2 E_f (2\pi)^3} \left[ 1\pm f_f (E_f) \right] \int \frac{d^3 \mathbf{p}_a}{2 \omega_a (2\pi)^3} \omega_a \nn \\
& & \times \frac{1}{\cal S}  \sum_{\text{spin and pol.}} |{\cal M}|^2 (2\pi)^4 \delta^4\left( \sum p_i - \sum p_f- p_a \right) \, ,\label{Q}
\eea
where $\cal S$ is a symmetrization factor for identical particles.
In \eq{Q}, the  medium is composed of the initial particles $i$ and final particles $f$ with the corresponding energy $\omega$ and momentum $\mathbf{p}$ and  with occupation number following the  distribution function (Fermi or Bose depending on the particles) ($k_B=1$):  
\be
n_j(E_j)  = g_j \int \frac{d^2 \mathbf{p}_j}{(2\pi)^3} f (E_j) \, , \label{dist}
\ee
where $g_j$ is the degeneracy number. The emitted axion carries energy $\omega_a$ and momentum $\mathbf{p}_a$.

Given the squared matrix element $\sum |{\cal M}|^2$ for the process of interest, the electron and nucleon \textit{Bremsstrahlung} in section~\ref{sec:con1},  the corresponding luminosity can be computed as
\be
L = \int d V {\cal Q} \,e^{-\tau}\, , \label{lumi}
\ee
where $\tau$ is an attenuation factor taking into account the optical depth of the emission, and compared to the observational data.


When the emitted particle mixes with the ordinary photon,   the approach above of computing the matrix element in the vacuum is no longer a reliable approximation and the full thermal field theory must be used. We follow \cite{Braaten:1993jw} in giving the essential formulas.

The electromagnetic polarization tensor is given by
\bea
\Pi^{\mu \nu} (k) &=& 16 \pi \alpha \int \frac{d^3 \mathbf{p}_i}{(2\pi)^3} \frac{1}{2\, E} \left[ n_e(E) + n_{\bar e} (E) \right] \nn \\
& \times & \frac{p\cdot k \, (p^\mu k^\nu + k^\mu p^\nu ) - k^2 p^\mu p^\nu - (p\cdot k)^2 g^{\mu\nu}} 
{(p\cdot k)^2 - (k^2)^2/4}\, ,
\eea
where $k^\mu=(\omega,\, \mathbf{k})$ and $p^\mu=(E,\, \mathbf{p})$. The transverse and longitudinal polarizations are
defined as 
\be
\Pi_T (\omega,\, \mathbf{k}) = \frac{1}{2} \left( \delta^{ij} - k^i k^j \right) \Pi^{ij} (\omega,\, \mathbf{k})
\ee
and
\be
\Pi_L (\omega,\, \mathbf{k}) = \Pi^{00} (\omega,\, \mathbf{k})\, .
\ee

The effective propagator of the photon (in the Coulomb gauge) has components
\be
D^{00}(\omega,\, k) = \frac{1}{k^2 - \Pi_L (\omega,\, k)}
\ee
and
\be
D^{ij}(\omega,\, k) = \frac{1}{k^2 - \Pi_T (\omega,\, k)} \left( \delta^{ij} - k^i k^j \right) \, .
\ee
The dispersion relationships are defined by the solutions of the equations
\be
\omega_T^2 = k^2 +  \Pi_T (\omega_T,\, k) \quad
\mbox{and} \quad \omega_L^2 =\frac{\omega_L^2}{ k^2 }+  \Pi_T(\omega_L,\, k)
\ee

In the degenerate limit, the distribution functions in \eq{dist} reduce to step functions at the Fermi momentum $p_F=\sqrt{3 \pi^2 n_e}$  and we have
\be
\Pi_T (\omega,\, \mathbf{k})  = \omega_P^2  \frac{3 \omega^2}{2 v_F^2 k^2} \left( 1 - \frac{\omega^2 - v_F^2 k^2}{2 v_F \omega k }  \log \frac{\omega + v_F k}{\omega - v_F k} \right)
\ee
and
\be
\Pi_L (\omega,\, \mathbf{k}) = \omega_P^2  \frac{3 \omega}{2 v_F^3 k} \left(  \frac{\omega}{2 v_F  k }  \log \frac{\omega + v_F k}{\omega - v_F k}-1  \right)
\ee
where $\omega_P=4 \alpha p_F^2 v_F/3 \pi$ is the plasma frequency.

The energy loss rate ${\cal Q}$ (energy per volume and unit time)  is written in terms  of the imaginary part of the polarization of the photon in the medium of charged particles. The contribution of the longitudinal and transverse modes is obtained (by the optical theorem) as
\be
{\cal Q}= - \int \frac{d^3\mathbf{k}}{(2 \pi)^3}   \frac{\text{Im} \, \Pi_{L}(\omega,\, \mathbf{k}) +\text{Im} \, \Pi_{T}(\omega,\, \mathbf{k}) }{\omega ( e^{\omega/T } -1)} \, ,
\ee
from which  the luminosity in \eq{lumi} can be computed and compared with the astrophysical limit of interest. The expression for $\text{Im} \, \Pi_{L,T}(\omega,\, \mathbf{k})$ for the massive dark photon can be found in \cite{An:2013yfc,Redondo:2013lna}   and \cite{Hardy:2016kme} where plasma effects are also included.




%
%

{\small
\bibliographystyle{utphys.bst}
\bibliography{review}

\providecommand{\href}[2]{#2}\begingroup\raggedright\begin{thebibliography}{100}

\bibitem{Holdom:1985ag}
B.~Holdom, ``{Two U(1)'s and Epsilon Charge Shifts},''
\href{http://dx.doi.org/10.1016/0370-2693(86)91377-8}{{\em Phys. Lett.}
  {\bfseries 166B} (1986) 196--198}.

\bibitem{Fayet:1990wx}
P.~Fayet, ``{Extra U(1)'s and New Forces},''
\href{http://dx.doi.org/10.1016/0550-3213(90)90381-M}{{\em Nucl. Phys.}
  {\bfseries B347} (1990) 743--768}.

\bibitem{Fayet:1980ad}
P.~Fayet, ``{Effects of the Spin 1 Partner of the Goldstino (Gravitino) on
  Neutral Current Phenomenology},''
  \href{http://dx.doi.org/10.1016/0370-2693(80)90488-8}{{\em Phys. Lett. B}
  {\bfseries 95} (1980) 285--289}.

\bibitem{Fayet:1980rr}
P.~Fayet, ``{On the Search for a New Spin 1 Boson},''
  \href{http://dx.doi.org/10.1016/0550-3213(81)90122-X}{{\em Nucl. Phys. B}
  {\bfseries 187} (1981) 184--204}.

\bibitem{Okun:1982xi}
L.~B. Okun, ``{LIMITS OF ELECTRODYNAMICS: PARAPHOTONS?},'' {\em Sov. Phys.
  JETP} {\bfseries 56} (1982) 502.
[Zh. Eksp. Teor. Fiz.83,892(1982)].

\bibitem{Georgi:1983sy}
H.~Georgi, P.~H. Ginsparg, and S.~L. Glashow, ``{Photon Oscillations and the
  Cosmic Background Radiation},''
\href{http://dx.doi.org/10.1038/306765a0}{{\em Nature} {\bfseries 306} (1983)
  765--766}.

\bibitem{Hewett:2012ns}
J.~L. Hewett {\em et~al.},
  \href{http://dx.doi.org/10.2172/1042577}{``{Fundamental Physics at the
  Intensity Frontier},''}
\newblock 2012.
\newblock \href{http://arxiv.org/abs/1205.2671}{{\ttfamily arXiv:1205.2671
  [hep-ex]}}.
\newblock
\url{http://lss.fnal.gov/archive/preprint/fermilab-conf-12-879-ppd.shtml}.
\newblock

\bibitem{Essig:2013lka}
R.~Essig {\em et~al.}, ``{Working Group Report: New Light Weakly Coupled
  Particles},'' in {\em {Proceedings, 2013 Community Summer Study on the Future
  of U.S. Particle Physics: Snowmass on the Mississippi (CSS2013): Minneapolis,
  MN, USA, July 29-August 6, 2013}}.
\newblock 2013.
\newblock \href{http://arxiv.org/abs/1311.0029}{{\ttfamily arXiv:1311.0029
  [hep-ph]}}.
\newblock
\url{http://www.slac.stanford.edu/econf/C1307292/docs/IntensityFrontier/NewLight-17.pdf}.
\newblock

\bibitem{Raggi:2015yfk}
M.~Raggi and V.~Kozhuharov, ``{Results and perspectives in dark photon
  physics},''
\href{http://dx.doi.org/10.1393/ncr/i2015-10117-9}{{\em Riv. Nuovo Cim.}
  {\bfseries 38} no.~10, (2015) 449--505}.

\bibitem{Deliyergiyev:2015oxa}
M.~A. Deliyergiyev, ``{Recent Progress in Search for Dark Sector Signatures},''
  \href{http://dx.doi.org/10.1515/phys-2016-0034}{{\em Open Phys.} {\bfseries
  14} no.~1, (2016) 281--303},
\href{http://arxiv.org/abs/1510.06927}{{\ttfamily arXiv:1510.06927 [hep-ph]}}.

\bibitem{Alekhin:2015byh}
S.~Alekhin {\em et~al.}, ``{A facility to Search for Hidden Particles at the
  CERN SPS: the SHiP physics case},''
  \href{http://dx.doi.org/10.1088/0034-4885/79/12/124201}{{\em Rept. Prog.
  Phys.} {\bfseries 79} no.~12, (2016) 124201},
\href{http://arxiv.org/abs/1504.04855}{{\ttfamily arXiv:1504.04855 [hep-ph]}}.

\bibitem{Curciarello:2016jbz}
F.~Curciarello, ``{Review on Dark Photon},''
\href{http://dx.doi.org/10.1051/epjconf/201611801008}{{\em EPJ Web Conf.}
  {\bfseries 118} (2016) 01008}.

\bibitem{Alexander:2016aln}
J.~Alexander {\em et~al.}, ``{Dark Sectors 2016 Workshop: Community Report},''
\newblock 2016.
\newblock \href{http://arxiv.org/abs/1608.08632}{{\ttfamily arXiv:1608.08632
  [hep-ph]}}.
\newblock
\url{http://lss.fnal.gov/archive/2016/conf/fermilab-conf-16-421.pdf}.
\newblock

\bibitem{Beacham:2019nyx}
J.~Beacham {\em et~al.}, ``{Physics Beyond Colliders at CERN: Beyond the
  Standard Model Working Group Report},''
  \href{http://dx.doi.org/10.1088/1361-6471/ab4cd2}{{\em J. Phys. G} {\bfseries
  47} no.~1, (2020) 010501}, \href{http://arxiv.org/abs/1901.09966}{{\ttfamily
  arXiv:1901.09966 [hep-ex]}}.

\bibitem{Dobrescu:2004wz}
B.~A. Dobrescu, ``{Massless gauge bosons other than the photon},''
  \href{http://dx.doi.org/10.1103/PhysRevLett.94.151802}{{\em Phys. Rev. Lett.}
  {\bfseries 94} (2005) 151802},
\href{http://arxiv.org/abs/hep-ph/0411004}{{\ttfamily arXiv:hep-ph/0411004
  [hep-ph]}}.

\bibitem{delAguila:1995rb}
F.~del Aguila, M.~Masip, and M.~Perez-Victoria, ``{Physical parameters and
  renormalization of U(1)-a x U(1)-b models},''
  \href{http://dx.doi.org/10.1016/0550-3213(95)00511-6}{{\em Nucl. Phys.}
  {\bfseries B456} (1995) 531--549},
\href{http://arxiv.org/abs/hep-ph/9507455}{{\ttfamily arXiv:hep-ph/9507455
  [hep-ph]}}.

\bibitem{Feldman:2007wj}
D.~Feldman, Z.~Liu, and P.~Nath, ``{The Stueckelberg Z-prime Extension with
  Kinetic Mixing and Milli-Charged Dark Matter From the Hidden Sector},''
  \href{http://dx.doi.org/10.1103/PhysRevD.75.115001}{{\em Phys. Rev.}
  {\bfseries D75} (2007) 115001},
\href{http://arxiv.org/abs/hep-ph/0702123}{{\ttfamily arXiv:hep-ph/0702123
  [HEP-PH]}}.

\bibitem{Davidson:2000hf}
S.~Davidson, S.~Hannestad, and G.~Raffelt, ``{Updated bounds on millicharged
  particles},'' \href{http://dx.doi.org/10.1088/1126-6708/2000/05/003}{{\em
  JHEP} {\bfseries 05} (2000) 003},
\href{http://arxiv.org/abs/hep-ph/0001179}{{\ttfamily arXiv:hep-ph/0001179
  [hep-ph]}}.

\bibitem{Ruegg:2003ps}
H.~Ruegg and M.~Ruiz-Altaba, ``{The Stueckelberg field},''
  \href{http://dx.doi.org/10.1142/S0217751X04019755}{{\em Int. J. Mod. Phys.}
  {\bfseries A19} (2004) 3265--3348},
\href{http://arxiv.org/abs/hep-th/0304245}{{\ttfamily arXiv:hep-th/0304245
  [hep-th]}}.

\bibitem{Appelquist:2002mw}
T.~Appelquist, B.~A. Dobrescu, and A.~R. Hopper, ``{Nonexotic Neutral Gauge
  Bosons},'' \href{http://dx.doi.org/10.1103/PhysRevD.68.035012}{{\em Phys.
  Rev.} {\bfseries D68} (2003) 035012},
\href{http://arxiv.org/abs/hep-ph/0212073}{{\ttfamily arXiv:hep-ph/0212073
  [hep-ph]}}.

\bibitem{Galison:1983pa}
P.~Galison and A.~Manohar, ``{TWO Z's OR NOT TWO Z's?},''
\href{http://dx.doi.org/10.1016/0370-2693(84)91161-4}{{\em Phys. Lett.}
  {\bfseries 136B} (1984) 279--283}.

\bibitem{He:1991qd}
X.-G. He, G.~C. Joshi, H.~Lew, and R.~R. Volkas, ``{Simplest Z-prime model},''
\href{http://dx.doi.org/10.1103/PhysRevD.44.2118}{{\em Phys. Rev.} {\bfseries
  D44} (1991) 2118--2132}.

\bibitem{Babu:1997st}
K.~S. Babu, C.~F. Kolda, and J.~March-Russell, ``{Implications of generalized Z
  - Z-prime mixing},'' \href{http://dx.doi.org/10.1103/PhysRevD.57.6788}{{\em
  Phys. Rev.} {\bfseries D57} (1998) 6788--6792},
\href{http://arxiv.org/abs/hep-ph/9710441}{{\ttfamily arXiv:hep-ph/9710441
  [hep-ph]}}.

\bibitem{Davoudiasl:2012ag}
H.~Davoudiasl, H.-S. Lee, and W.~J. Marciano, ``{'Dark' Z implications for
  Parity Violation, Rare Meson Decays, and Higgs Physics},''
  \href{http://dx.doi.org/10.1103/PhysRevD.85.115019}{{\em Phys. Rev. D}
  {\bfseries 85} (2012) 115019},
  \href{http://arxiv.org/abs/1203.2947}{{\ttfamily arXiv:1203.2947 [hep-ph]}}.

\bibitem{Heeck:2014zfa}
J.~Heeck, ``{Unbroken B ? L symmetry},''
  \href{http://dx.doi.org/10.1016/j.physletb.2014.10.067}{{\em Phys. Lett.}
  {\bfseries B739} (2014) 256--262},
\href{http://arxiv.org/abs/1408.6845}{{\ttfamily arXiv:1408.6845 [hep-ph]}}.

\bibitem{Bauer:2018onh}
M.~Bauer, P.~Foldenauer, and J.~Jaeckel, ``{Hunting All the Hidden Photons},''
  \href{http://dx.doi.org/10.1007/JHEP07(2018)094}{{\em JHEP} {\bfseries 07}
  (2018) 094}, \href{http://arxiv.org/abs/1803.05466}{{\ttfamily
  arXiv:1803.05466 [hep-ph]}}.
[JHEP18,094(2020)].

\bibitem{Fayet:2016nyc}
P.~Fayet, ``{The light $U$ boson as the mediator of a new force, coupled to a
  combination of $Q,B,L$ and dark matter},''
  \href{http://dx.doi.org/10.1140/epjc/s10052-016-4568-9}{{\em Eur. Phys. J.}
  {\bfseries C77} no.~1, (2017) 53},
\href{http://arxiv.org/abs/1611.05357}{{\ttfamily arXiv:1611.05357 [hep-ph]}}.

\bibitem{Rizzo:2018ntg}
T.~G. Rizzo, ``{Kinetic mixing, dark photons and an extra dimension. Part I},''
  \href{http://dx.doi.org/10.1007/JHEP07(2018)118}{{\em JHEP} {\bfseries 07}
  (2018) 118},
\href{http://arxiv.org/abs/1801.08525}{{\ttfamily arXiv:1801.08525 [hep-ph]}}.

\bibitem{Bertuzzo:2018ftf}
E.~Bertuzzo, S.~Jana, P.~A. Machado, and R.~Zukanovich~Funchal, ``{Neutrino
  Masses and Mixings Dynamically Generated by a Light Dark Sector},''
  \href{http://dx.doi.org/10.1016/j.physletb.2019.02.023}{{\em Phys. Lett. B}
  {\bfseries 791} (2019) 210--214},
  \href{http://arxiv.org/abs/1808.02500}{{\ttfamily arXiv:1808.02500
  [hep-ph]}}.

\bibitem{Essig:2009nc}
R.~Essig, P.~Schuster, and N.~Toro, ``{Probing Dark Forces and Light Hidden
  Sectors at Low-Energy e+e- Colliders},''
  \href{http://dx.doi.org/10.1103/PhysRevD.80.015003}{{\em Phys. Rev.}
  {\bfseries D80} (2009) 015003},
\href{http://arxiv.org/abs/0903.3941}{{\ttfamily arXiv:0903.3941 [hep-ph]}}.

\bibitem{Koren:2019iuv}
S.~Koren and R.~McGehee, ``{Freezing-in twin dark matter},''
  \href{http://dx.doi.org/10.1103/PhysRevD.101.055024}{{\em Phys. Rev. D}
  {\bfseries 101} no.~5, (2020) 055024},
  \href{http://arxiv.org/abs/1908.03559}{{\ttfamily arXiv:1908.03559
  [hep-ph]}}.

\bibitem{Gherghetta:2019coi}
T.~Gherghetta, J.~Kersten, K.~Olive, and M.~Pospelov, ``{Evaluating the price
  of tiny kinetic mixing},''
  \href{http://dx.doi.org/10.1103/PhysRevD.100.095001}{{\em Phys. Rev. D}
  {\bfseries 100} no.~9, (2019) 095001},
  \href{http://arxiv.org/abs/1909.00696}{{\ttfamily arXiv:1909.00696
  [hep-ph]}}.

\bibitem{Dienes:1996zr}
K.~R. Dienes, C.~F. Kolda, and J.~March-Russell, ``{Kinetic mixing and the
  supersymmetric gauge hierarchy},''
  \href{http://dx.doi.org/10.1016/S0550-3213(97)80028-4,
  10.1016/S0550-3213(97)00173-9}{{\em Nucl. Phys.} {\bfseries B492} (1997)
  104--118},
\href{http://arxiv.org/abs/hep-ph/9610479}{{\ttfamily arXiv:hep-ph/9610479
  [hep-ph]}}.

\bibitem{Abel:2003ue}
S.~A. Abel and B.~W. Schofield, ``{Brane anti-brane kinetic mixing,
  millicharged particles and SUSY breaking},''
  \href{http://dx.doi.org/10.1016/j.nuclphysb.2004.02.037}{{\em Nucl. Phys.}
  {\bfseries B685} (2004) 150--170},
\href{http://arxiv.org/abs/hep-th/0311051}{{\ttfamily arXiv:hep-th/0311051
  [hep-th]}}.

\bibitem{Abel:2006qt}
S.~A. Abel, J.~Jaeckel, V.~V. Khoze, and A.~Ringwald, ``{Illuminating the
  Hidden Sector of String Theory by Shining Light through a Magnetic Field},''
  \href{http://dx.doi.org/10.1016/j.physletb.2008.03.076}{{\em Phys. Lett.}
  {\bfseries B666} (2008) 66--70},
\href{http://arxiv.org/abs/hep-ph/0608248}{{\ttfamily arXiv:hep-ph/0608248
  [hep-ph]}}.

\bibitem{Goodsell:2009pi}
M.~Goodsell,
  \href{http://dx.doi.org/10.3204/DESY-PROC-2009-05/goodsell_mark}{``{Light
  Hidden U(1)s from String Theory},''} in {\em {Proceedings, 5th Patras
  Workshop on Axions, WIMPs and WISPs (AXION-WIMP 2009): Durham, UK, July
  13-17, 2009}}, pp.~165--168.
\newblock 2009.
\newblock
\href{http://arxiv.org/abs/0912.4206}{{\ttfamily arXiv:0912.4206 [hep-th]}}.
\newblock

\bibitem{Goodsell:2009xc}
M.~Goodsell, J.~Jaeckel, J.~Redondo, and A.~Ringwald, ``{Naturally Light Hidden
  Photons in LARGE Volume String Compactifications},''
  \href{http://dx.doi.org/10.1088/1126-6708/2009/11/027}{{\em JHEP} {\bfseries
  11} (2009) 027},
\href{http://arxiv.org/abs/0909.0515}{{\ttfamily arXiv:0909.0515 [hep-ph]}}.

\bibitem{Heckman:2010fh}
J.~J. Heckman and C.~Vafa, ``{An Exceptional Sector for F-theory GUTs},''
  \href{http://dx.doi.org/10.1103/PhysRevD.83.026006}{{\em Phys. Rev.}
  {\bfseries D83} (2011) 026006},
\href{http://arxiv.org/abs/1006.5459}{{\ttfamily arXiv:1006.5459 [hep-th]}}.

\bibitem{ArkaniHamed:2008qp}
N.~Arkani-Hamed and N.~Weiner, ``{LHC Signals for a SuperUnified Theory of Dark
  Matter},'' \href{http://dx.doi.org/10.1088/1126-6708/2008/12/104}{{\em JHEP}
  {\bfseries 12} (2008) 104},
\href{http://arxiv.org/abs/0810.0714}{{\ttfamily arXiv:0810.0714 [hep-ph]}}.

\bibitem{Chan:2011aa}
Y.~F. Chan, M.~Low, D.~E. Morrissey, and A.~P. Spray, ``{LHC Signatures of a
  Minimal Supersymmetric Hidden Valley},''
  \href{http://dx.doi.org/10.1007/JHEP05(2012)155}{{\em JHEP} {\bfseries 05}
  (2012) 155},
\href{http://arxiv.org/abs/1112.2705}{{\ttfamily arXiv:1112.2705 [hep-ph]}}.

\bibitem{Grzadkowski:2010es}
B.~Grzadkowski, M.~Iskrzynski, M.~Misiak, and J.~Rosiek, ``{Dimension-Six Terms
  in the Standard Model Lagrangian},''
  \href{http://dx.doi.org/10.1007/JHEP10(2010)085}{{\em JHEP} {\bfseries 10}
  (2010) 085},
\href{http://arxiv.org/abs/1008.4884}{{\ttfamily arXiv:1008.4884 [hep-ph]}}.

\bibitem{Goldberg:1986nk}
H.~Goldberg and L.~J. Hall, ``{A New Candidate for Dark Matter},''
  \href{http://dx.doi.org/10.1016/0370-2693(86)90731-8}{{\em Phys. Lett.}
  {\bfseries B174} (1986) 151}.
[,467(1986)].

\bibitem{Holdom:1986eq}
B.~Holdom, ``{Searching for $\epsilon$ Charges and a New U(1)},''
\href{http://dx.doi.org/10.1016/0370-2693(86)90470-3}{{\em Phys. Lett.}
  {\bfseries B178} (1986) 65--70}.

\bibitem{Gradwohl:1992ue}
B.-A. Gradwohl and J.~A. Frieman, ``{Dark matter, long range forces, and large
  scale structure},''
\href{http://dx.doi.org/10.1086/171865}{{\em Astrophys. J.} {\bfseries 398}
  (1992) 407--424}.

\bibitem{Carlson:1992fn}
E.~D. Carlson, M.~E. Machacek, and L.~J. Hall, ``{Self-interacting dark
  matter},''
\href{http://dx.doi.org/10.1086/171833}{{\em Astrophys. J.} {\bfseries 398}
  (1992) 43--52}.

\bibitem{Foot:2004pa}
R.~Foot, ``{Mirror matter-type dark matter},''
  \href{http://dx.doi.org/10.1142/S0218271804006449}{{\em Int. J. Mod. Phys.}
  {\bfseries D13} (2004) 2161--2192},
\href{http://arxiv.org/abs/astro-ph/0407623}{{\ttfamily arXiv:astro-ph/0407623
  [astro-ph]}}.

\bibitem{Feng:2008mu}
J.~L. Feng, H.~Tu, and H.-B. Yu, ``{Thermal Relics in Hidden Sectors},''
  \href{http://dx.doi.org/10.1088/1475-7516/2008/10/043}{{\em JCAP} {\bfseries
  0810} (2008) 043},
\href{http://arxiv.org/abs/0808.2318}{{\ttfamily arXiv:0808.2318 [hep-ph]}}.

\bibitem{Ackerman:mha}
L.~Ackerman, M.~R. Buckley, S.~M. Carroll, and M.~Kamionkowski, ``{Dark Matter
  and Dark Radiation},'' \href{http://dx.doi.org/10.1103/PhysRevD.79.023519,
  10.1142/9789814293792_0021}{{\em Phys. Rev.} {\bfseries D79} (2009) 023519},
  \href{http://arxiv.org/abs/0810.5126}{{\ttfamily arXiv:0810.5126 [hep-ph]}}.
[,277(2008)].

\bibitem{Feng:2009mn}
J.~L. Feng, M.~Kaplinghat, H.~Tu, and H.-B. Yu, ``{Hidden Charged Dark
  Matter},'' \href{http://dx.doi.org/10.1088/1475-7516/2009/07/004}{{\em JCAP}
  {\bfseries 0907} (2009) 004},
\href{http://arxiv.org/abs/0905.3039}{{\ttfamily arXiv:0905.3039 [hep-ph]}}.

\bibitem{ArkaniHamed:2008qn}
N.~Arkani-Hamed, D.~P. Finkbeiner, T.~R. Slatyer, and N.~Weiner, ``{A Theory of
  Dark Matter},'' \href{http://dx.doi.org/10.1103/PhysRevD.79.015014}{{\em
  Phys. Rev.} {\bfseries D79} (2009) 015014},
\href{http://arxiv.org/abs/0810.0713}{{\ttfamily arXiv:0810.0713 [hep-ph]}}.

\bibitem{Kaplan:2009de}
D.~E. Kaplan, G.~Z. Krnjaic, K.~R. Rehermann, and C.~M. Wells, ``{Atomic Dark
  Matter},'' \href{http://dx.doi.org/10.1088/1475-7516/2010/05/021}{{\em JCAP}
  {\bfseries 1005} (2010) 021},
\href{http://arxiv.org/abs/0909.0753}{{\ttfamily arXiv:0909.0753 [hep-ph]}}.

\bibitem{Buckley:2009in}
M.~R. Buckley and P.~J. Fox, ``{Dark Matter Self-Interactions and Light Force
  Carriers},'' \href{http://dx.doi.org/10.1103/PhysRevD.81.083522}{{\em Phys.
  Rev.} {\bfseries D81} (2010) 083522},
\href{http://arxiv.org/abs/0911.3898}{{\ttfamily arXiv:0911.3898 [hep-ph]}}.

\bibitem{Hooper:2012cw}
D.~Hooper, N.~Weiner, and W.~Xue, ``{Dark Forces and Light Dark Matter},''
  \href{http://dx.doi.org/10.1103/PhysRevD.86.056009}{{\em Phys. Rev.}
  {\bfseries D86} (2012) 056009},
\href{http://arxiv.org/abs/1206.2929}{{\ttfamily arXiv:1206.2929 [hep-ph]}}.

\bibitem{Aarssen:2012fx}
L.~G. van~den Aarssen, T.~Bringmann, and C.~Pfrommer, ``{Is dark matter with
  long-range interactions a solution to all small-scale problems of $\Lambda$
  CDM cosmology?},''
  \href{http://dx.doi.org/10.1103/PhysRevLett.109.231301}{{\em Phys. Rev.
  Lett.} {\bfseries 109} (2012) 231301},
\href{http://arxiv.org/abs/1205.5809}{{\ttfamily arXiv:1205.5809
  [astro-ph.CO]}}.

\bibitem{Cline:2012is}
J.~M. Cline, Z.~Liu, and W.~Xue, ``{Millicharged Atomic Dark Matter},''
  \href{http://dx.doi.org/10.1103/PhysRevD.85.101302}{{\em Phys. Rev.}
  {\bfseries D85} (2012) 101302},
\href{http://arxiv.org/abs/1201.4858}{{\ttfamily arXiv:1201.4858 [hep-ph]}}.

\bibitem{Tulin:2013teo}
S.~Tulin, H.-B. Yu, and K.~M. Zurek, ``{Beyond Collisionless Dark Matter:
  Particle Physics Dynamics for Dark Matter Halo Structure},''
  \href{http://dx.doi.org/10.1103/PhysRevD.87.115007}{{\em Phys. Rev.}
  {\bfseries D87} no.~11, (2013) 115007},
\href{http://arxiv.org/abs/1302.3898}{{\ttfamily arXiv:1302.3898 [hep-ph]}}.

\bibitem{Gabrielli:2013jka}
E.~Gabrielli and M.~Raidal, ``{Exponentially spread dynamical Yukawa couplings
  from nonperturbative chiral symmetry breaking in the dark sector},''
  \href{http://dx.doi.org/10.1103/PhysRevD.89.015008}{{\em Phys. Rev.}
  {\bfseries D89} no.~1, (2014) 015008},
\href{http://arxiv.org/abs/1310.1090}{{\ttfamily arXiv:1310.1090 [hep-ph]}}.

\bibitem{Baldi:2012ua}
M.~Baldi, ``{Structure formation in Multiple Dark Matter cosmologies with
  long-range scalar interactions},''
  \href{http://dx.doi.org/10.1093/mnras/sts169}{{\em Mon. Not. Roy. Astron.
  Soc.} {\bfseries 428} (2013) 2074},
\href{http://arxiv.org/abs/1206.2348}{{\ttfamily arXiv:1206.2348
  [astro-ph.CO]}}.

\bibitem{CyrRacine:2012fz}
F.-Y. Cyr-Racine and K.~Sigurdson, ``{Cosmology of atomic dark matter},''
  \href{http://dx.doi.org/10.1103/PhysRevD.87.103515}{{\em Phys. Rev.}
  {\bfseries D87} no.~10, (2013) 103515},
\href{http://arxiv.org/abs/1209.5752}{{\ttfamily arXiv:1209.5752
  [astro-ph.CO]}}.

\bibitem{Cline:2013zca}
J.~M. Cline, Z.~Liu, G.~Moore, and W.~Xue, ``{Composite strongly interacting
  dark matter},'' \href{http://dx.doi.org/10.1103/PhysRevD.90.015023}{{\em
  Phys. Rev.} {\bfseries D90} no.~1, (2014) 015023},
\href{http://arxiv.org/abs/1312.3325}{{\ttfamily arXiv:1312.3325 [hep-ph]}}.

\bibitem{Chu:2014lja}
X.~Chu and B.~Dasgupta, ``{Dark Radiation Alleviates Problems with Dark Matter
  Halos},'' \href{http://dx.doi.org/10.1103/PhysRevLett.113.161301}{{\em Phys.
  Rev. Lett.} {\bfseries 113} no.~16, (2014) 161301},
\href{http://arxiv.org/abs/1404.6127}{{\ttfamily arXiv:1404.6127 [hep-ph]}}.

\bibitem{Boddy:2014yra}
K.~K. Boddy, J.~L. Feng, M.~Kaplinghat, and T.~M.~P. Tait, ``{Self-Interacting
  Dark Matter from a Non-Abelian Hidden Sector},''
  \href{http://dx.doi.org/10.1103/PhysRevD.89.115017}{{\em Phys. Rev.}
  {\bfseries D89} no.~11, (2014) 115017},
\href{http://arxiv.org/abs/1402.3629}{{\ttfamily arXiv:1402.3629 [hep-ph]}}.

\bibitem{Buen-Abad:2015ova}
M.~A. Buen-Abad, G.~Marques-Tavares, and M.~Schmaltz, ``{Non-Abelian dark
  matter and dark radiation},''
  \href{http://dx.doi.org/10.1103/PhysRevD.92.023531}{{\em Phys. Rev.}
  {\bfseries D92} no.~2, (2015) 023531},
\href{http://arxiv.org/abs/1505.03542}{{\ttfamily arXiv:1505.03542 [hep-ph]}}.

\bibitem{Agrawal:2016quu}
P.~Agrawal, F.-Y. Cyr-Racine, L.~Randall, and J.~Scholtz, ``{Make Dark Matter
  Charged Again},'' \href{http://dx.doi.org/10.1088/1475-7516/2017/05/022}{{\em
  JCAP} {\bfseries 1705} no.~05, (2017) 022},
\href{http://arxiv.org/abs/1610.04611}{{\ttfamily arXiv:1610.04611 [hep-ph]}}.

\bibitem{Clowe:2006eq}
D.~Clowe, M.~Bradac, A.~H. Gonzalez, M.~Markevitch, S.~W. Randall, C.~Jones,
  and D.~Zaritsky, ``{A direct empirical proof of the existence of dark
  matter},'' \href{http://dx.doi.org/10.1086/508162}{{\em Astrophys. J.}
  {\bfseries 648} (2006) L109--L113},
\href{http://arxiv.org/abs/astro-ph/0608407}{{\ttfamily arXiv:astro-ph/0608407
  [astro-ph]}}.

\bibitem{Feng:2009hw}
J.~L. Feng, M.~Kaplinghat, and H.-B. Yu, ``{Halo Shape and Relic Density
  Exclusions of Sommerfeld-Enhanced Dark Matter Explanations of Cosmic Ray
  Excesses},'' \href{http://dx.doi.org/10.1103/PhysRevLett.104.151301}{{\em
  Phys. Rev. Lett.} {\bfseries 104} (2010) 151301},
\href{http://arxiv.org/abs/0911.0422}{{\ttfamily arXiv:0911.0422 [hep-ph]}}.

\bibitem{Lin:2011gj}
T.~Lin, H.-B. Yu, and K.~M. Zurek, ``{On Symmetric and Asymmetric Light Dark
  Matter},'' \href{http://dx.doi.org/10.1103/PhysRevD.85.063503}{{\em Phys.
  Rev.} {\bfseries D85} (2012) 063503},
\href{http://arxiv.org/abs/1111.0293}{{\ttfamily arXiv:1111.0293 [hep-ph]}}.

\bibitem{Peter:2012jh}
A.~H.~G. Peter, M.~Rocha, J.~S. Bullock, and M.~Kaplinghat, ``{Cosmological
  Simulations with Self-Interacting Dark Matter II: Halo Shapes vs.
  Observations},'' \href{http://dx.doi.org/10.1093/mnras/sts535}{{\em Mon. Not.
  Roy. Astron. Soc.} {\bfseries 430} (2013) 105},
\href{http://arxiv.org/abs/1208.3026}{{\ttfamily arXiv:1208.3026
  [astro-ph.CO]}}.

\bibitem{Pospelov:2000bq}
M.~Pospelov and T.~ter Veldhuis, ``{Direct and indirect limits on the
  electromagnetic form-factors of WIMPs},''
  \href{http://dx.doi.org/10.1016/S0370-2693(00)00358-0}{{\em Phys.\ Lett.\ B}
  {\bfseries 480} (2000) 181--186},
  \href{http://arxiv.org/abs/hep-ph/0003010}{{\ttfamily arXiv:hep-ph/0003010}}.

\bibitem{Sigurdson:2004zp}
K.~Sigurdson, M.~Doran, A.~Kurylov, R.~R. Caldwell, and M.~Kamionkowski,
  ``{Dark-matter electric and magnetic dipole moments},''
  \href{http://dx.doi.org/10.1103/PhysRevD.70.083501,
  10.1103/PhysRevD.73.089903}{{\em Phys. Rev.} {\bfseries D70} (2004) 083501},
  \href{http://arxiv.org/abs/astro-ph/0406355}{{\ttfamily
  arXiv:astro-ph/0406355 [astro-ph]}}.
[Erratum: Phys. Rev.D73,089903(2006)].

\bibitem{Banks:2010eh}
T.~Banks, J.-F. Fortin, and S.~Thomas, ``{Direct Detection of Dark Matter
  Electromagnetic Dipole Moments},''
\href{http://arxiv.org/abs/1007.5515}{{\ttfamily arXiv:1007.5515 [hep-ph]}}.

\bibitem{Barger:2010gv}
V.~Barger, W.-Y. Keung, and D.~Marfatia, ``{Electromagnetic properties of dark
  matter: Dipole moments and charge form factor},''
  \href{http://dx.doi.org/10.1016/j.physletb.2010.12.008}{{\em Phys.\ Lett.\ B}
  {\bfseries 696} (2011) 74--78},
  \href{http://arxiv.org/abs/1007.4345}{{\ttfamily arXiv:1007.4345 [hep-ph]}}.

\bibitem{Fornengo:2011sz}
N.~Fornengo, P.~Panci, and M.~Regis, ``{Long-Range Forces in Direct Dark Matter
  Searches},'' \href{http://dx.doi.org/10.1103/PhysRevD.84.115002}{{\em Phys.\
  Rev.\ D} {\bfseries 84} (2011) 115002},
  \href{http://arxiv.org/abs/1108.4661}{{\ttfamily arXiv:1108.4661 [hep-ph]}}.

\bibitem{DelNobile:2012tx}
E.~Del~Nobile, C.~Kouvaris, P.~Panci, F.~Sannino, and J.~Virkajarvi, ``{Light
  Magnetic Dark Matter in Direct Detection Searches},''
  \href{http://dx.doi.org/10.1088/1475-7516/2012/08/010}{{\em JCAP} {\bfseries
  08} (2012) 010}, \href{http://arxiv.org/abs/1203.6652}{{\ttfamily
  arXiv:1203.6652 [hep-ph]}}.

\bibitem{Chu:2020ysb}
X.~Chu, J.-L. Kuo, and J.~Pradler, ``{Dark sector-photon interactions in
  proton-beam experiments},''
  \href{http://dx.doi.org/10.1103/PhysRevD.101.075035}{{\em Phys. Rev. D}
  {\bfseries 101} (2020) 075035},
  \href{http://arxiv.org/abs/2001.06042}{{\ttfamily arXiv:2001.06042
  [hep-ph]}}.

\bibitem{Fitzpatrick:2012ix}
A.~L. Fitzpatrick, W.~Haxton, E.~Katz, N.~Lubbers, and Y.~Xu, ``{The Effective
  Field Theory of Dark Matter Direct Detection},''
  \href{http://dx.doi.org/10.1088/1475-7516/2013/02/004}{{\em JCAP} {\bfseries
  1302} (2013) 004},
\href{http://arxiv.org/abs/1203.3542}{{\ttfamily arXiv:1203.3542 [hep-ph]}}.

\bibitem{Liem:2016xpm}
S.~Liem, G.~Bertone, F.~Calore, R.~Ruiz~de Austri, T.~M.~P. Tait, R.~Trotta,
  and C.~Weniger, ``{Effective field theory of dark matter: a global
  analysis},'' \href{http://dx.doi.org/10.1007/JHEP09(2016)077}{{\em JHEP}
  {\bfseries 09} (2016) 077}, \href{http://arxiv.org/abs/1603.05994}{{\ttfamily
  arXiv:1603.05994 [hep-ph]}}.

\bibitem{Brod:2017bsw}
J.~Brod, A.~Gootjes-Dreesbach, M.~Tammaro, and J.~Zupan, ``{Effective Field
  Theory for Dark Matter Direct Detection up to Dimension Seven},''
  \href{http://dx.doi.org/10.1007/JHEP10(2018)065}{{\em JHEP} {\bfseries 10}
  (2018) 065},
\href{http://arxiv.org/abs/1710.10218}{{\ttfamily arXiv:1710.10218 [hep-ph]}}.

\bibitem{Boehm:2002yz}
C.~Boehm, T.~Ensslin, and J.~Silk, ``{Can Annihilating dark matter be lighter
  than a few GeVs?},'' \href{http://dx.doi.org/10.1088/0954-3899/30/3/004}{{\em
  J. Phys. G} {\bfseries 30} (2004) 279--286},
  \href{http://arxiv.org/abs/astro-ph/0208458}{{\ttfamily
  arXiv:astro-ph/0208458}}.

\bibitem{Knapen:2017xzo}
S.~Knapen, T.~Lin, and K.~M. Zurek, ``{Light Dark Matter: Models and
  Constraints},'' \href{http://dx.doi.org/10.1103/PhysRevD.96.115021}{{\em
  Phys. Rev.} {\bfseries D96} no.~11, (2017) 115021},
\href{http://arxiv.org/abs/1709.07882}{{\ttfamily arXiv:1709.07882 [hep-ph]}}.

\bibitem{Essig:2011nj}
R.~Essig, J.~Mardon, and T.~Volansky, ``{Direct Detection of Sub-GeV Dark
  Matter},'' \href{http://dx.doi.org/10.1103/PhysRevD.85.076007}{{\em Phys.
  Rev.} {\bfseries D85} (2012) 076007},
\href{http://arxiv.org/abs/1108.5383}{{\ttfamily arXiv:1108.5383 [hep-ph]}}.

\bibitem{Boehm:2003hm}
C.~Boehm and P.~Fayet, ``{Scalar dark matter candidates},''
  \href{http://dx.doi.org/10.1016/j.nuclphysb.2004.01.015}{{\em Nucl. Phys. B}
  {\bfseries 683} (2004) 219--263},
  \href{http://arxiv.org/abs/hep-ph/0305261}{{\ttfamily arXiv:hep-ph/0305261}}.

\bibitem{Hambye:2019dwd}
T.~Hambye, M.~H. Tytgat, J.~Vandecasteele, and L.~Vanderheyden, ``{Dark matter
  from dark photons: a taxonomy of dark matter production},''
  \href{http://dx.doi.org/10.1103/PhysRevD.100.095018}{{\em Phys. Rev. D}
  {\bfseries 100} no.~9, (2019) 095018},
  \href{http://arxiv.org/abs/1908.09864}{{\ttfamily arXiv:1908.09864
  [hep-ph]}}.

\bibitem{Izaguirre:2015yja}
E.~Izaguirre, G.~Krnjaic, P.~Schuster, and N.~Toro, ``{Analyzing the Discovery
  Potential for Light Dark Matter},''
  \href{http://dx.doi.org/10.1103/PhysRevLett.115.251301}{{\em Phys. Rev.
  Lett.} {\bfseries 115} no.~25, (2015) 251301},
\href{http://arxiv.org/abs/1505.00011}{{\ttfamily arXiv:1505.00011 [hep-ph]}}.

\bibitem{Mondino:2020lsc}
C.~Mondino, M.~Pospelov, J.~T. Ruderman, and O.~Slone, ``{Dark Higgs Dark
  Matter},'' \href{http://arxiv.org/abs/2005.02397}{{\ttfamily arXiv:2005.02397
  [hep-ph]}}.

\bibitem{Preskill:1982cy}
J.~Preskill, M.~B. Wise, and F.~Wilczek, ``{Cosmology of the Invisible
  Axion},''
\href{http://dx.doi.org/10.1016/0370-2693(83)90637-8}{{\em Phys. Lett.}
  {\bfseries 120B} (1983) 127--132}.

\bibitem{Abbott:1982af}
L.~F. Abbott and P.~Sikivie, ``{A Cosmological Bound on the Invisible Axion},''
\href{http://dx.doi.org/10.1016/0370-2693(83)90638-X}{{\em Phys. Lett.}
  {\bfseries 120B} (1983) 133--136}.

\bibitem{Dine:1982ah}
M.~Dine and W.~Fischler, ``{The Not So Harmless Axion},''
\href{http://dx.doi.org/10.1016/0370-2693(83)90639-1}{{\em Phys. Lett.}
  {\bfseries 120B} (1983) 137--141}.

\bibitem{Nelson:2011sf}
A.~E. Nelson and J.~Scholtz, ``{Dark Light, Dark Matter and the Misalignment
  Mechanism},'' \href{http://dx.doi.org/10.1103/PhysRevD.84.103501}{{\em Phys.
  Rev.} {\bfseries D84} (2011) 103501},
\href{http://arxiv.org/abs/1105.2812}{{\ttfamily arXiv:1105.2812 [hep-ph]}}.

\bibitem{Arias:2012az}
P.~Arias, D.~Cadamuro, M.~Goodsell, J.~Jaeckel, J.~Redondo, and A.~Ringwald,
  ``{WISPy Cold Dark Matter},''
  \href{http://dx.doi.org/10.1088/1475-7516/2012/06/013}{{\em JCAP} {\bfseries
  1206} (2012) 013},
\href{http://arxiv.org/abs/1201.5902}{{\ttfamily arXiv:1201.5902 [hep-ph]}}.

\bibitem{Graham:2015rva}
P.~W. Graham, J.~Mardon, and S.~Rajendran, ``{Vector Dark Matter from
  Inflationary Fluctuations},''
  \href{http://dx.doi.org/10.1103/PhysRevD.93.103520}{{\em Phys. Rev. D}
  {\bfseries 93} no.~10, (2016) 103520},
  \href{http://arxiv.org/abs/1504.02102}{{\ttfamily arXiv:1504.02102
  [hep-ph]}}.

\bibitem{Nakai:2020cfw}
Y.~Nakai, R.~Namba, and Z.~Wang, ``{Light Dark Photon Dark Matter from
  Inflation},'' \href{http://arxiv.org/abs/2004.10743}{{\ttfamily
  arXiv:2004.10743 [hep-ph]}}.

\bibitem{Pospelov:2008jk}
M.~Pospelov, A.~Ritz, and M.~B. Voloshin, ``{Bosonic super-WIMPs as keV-scale
  dark matter},'' \href{http://dx.doi.org/10.1103/PhysRevD.78.115012}{{\em
  Phys. Rev.} {\bfseries D78} (2008) 115012},
\href{http://arxiv.org/abs/0807.3279}{{\ttfamily arXiv:0807.3279 [hep-ph]}}.

\bibitem{Bloch:2016sjj}
I.~M. Bloch, R.~Essig, K.~Tobioka, T.~Volansky, and T.-T. Yu, ``{Searching for
  Dark Absorption with Direct Detection Experiments},''
  \href{http://dx.doi.org/10.1007/JHEP06(2017)087}{{\em JHEP} {\bfseries 06}
  (2017) 087},
\href{http://arxiv.org/abs/1608.02123}{{\ttfamily arXiv:1608.02123 [hep-ph]}}.

\bibitem{Hoffmann:1987et}
S.~Hoffmann, ``{Paraphotons and Axions: Similarities in Stellar Emission and
  Detection},''
\href{http://dx.doi.org/10.1016/0370-2693(87)90467-9}{{\em Phys. Lett.}
  {\bfseries B193} (1987) 117--122}.

\bibitem{Raffelt:1996wa}
G.~G. Raffelt, {\em {Stars as laboratories for fundamental physics}}.
\newblock Chicago, USA: Univ. Pr. (1996) 664 p, 1996.
\newblock
\url{http://wwwth.mpp.mpg.de/members/raffelt/mypapers/199613.pdf}.
\newblock

\bibitem{Carlson:1986cu}
E.~D. Carlson, ``{LIMITS ON A NEW U(1) COUPLING},''
  \href{http://dx.doi.org/10.1016/0550-3213(87)90446-9}{{\em Nucl.\ Phys.\ B}
  {\bfseries 286} (1987) 378--398}.

\bibitem{Nakagawa:1987pga}
M.~Nakagawa, Y.~Kohyama, and N.~Itoh, ``{Axion Bremsstrahlung in Dense
  Stars},'' \href{http://dx.doi.org/10.1086/165724}{{\em Astrophys.\ J.}
  {\bfseries 322} (1987) 291}.

\bibitem{Raffelt:1989zt}
G.~G. Raffelt, ``{Axion bremsstrahlung in red giants},''
  \href{http://dx.doi.org/10.1103/PhysRevD.41.1324}{{\em Phys.\ Rev.\ D}
  {\bfseries 41} (1990) 1324--1326}.

\bibitem{Bertolami:2014wua}
M.~M. Miller~Bertolami, B.~E. Melendez, L.~G. Althaus, and J.~Isern,
  ``{Revisiting the axion bounds from the Galactic white dwarf luminosity
  function},'' \href{http://dx.doi.org/10.1088/1475-7516/2014/10/069}{{\em
  JCAP} {\bfseries 1410} no.~10, (2014) 069},
\href{http://arxiv.org/abs/1406.7712}{{\ttfamily arXiv:1406.7712 [hep-ph]}}.

\bibitem{Viaux:2013lha}
N.~Viaux, M.~Catelan, P.~B. Stetson, G.~Raffelt, J.~Redondo, A.~A.~R. Valcarce,
  and A.~Weiss, ``{Neutrino and axion bounds from the globular cluster M5 (NGC
  5904)},'' \href{http://dx.doi.org/10.1103/PhysRevLett.111.231301}{{\em Phys.
  Rev. Lett.} {\bfseries 111} (2013) 231301},
\href{http://arxiv.org/abs/1311.1669}{{\ttfamily arXiv:1311.1669
  [astro-ph.SR]}}.

\bibitem{Giannotti:2015kwo}
M.~Giannotti, I.~Irastorza, J.~Redondo, and A.~Ringwald, ``{Cool WISPs for
  stellar cooling excesses},''
  \href{http://dx.doi.org/10.1088/1475-7516/2016/05/057}{{\em JCAP} {\bfseries
  1605} no.~05, (2016) 057},
\href{http://arxiv.org/abs/1512.08108}{{\ttfamily arXiv:1512.08108
  [astro-ph.HE]}}.

\bibitem{Brinkmann:1988vi}
R.~P. Brinkmann and M.~S. Turner, ``{Numerical Rates for Nucleon-Nucleon Axion
  Bremsstrahlung},'' \href{http://dx.doi.org/10.1103/PhysRevD.38.2338}{{\em
  Phys.\ Rev.\ D} {\bfseries 38} (1988) 2338}.

\bibitem{Raffelt:1993ix}
G.~Raffelt and D.~Seckel, ``{A selfconsistent approach to neutral current
  processes in supernova cores},''
  \href{http://dx.doi.org/10.1103/PhysRevD.52.1780}{{\em Phys.\ Rev.\ D}
  {\bfseries 52} (1995) 1780--1799},
  \href{http://arxiv.org/abs/astro-ph/9312019}{{\ttfamily
  arXiv:astro-ph/9312019}}.

\bibitem{Iwamoto:1984ir}
N.~Iwamoto, ``{Axion Emission from Neutron Stars},''
  \href{http://dx.doi.org/10.1103/PhysRevLett.53.1198}{{\em Phys.\ Rev.\ Lett.}
  {\bfseries 53} (1984) 1198--1201}.

\bibitem{Keil:1996ju}
W.~Keil, H.-T. Janka, D.~N. Schramm, G.~Sigl, M.~S. Turner, and J.~R. Ellis,
  ``{A Fresh look at axions and SN-1987A},''
  \href{http://dx.doi.org/10.1103/PhysRevD.56.2419}{{\em Phys. Rev.} {\bfseries
  D56} (1997) 2419--2432},
\href{http://arxiv.org/abs/astro-ph/9612222}{{\ttfamily arXiv:astro-ph/9612222
  [astro-ph]}}.

\bibitem{Carenza:2019pxu}
P.~Carenza, T.~Fischer, M.~Giannotti, G.~Guo, G.~Mart{\'i}nez-Pinedo, and
  A.~Mirizzi, ``{Improved axion emissivity from a supernova via nucleon-nucleon
  bremsstrahlung},''
  \href{http://dx.doi.org/10.1088/1475-7516/2019/10/016}{{\em JCAP} {\bfseries
  1910} no.~10, (2019) 016},
\href{http://arxiv.org/abs/1906.11844}{{\ttfamily arXiv:1906.11844 [hep-ph]}}.

\bibitem{Fields:2019pfx}
B.~D. Fields, K.~A. Olive, T.-H. Yeh, and C.~Young, ``{Big-Bang Nucleosynthesis
  After Planck},'' \href{http://dx.doi.org/10.1088/1475-7516/2020/03/010}{{\em
  JCAP} {\bfseries 2003} no.~03, (2020) 010},
\href{http://arxiv.org/abs/1912.01132}{{\ttfamily arXiv:1912.01132
  [astro-ph.CO]}}.

\bibitem{Dobrescu:2006au}
B.~A. Dobrescu and I.~Mocioiu, ``{Spin-dependent macroscopic forces from new
  particle exchange},''
  \href{http://dx.doi.org/10.1088/1126-6708/2006/11/005}{{\em JHEP} {\bfseries
  11} (2006) 005},
\href{http://arxiv.org/abs/hep-ph/0605342}{{\ttfamily arXiv:hep-ph/0605342
  [hep-ph]}}.

\bibitem{Ficek:2016qwp}
F.~Ficek, D.~F.~J. Kimball, M.~Kozlov, N.~Leefer, S.~Pustelny, and D.~Budker,
  ``{Constraints on exotic spin-dependent interactions between electrons from
  helium fine-structure spectroscopy},''
  \href{http://dx.doi.org/10.1103/PhysRevA.95.032505}{{\em Phys. Rev.}
  {\bfseries A95} no.~3, (2017) 032505},
\href{http://arxiv.org/abs/1608.05779}{{\ttfamily arXiv:1608.05779
  [physics.atom-ph]}}.

\bibitem{Ni:1999di}
W.-T. Ni, S.-S. Pan, H.-C. Yeh, L.-S. Hou, and J.-L. Wan, ``{Search for an
  axionlike spin coupling using a paramagnetic salt with a dc SQUID},''
\href{http://dx.doi.org/10.1103/PhysRevLett.82.2439}{{\em Phys. Rev. Lett.}
  {\bfseries 82} (1999) 2439--2442}.

\bibitem{Wineland:1991zz}
D.~J. Wineland, J.~J. Bollinger, D.~J. Heinzen, W.~M. Itano, and M.~G. Raizen,
  ``{Search for anomalous spin-dependent forces using stored-ion
  spectroscopy},''
\href{http://dx.doi.org/10.1103/PhysRevLett.67.1735}{{\em Phys. Rev. Lett.}
  {\bfseries 67} (1991) 1735--1738}.

\bibitem{Hanneke:2008tm}
D.~Hanneke, S.~Fogwell, and G.~Gabrielse, ``{New Measurement of the Electron
  Magnetic Moment and the Fine Structure Constant},''
  \href{http://dx.doi.org/10.1103/PhysRevLett.100.120801}{{\em Phys. Rev.
  Lett.} {\bfseries 100} (2008) 120801},
\href{http://arxiv.org/abs/0801.1134}{{\ttfamily arXiv:0801.1134
  [physics.atom-ph]}}.

\bibitem{Giudice:2012ms}
G.~Giudice, P.~Paradisi, and M.~Passera, ``{Testing new physics with the
  electron g-2},'' \href{http://dx.doi.org/10.1007/JHEP11(2012)113}{{\em JHEP}
  {\bfseries 11} (2012) 113}, \href{http://arxiv.org/abs/1208.6583}{{\ttfamily
  arXiv:1208.6583 [hep-ph]}}.

\bibitem{Bennett:2006fi}
{\bfseries Muon g-2} Collaboration, G.~W. Bennett {\em et~al.}, ``{Final Report
  of the Muon E821 Anomalous Magnetic Moment Measurement at BNL},''
  \href{http://dx.doi.org/10.1103/PhysRevD.73.072003}{{\em Phys. Rev.}
  {\bfseries D73} (2006) 072003},
\href{http://arxiv.org/abs/hep-ex/0602035}{{\ttfamily arXiv:hep-ex/0602035
  [hep-ex]}}.

\bibitem{Blum:2018mom}
{\bfseries RBC, UKQCD} Collaboration, T.~Blum, P.~Boyle, V.~G{\"u}lpers,
  T.~Izubuchi, L.~Jin, C.~Jung, A.~J{\"u}ttner, C.~Lehner, A.~Portelli, and
  J.~Tsang, ``{Calculation of the hadronic vacuum polarization contribution to
  the muon anomalous magnetic moment},''
  \href{http://dx.doi.org/10.1103/PhysRevLett.121.022003}{{\em Phys. Rev.
  Lett.} {\bfseries 121} no.~2, (2018) 022003},
  \href{http://arxiv.org/abs/1801.07224}{{\ttfamily arXiv:1801.07224
  [hep-lat]}}.

\bibitem{Bayes:2014lxz}
{\bfseries TWIST} Collaboration, R.~Bayes {\em et~al.}, ``{Search for two body
  muon decay signals},''
  \href{http://dx.doi.org/10.1103/PhysRevD.91.052020}{{\em Phys. Rev. D}
  {\bfseries 91} no.~5, (2015) 052020},
  \href{http://arxiv.org/abs/1409.0638}{{\ttfamily arXiv:1409.0638 [hep-ex]}}.

\bibitem{Engelfried:2019pva}
{\bfseries NA62} Collaboration, J.~Engelfried, ``{Search for $K^+\rightarrow
  \pi ^+\nu \overline{\nu }$: First NA62 Results},''
  \href{http://dx.doi.org/10.1007/978-3-030-29622-3\_19}{{\em Springer Proc.
  Phys.} {\bfseries 234} (2019) 135--141}.

\bibitem{Engel:1990zd}
J.~Engel, D.~Seckel, and A.~C. Hayes, ``{Emission and detectability of hadronic
  axions from SN1987A},''
\href{http://dx.doi.org/10.1103/PhysRevLett.65.960}{{\em Phys. Rev. Lett.}
  {\bfseries 65} (1990) 960--963}.

\bibitem{Abbiendi:1998yu}
{\bfseries OPAL} Collaboration, G.~Abbiendi {\em et~al.}, ``{Search for
  anomalous photonic events with missing energy in $e^+ e^-$ collisions at
  $\sqrt{s}$ = 130-GeV, 136-GeV and 183-GeV},''
  \href{http://dx.doi.org/10.1007/s100520050442}{{\em Eur. Phys. J.} {\bfseries
  C8} (1999) 23--40},
\href{http://arxiv.org/abs/hep-ex/9810021}{{\ttfamily arXiv:hep-ex/9810021
  [hep-ex]}}.

\bibitem{Acciarri:1999kp}
{\bfseries L3} Collaboration, M.~Acciarri {\em et~al.}, ``{Single and
  multiphoton events with missing energy in $e^{+} e^{-}$ collisions at
  $\sqrt{S}$ - 189-GeV},''
  \href{http://dx.doi.org/10.1016/S0370-2693(99)01286-1}{{\em Phys. Lett.}
  {\bfseries B470} (1999) 268--280},
\href{http://arxiv.org/abs/hep-ex/9910009}{{\ttfamily arXiv:hep-ex/9910009
  [hep-ex]}}.

\bibitem{Heister:2002ut}
{\bfseries ALEPH} Collaboration, A.~Heister {\em et~al.}, ``{Single photon and
  multiphoton production in $e^{+} e^{-}$ collisions at $\sqrt{s}$ up to
  209-GeV},''
\href{http://dx.doi.org/10.1140/epjc/s2002-01129-7}{{\em Eur. Phys. J.}
  {\bfseries C28} (2003) 1--13}.

\bibitem{Aaboud:2016uro}
{\bfseries ATLAS} Collaboration, M.~Aaboud {\em et~al.}, ``{Search for new
  phenomena in events with a photon and missing transverse momentum in $pp$
  collisions at $\sqrt{s}=13$ TeV with the ATLAS detector},''
  \href{http://dx.doi.org/10.1007/JHEP06(2016)059}{{\em JHEP} {\bfseries 06}
  (2016) 059},
\href{http://arxiv.org/abs/1604.01306}{{\ttfamily arXiv:1604.01306 [hep-ex]}}.

\bibitem{Sirunyan:2018dsf}
{\bfseries CMS} Collaboration, A.~M. Sirunyan {\em et~al.}, ``{Search for new
  physics in final states with a single photon and missing transverse momentum
  in proton-proton collisions at $\sqrt{s} =$ 13 TeV},''
  \href{http://dx.doi.org/10.1007/JHEP02(2019)074}{{\em JHEP} {\bfseries 02}
  (2019) 074},
\href{http://arxiv.org/abs/1810.00196}{{\ttfamily arXiv:1810.00196 [hep-ex]}}.

\bibitem{Vogel:2013raa}
H.~Vogel and J.~Redondo, ``{Dark Radiation constraints on minicharged particles
  in models with a hidden photon},''
  \href{http://dx.doi.org/10.1088/1475-7516/2014/02/029}{{\em JCAP} {\bfseries
  1402} (2014) 029},
\href{http://arxiv.org/abs/1311.2600}{{\ttfamily arXiv:1311.2600 [hep-ph]}}.

\bibitem{Chang:2018rso}
J.~H. Chang, R.~Essig, and S.~D. McDermott, ``{Supernova 1987A Constraints on
  Sub-GeV Dark Sectors, Millicharged Particles, the QCD Axion, and an
  Axion-like Particle},'' \href{http://dx.doi.org/10.1007/JHEP09(2018)051}{{\em
  JHEP} {\bfseries 09} (2018) 051},
\href{http://arxiv.org/abs/1803.00993}{{\ttfamily arXiv:1803.00993 [hep-ph]}}.

\bibitem{Badertscher:2006fm}
A.~Badertscher, P.~Crivelli, W.~Fetscher, U.~Gendotti, S.~Gninenko, V.~Postoev,
  A.~Rubbia, V.~Samoylenko, and D.~Sillou, ``{An Improved Limit on Invisible
  Decays of Positronium},''
  \href{http://dx.doi.org/10.1103/PhysRevD.75.032004}{{\em Phys. Rev.}
  {\bfseries D75} (2007) 032004},
\href{http://arxiv.org/abs/hep-ex/0609059}{{\ttfamily arXiv:hep-ex/0609059
  [hep-ex]}}.

\bibitem{Prinz:1998ua}
A.~A. Prinz {\em et~al.}, ``{Search for millicharged particles at SLAC},''
  \href{http://dx.doi.org/10.1103/PhysRevLett.81.1175}{{\em Phys. Rev. Lett.}
  {\bfseries 81} (1998) 1175--1178},
\href{http://arxiv.org/abs/hep-ex/9804008}{{\ttfamily arXiv:hep-ex/9804008
  [hep-ex]}}.

\bibitem{Magill:2018tbb}
G.~Magill, R.~Plestid, M.~Pospelov, and Y.-D. Tsai, ``{Millicharged particles
  in neutrino experiments},''
  \href{http://dx.doi.org/10.1103/PhysRevLett.122.071801}{{\em Phys. Rev.
  Lett.} {\bfseries 122} no.~7, (2019) 071801},
\href{http://arxiv.org/abs/1806.03310}{{\ttfamily arXiv:1806.03310 [hep-ph]}}.

\bibitem{Jaeckel:2012yz}
J.~Jaeckel, M.~Jankowiak, and M.~Spannowsky, ``{LHC probes the hidden
  sector},'' \href{http://dx.doi.org/10.1016/j.dark.2013.06.001}{{\em Phys.
  Dark Univ.} {\bfseries 2} (2013) 111--117},
\href{http://arxiv.org/abs/1212.3620}{{\ttfamily arXiv:1212.3620 [hep-ph]}}.

\bibitem{NA64:eplus}
D.~Banerjee {\em et~al.}, ``{Addendum to the NA64 Proposal: search for
  $A^{\prime} \to $ invisible and $X \to e^+ e^-$ decays in 2021},''.
  \url{http://cds.cern.ch/record/2300189?ln=en}.

\bibitem{NA64:2018iqr}
D.~Banerjee {\em et~al.}, ``{Addendum to the Proposal P348: Search for dark
  sector particles weakly coupled to muon with NA64 $\mu$},''.
  \url{http://cds.cern.ch/record/2640930?ln=en}.

\bibitem{Kelly:2018brz}
K.~J. Kelly and Y.-D. Tsai, ``{Proton fixed-target scintillation experiment to
  search for millicharged dark matter},''
  \href{http://dx.doi.org/10.1103/PhysRevD.100.015043}{{\em Phys. Rev.}
  {\bfseries D100} no.~1, (2019) 015043},
\href{http://arxiv.org/abs/1812.03998}{{\ttfamily arXiv:1812.03998 [hep-ph]}}.

\bibitem{Ball:2016zrp}
A.~Ball {\em et~al.}, ``{A Letter of Intent to Install a milli-charged Particle
  Detector at LHC P5},''
\href{http://arxiv.org/abs/1607.04669}{{\ttfamily arXiv:1607.04669
  [physics.ins-det]}}.

\bibitem{Akesson:2018vlm}
{\bfseries LDMX} Collaboration, T.~Akesson {\em et~al.}, ``{Light Dark Matter
  eXperiment (LDMX)},''
\href{http://arxiv.org/abs/1808.05219}{{\ttfamily arXiv:1808.05219 [hep-ex]}}.

\bibitem{Hagley:1994zz}
E.~W. Hagley and F.~M. Pipkin, ``{Separated oscillatory field measurement of
  hydrogen S-21/2- P-23/2 fine structure interval},''
\href{http://dx.doi.org/10.1103/PhysRevLett.72.1172}{{\em Phys. Rev. Lett.}
  {\bfseries 72} (1994) 1172--1175}.

\bibitem{Kovetz:2018zan}
E.~D. Kovetz, V.~Poulin, V.~Gluscevic, K.~K. Boddy, R.~Barkana, and
  M.~Kamionkowski, ``{Tighter limits on dark matter explanations of the
  anomalous EDGES 21 cm signal},''
  \href{http://dx.doi.org/10.1103/PhysRevD.98.103529}{{\em Phys. Rev.}
  {\bfseries D98} no.~10, (2018) 103529},
\href{http://arxiv.org/abs/1807.11482}{{\ttfamily arXiv:1807.11482
  [astro-ph.CO]}}.

\bibitem{Liu:2019knx}
H.~Liu, N.~J. Outmezguine, D.~Redigolo, and T.~Volansky, ``{Reviving
  Millicharged Dark Matter for 21-cm Cosmology},''
  \href{http://dx.doi.org/10.1103/PhysRevD.100.123011}{{\em Phys. Rev. D}
  {\bfseries 100} no.~12, (2019) 123011},
  \href{http://arxiv.org/abs/1908.06986}{{\ttfamily arXiv:1908.06986
  [hep-ph]}}.

\bibitem{Monsalve:2018fno}
R.~A. Monsalve, B.~Greig, J.~D. Bowman, A.~Mesinger, A.~E.~E. Rogers, T.~J.
  Mozdzen, N.~S. Kern, and N.~Mahesh, ``{Results from EDGES High-Band: II.
  Constraints on Parameters of Early Galaxies},''
  \href{http://dx.doi.org/10.3847/1538-4357/aace54}{{\em Astrophys. J.}
  {\bfseries 863} no.~1, (2018) 11},
\href{http://arxiv.org/abs/1806.07774}{{\ttfamily arXiv:1806.07774
  [astro-ph.CO]}}.

\bibitem{LDMX:CERN}
T.~Akesson {\em et~al.}, ``{Dark Sector Physics with a Primary Electron Beam
  Facility at CERN},''. \url{http://cds.cern.ch/record/2640784?ln=en}.

\bibitem{Raubenheimer:2018mwt}
T.~Raubenheimer, A.~Beukers, A.~Fry, C.~Hast, T.~Markiewicz, Y.~Nosochkov,
  N.~Phinney, P.~Schuster, and N.~Toro, ``{DASEL: Dark Sector Experiments at
  LCLS-II},'' \href{http://arxiv.org/abs/1801.07867}{{\ttfamily
  arXiv:1801.07867 [physics.acc-ph]}}.

\bibitem{Gabrielli:2016vbb}
E.~Gabrielli, L.~Marzola, and M.~Raidal, ``{Radiative Yukawa Couplings in the
  Simplest Left-Right Symmetric Model},''
  \href{http://dx.doi.org/10.1103/PhysRevD.95.035005}{{\em Phys. Rev.}
  {\bfseries D95} no.~3, (2017) 035005},
\href{http://arxiv.org/abs/1611.00009}{{\ttfamily arXiv:1611.00009 [hep-ph]}}.

\bibitem{Gabrielli:2016cut}
E.~Gabrielli, B.~Mele, M.~Raidal, and E.~Venturini, ``{FCNC decays of standard
  model fermions into a dark photon},''
  \href{http://dx.doi.org/10.1103/PhysRevD.94.115013}{{\em Phys. Rev.}
  {\bfseries D94} no.~11, (2016) 115013},
\href{http://arxiv.org/abs/1607.05928}{{\ttfamily arXiv:1607.05928 [hep-ph]}}.

\bibitem{Acuna:2020ccz}
J.~T. Acu{\~n}a, M.~Fabbrichesi, and P.~Ullio, ``{Phenomenological consequences
  of an interacting multicomponent dark sector},''
  \href{http://arxiv.org/abs/2005.04146}{{\ttfamily arXiv:2005.04146
  [hep-ph]}}.

\bibitem{Aaboud:2017vwy}
{\bfseries ATLAS} Collaboration, M.~Aaboud {\em et~al.}, ``{Search for squarks
  and gluinos in final states with jets and missing transverse momentum using
  36 fb$^{-1}$ of $\sqrt{s}=13$ TeV pp collision data with the ATLAS
  detector},'' \href{http://dx.doi.org/10.1103/PhysRevD.97.112001}{{\em Phys.
  Rev.} {\bfseries D97} no.~11, (2018) 112001},
\href{http://arxiv.org/abs/1712.02332}{{\ttfamily arXiv:1712.02332 [hep-ex]}}.

\bibitem{Barducci:2018rlx}
D.~Barducci, M.~Fabbrichesi, and E.~Gabrielli, ``{Neutral Hadrons Disappearing
  into the Darkness},''
  \href{http://dx.doi.org/10.1103/PhysRevD.98.035049}{{\em Phys. Rev.}
  {\bfseries D98} no.~3, (2018) 035049},
\href{http://arxiv.org/abs/1806.05678}{{\ttfamily arXiv:1806.05678 [hep-ph]}}.

\bibitem{Sirunyan:2018nwe}
{\bfseries CMS} Collaboration, A.~M. Sirunyan {\em et~al.}, ``{Search for
  supersymmetric partners of electrons and muons in proton-proton collisions at
  $\sqrt{s}=$ 13 TeV},''
  \href{http://dx.doi.org/10.1016/j.physletb.2019.01.005}{{\em Phys. Lett.}
  {\bfseries B790} (2019) 140--166},
\href{http://arxiv.org/abs/1806.05264}{{\ttfamily arXiv:1806.05264 [hep-ex]}}.

\bibitem{TheMEG:2016wtm}
{\bfseries MEG} Collaboration, A.~Baldini {\em et~al.}, ``{Search for the
  lepton flavour violating decay $\mu ^+ \rightarrow \mathrm {e}^+ \gamma $
  with the full dataset of the MEG experiment},''
  \href{http://dx.doi.org/10.1140/epjc/s10052-016-4271-x}{{\em Eur. Phys. J. C}
  {\bfseries 76} no.~8, (2016) 434},
  \href{http://arxiv.org/abs/1605.05081}{{\ttfamily arXiv:1605.05081
  [hep-ex]}}.

\bibitem{Lees:2012ym}
{\bfseries BaBar} Collaboration, J.~Lees {\em et~al.}, ``{Precision Measurement
  of the $B \to X_s \gamma$ Photon Energy Spectrum, Branching Fraction, and
  Direct CP Asymmetry $A_{CP}(B \to X_{s+d}\gamma)$},''
  \href{http://dx.doi.org/10.1103/PhysRevLett.109.191801}{{\em Phys. Rev.
  Lett.} {\bfseries 109} (2012) 191801},
  \href{http://arxiv.org/abs/1207.2690}{{\ttfamily arXiv:1207.2690 [hep-ex]}}.

\bibitem{Misiak:2006zs}
M.~Misiak {\em et~al.}, ``{Estimate of $\mathcal{B} (\bar B \to X_s \gamma)$ at
  $O(\alpha_s^2)$},''
  \href{http://dx.doi.org/10.1103/PhysRevLett.98.022002}{{\em Phys. Rev. Lett.}
  {\bfseries 98} (2007) 022002},
  \href{http://arxiv.org/abs/hep-ph/0609232}{{\ttfamily arXiv:hep-ph/0609232}}.

\bibitem{Fabbrichesi:2017vma}
M.~Fabbrichesi, E.~Gabrielli, and B.~Mele, ``{Hunting down massless dark
  photons in kaon physics},''
  \href{http://dx.doi.org/10.1103/PhysRevLett.119.031801}{{\em Phys. Rev.
  Lett.} {\bfseries 119} no.~3, (2017) 031801},
\href{http://arxiv.org/abs/1705.03470}{{\ttfamily arXiv:1705.03470 [hep-ph]}}.

\bibitem{Tanabashi:2018oca}
{\bfseries Particle Data Group} Collaboration, M.~Tanabashi {\em et~al.},
  ``{Review of Particle Physics},''
  \href{http://dx.doi.org/10.1103/PhysRevD.98.030001}{{\em Phys. Rev. D}
  {\bfseries 98} no.~3, (2018) 030001}.

\bibitem{NA62:2017rwk}
{\bfseries NA62} Collaboration, E.~Cortina~Gil {\em et~al.}, ``{The Beam and
  detector of the NA62 experiment at CERN},''
  \href{http://dx.doi.org/10.1088/1748-0221/12/05/P05025}{{\em JINST}
  {\bfseries 12} no.~05, (2017) P05025},
\href{http://arxiv.org/abs/1703.08501}{{\ttfamily arXiv:1703.08501
  [physics.ins-det]}}.

\bibitem{CortinaGil:2018fkc}
{\bfseries NA62} Collaboration, E.~Cortina~Gil {\em et~al.}, ``{First search
  for $K^+\rightarrow\pi^+\nu\bar{\nu}$ using the decay-in-flight technique},''
  \href{http://dx.doi.org/10.1016/j.physletb.2019.01.067}{{\em Phys.\ Lett.\ B}
  {\bfseries 791} (2019) 156--166},
  \href{http://arxiv.org/abs/1811.08508}{{\ttfamily arXiv:1811.08508
  [hep-ex]}}.

\bibitem{Ahn:2018mvc}
{\bfseries KOTO} Collaboration, J.~Ahn {\em et~al.}, ``{Search for the $K_L
  \!\to\! \pi^0 \nu \overline{\nu}$ and $K_L \!\to\! \pi^0 X^0$ decays at the
  J-PARC KOTO experiment},''
  \href{http://dx.doi.org/10.1103/PhysRevLett.122.021802}{{\em Phys.\ Rev.\
  Lett.} {\bfseries 122} no.~2, (2019) 021802},
  \href{http://arxiv.org/abs/1810.09655}{{\ttfamily arXiv:1810.09655
  [hep-ex]}}.

\bibitem{Fabbrichesi:2019bmo}
M.~Fabbrichesi and E.~Gabrielli, ``{Dark-sector physics in the search for the
  rare decays $K^+\rightarrow \pi ^+ \nu {\bar{\nu }}$ and $K_L\rightarrow \pi
  ^0 \nu {\bar{\nu }}$},''
  \href{http://dx.doi.org/10.1140/epjc/s10052-020-8103-7}{{\em Eur. Phys. J. C}
  {\bfseries 80} no.~6, (2020) 532},
  \href{http://arxiv.org/abs/1911.03755}{{\ttfamily arXiv:1911.03755
  [hep-ph]}}.

\bibitem{Su:2019ipw}
J.-Y. Su and J.~Tandean, ``{Searching for dark photons in hyperon decays},''
\href{http://arxiv.org/abs/1911.13301}{{\ttfamily arXiv:1911.13301 [hep-ph]}}.

\bibitem{1795128}
J.-Y. Su and J.~Tandean, ``{Seeking massless dark photons in the decays of
  charmed hadrons},'' \href{http://arxiv.org/abs/2005.05297}{{\ttfamily
  arXiv:2005.05297 [hep-ph]}}.

\bibitem{Lees:2012wv}
{\bfseries BaBar} Collaboration, J.~P. Lees {\em et~al.}, ``{Improved Limits on
  $B^0$ Decays to Invisible Final States and to $\nu \bar{\nu} \gamma$},''
  \href{http://dx.doi.org/10.1103/PhysRevD.86.051105}{{\em Phys. Rev.}
  {\bfseries D86} (2012) 051105},
\href{http://arxiv.org/abs/1206.2543}{{\ttfamily arXiv:1206.2543 [hep-ex]}}.

\bibitem{Hsu:2012uh}
{\bfseries Belle} Collaboration, C.~L. Hsu {\em et~al.}, ``{Search for $B^{0}$
  decays to invisible final states},''
  \href{http://dx.doi.org/10.1103/PhysRevD.86.032002}{{\em Phys. Rev.}
  {\bfseries D86} (2012) 032002},
\href{http://arxiv.org/abs/1206.5948}{{\ttfamily arXiv:1206.5948 [hep-ex]}}.

\bibitem{Gninenko:2015mea}
S.~N. Gninenko and N.~V. Krasnikov, ``{Invisible $K_L$ decays as a probe of new
  physics},'' \href{http://dx.doi.org/10.1103/PhysRevD.92.034009}{{\em Phys.
  Rev.} {\bfseries D92} no.~3, (2015) 034009},
\href{http://arxiv.org/abs/1503.01595}{{\ttfamily arXiv:1503.01595 [hep-ph]}}.

\bibitem{Gninenko:2014sxa}
S.~N. Gninenko, ``{Search for invisible decays of $\pi^0, \eta, \eta', K_S$ and
  $K_L$: A probe of new physics and tests using the Bell-Steinberger
  relation},'' \href{http://dx.doi.org/10.1103/PhysRevD.91.015004}{{\em Phys.
  Rev.} {\bfseries D91} no.~1, (2015) 015004},
\href{http://arxiv.org/abs/1409.2288}{{\ttfamily arXiv:1409.2288 [hep-ph]}}.

\bibitem{sommerfeld}
A.~Sommerfeld, ``{\"U}ber die beugung und bremsung der elektronen,'' {\em
  Annalen der {P}hysik} {\bfseries 403} (1931) 257.

\bibitem{fermi}
E.~Fermi, ``An attempt of a theory of beta radiation. 1.'' {\em Zeitschrift
  f{\"u}r Physik} {\bfseries 88} (1934) 161.

\bibitem{Gabrielli:2014oya}
E.~Gabrielli, M.~Heikinheimo, B.~Mele, and M.~Raidal, ``{Dark photons and
  resonant monophoton signatures in Higgs boson decays at the LHC},''
  \href{http://dx.doi.org/10.1103/PhysRevD.90.055032}{{\em Phys. Rev.}
  {\bfseries D90} no.~5, (2014) 055032},
\href{http://arxiv.org/abs/1405.5196}{{\ttfamily arXiv:1405.5196 [hep-ph]}}.

\bibitem{Biswas:2015sha}
S.~Biswas, E.~Gabrielli, M.~Heikinheimo, and B.~Mele, ``{Higgs-boson production
  in association with a dark photon in e$^{+}$e$^{?}$ collisions},''
  \href{http://dx.doi.org/10.1007/JHEP06(2015)102}{{\em JHEP} {\bfseries 06}
  (2015) 102},
\href{http://arxiv.org/abs/1503.05836}{{\ttfamily arXiv:1503.05836 [hep-ph]}}.

\bibitem{Biswas:2017lyg}
S.~Biswas, E.~Gabrielli, M.~Heikinheimo, and B.~Mele, ``{Dark-photon searches
  via $ZH$ production at $e^+e^-$ colliders},''
  \href{http://dx.doi.org/10.1103/PhysRevD.96.055012}{{\em Phys. Rev.}
  {\bfseries D96} no.~5, (2017) 055012},
\href{http://arxiv.org/abs/1703.00402}{{\ttfamily arXiv:1703.00402 [hep-ph]}}.

\bibitem{Fabbrichesi:2017zsc}
M.~Fabbrichesi, E.~Gabrielli, and B.~Mele, ``{$Z$ Boson Decay into Light and
  Darkness},'' \href{http://dx.doi.org/10.1103/PhysRevLett.120.171803}{{\em
  Phys. Rev. Lett.} {\bfseries 120} no.~17, (2018) 171803},
\href{http://arxiv.org/abs/1712.05412}{{\ttfamily arXiv:1712.05412 [hep-ph]}}.

\bibitem{Cobal:2020hmk}
M.~Cobal, C.~De~Dominicis, M.~Fabbrichesi, E.~Gabrielli, J.~Magro, B.~Mele, and
  G.~Panizzo, ``{$Z$-boson decays into an invisible dark photon at the LHC,
  HL-LHC and future lepton colliders},''
  \href{http://dx.doi.org/10.1103/PhysRevD.102.035027}{{\em Phys. Rev. D}
  {\bfseries 102} no.~3, (2020) 035027},
  \href{http://arxiv.org/abs/2006.15945}{{\ttfamily arXiv:2006.15945
  [hep-ph]}}.

\bibitem{Barbieri:2016vwg}
R.~Barbieri, C.~Braggio, G.~Carugno, C.~S. Gallo, A.~Lombardi, A.~Ortolan,
  R.~Pengo, G.~Ruoso, and C.~C. Speake, ``{Searching for galactic axions
  through magnetized media: the QUAX proposal},''
  \href{http://dx.doi.org/10.1016/j.dark.2017.01.003}{{\em Phys. Dark Univ.}
  {\bfseries 15} (2017) 135--141},
\href{http://arxiv.org/abs/1606.02201}{{\ttfamily arXiv:1606.02201 [hep-ph]}}.

\bibitem{Chigusa:2020gfs}
S.~Chigusa, T.~Moroi, and K.~Nakayama, ``{Detecting Light Boson Dark Matter
  through Conversion into Magnon},''
\href{http://arxiv.org/abs/2001.10666}{{\ttfamily arXiv:2001.10666 [hep-ph]}}.

\bibitem{Croon:2017zcu}
D.~Croon, A.~E. Nelson, C.~Sun, D.~G.~E. Walker, and Z.-Z. Xianyu,
  ``{Hidden-Sector Spectroscopy with Gravitational Waves from Binary Neutron
  Stars},'' \href{http://dx.doi.org/10.3847/2041-8213/aabe76}{{\em Astrophys.\
  J.} {\bfseries 858} no.~1, (2018) L2},
  \href{http://arxiv.org/abs/1711.02096}{{\ttfamily arXiv:1711.02096
  [hep-ph]}}.

\bibitem{Alexander:2018qzg}
S.~Alexander, E.~McDonough, R.~Sims, and N.~Yunes, ``{Hidden-Sector
  Modifications to Gravitational Waves From Binary Inspirals},''
  \href{http://dx.doi.org/10.1088/1361-6382/aaeb5c}{{\em Class.\ Quant.\ Grav.}
  {\bfseries 35} no.~23, (2018) 235012},
  \href{http://arxiv.org/abs/1808.05286}{{\ttfamily arXiv:1808.05286 [gr-qc]}}.

\bibitem{Kopp:2018jom}
J.~Kopp, R.~Laha, T.~Opferkuch, and W.~Shepherd, ``{Cuckoo's eggs in neutron
  stars: can LIGO hear chirps from the dark sector?},''
  \href{http://dx.doi.org/10.1007/JHEP11(2018)096}{{\em JHEP} {\bfseries 11}
  (2018) 096}, \href{http://arxiv.org/abs/1807.02527}{{\ttfamily
  arXiv:1807.02527 [hep-ph]}}.

\bibitem{Fabbrichesi:2019ema}
M.~Fabbrichesi and A.~Urbano, ``{Charged neutron stars and observational tests
  of a dark force weaker than gravity},''
  \href{http://arxiv.org/abs/1902.07914}{{\ttfamily arXiv:1902.07914
  [hep-ph]}}.

\bibitem{Merkel:2014avp}
H.~Merkel {\em et~al.}, ``{Search at the Mainz Microtron for Light Massive
  Gauge Bosons Relevant for the Muon g-2 Anomaly},''
  \href{http://dx.doi.org/10.1103/PhysRevLett.112.221802}{{\em Phys. Rev.
  Lett.} {\bfseries 112} no.~22, (2014) 221802},
\href{http://arxiv.org/abs/1404.5502}{{\ttfamily arXiv:1404.5502 [hep-ex]}}.

\bibitem{Aaij:2019bvg}
{\bfseries LHCb} Collaboration, R.~Aaij {\em et~al.}, ``{Search for
  $A'\to\mu^+\mu^-$ Decays},''
  \href{http://dx.doi.org/10.1103/PhysRevLett.124.041801}{{\em Phys. Rev.
  Lett.} {\bfseries 124} no.~4, (2020) 041801},
  \href{http://arxiv.org/abs/1910.06926}{{\ttfamily arXiv:1910.06926
  [hep-ex]}}.

\bibitem{CMS:2019kiy}
{\bfseries CMS} Collaboration, A.~M. Sirunyan {\em et~al.}, ``{Search for a
  narrow resonance decaying to a pair of muons in proton-proton collisions at
  13 TeV},''.

\bibitem{Lees:2014xha}
{\bfseries BaBar} Collaboration, J.~P. Lees {\em et~al.}, ``{Search for a Dark
  Photon in $e^+e^-$ Collisions at BaBar},''
  \href{http://dx.doi.org/10.1103/PhysRevLett.113.201801}{{\em Phys. Rev.
  Lett.} {\bfseries 113} no.~20, (2014) 201801},
\href{http://arxiv.org/abs/1406.2980}{{\ttfamily arXiv:1406.2980 [hep-ex]}}.

\bibitem{Archilli:2011zc}
{\bfseries KLOE-2} Collaboration, F.~Archilli {\em et~al.}, ``{Search for a
  vector gauge boson in $\phi$ meson decays with the KLOE detector},''
  \href{http://dx.doi.org/10.1016/j.physletb.2011.11.033}{{\em Phys. Lett.}
  {\bfseries B706} (2012) 251--255},
\href{http://arxiv.org/abs/1110.0411}{{\ttfamily arXiv:1110.0411 [hep-ex]}}.

\bibitem{Babusci:2012cr}
{\bfseries KLOE-2} Collaboration, D.~Babusci {\em et~al.}, ``{Limit on the
  production of a light vector gauge boson in phi meson decays with the KLOE
  detector},'' \href{http://dx.doi.org/10.1016/j.physletb.2013.01.067}{{\em
  Phys. Lett.} {\bfseries B720} (2013) 111--115},
\href{http://arxiv.org/abs/1210.3927}{{\ttfamily arXiv:1210.3927 [hep-ex]}}.

\bibitem{Babusci:2014sta}
{\bfseries KLOE-2} Collaboration, D.~Babusci {\em et~al.}, ``{Search for light
  vector boson production in $e^+e^- \rightarrow \mu^+ \mu^- \gamma$
  interactions with the KLOE experiment},''
  \href{http://dx.doi.org/10.1016/j.physletb.2014.08.005}{{\em Phys. Lett.}
  {\bfseries B736} (2014) 459--464},
\href{http://arxiv.org/abs/1404.7772}{{\ttfamily arXiv:1404.7772 [hep-ex]}}.

\bibitem{Anastasi:2016ktq}
{\bfseries KLOE-2} Collaboration, A.~Anastasi {\em et~al.}, ``{Limit on the
  production of a new vector boson in $\mathrm{e^+ e^-}\rightarrow {\rm
  U}\gamma$, U$\rightarrow \pi^+\pi^-$ with the KLOE experiment},''
  \href{http://dx.doi.org/10.1016/j.physletb.2016.04.019}{{\em Phys. Lett.}
  {\bfseries B757} (2016) 356--361},
\href{http://arxiv.org/abs/1603.06086}{{\ttfamily arXiv:1603.06086 [hep-ex]}}.

\bibitem{Batley:2015lha}
{\bfseries NA48/2} Collaboration, J.~R. Batley {\em et~al.}, ``{Search for the
  dark photon in $\pi^0$ decays},''
  \href{http://dx.doi.org/10.1016/j.physletb.2015.04.068}{{\em Phys. Lett.}
  {\bfseries B746} (2015) 178--185},
\href{http://arxiv.org/abs/1504.00607}{{\ttfamily arXiv:1504.00607 [hep-ex]}}.

\bibitem{Bross:1989mp}
A.~Bross, M.~Crisler, S.~H. Pordes, J.~Volk, S.~Errede, and J.~Wrbanek, ``{A
  Search for Shortlived Particles Produced in an Electron Beam Dump},''
\href{http://dx.doi.org/10.1103/PhysRevLett.67.2942}{{\em Phys. Rev. Lett.}
  {\bfseries 67} (1991) 2942--2945}.

\bibitem{Riordan:1987aw}
E.~M. Riordan {\em et~al.}, ``{A Search for Short Lived Axions in an Electron
  Beam Dump Experiment},''
\href{http://dx.doi.org/10.1103/PhysRevLett.59.755}{{\em Phys. Rev. Lett.}
  {\bfseries 59} (1987) 755}.

\bibitem{Bjorken:1988as}
J.~D. Bjorken, S.~Ecklund, W.~R. Nelson, A.~Abashian, C.~Church, B.~Lu, L.~W.
  Mo, T.~A. Nunamaker, and P.~Rassmann, ``{Search for Neutral Metastable
  Penetrating Particles Produced in the SLAC Beam Dump},''
\href{http://dx.doi.org/10.1103/PhysRevD.38.3375}{{\em Phys. Rev.} {\bfseries
  D38} (1988) 3375}.

\bibitem{Batell:2014mga}
B.~Batell, R.~Essig, and Z.~Surujon, ``{Strong Constraints on Sub-GeV Dark
  Sectors from SLAC Beam Dump E137},''
  \href{http://dx.doi.org/10.1103/PhysRevLett.113.171802}{{\em Phys.\ Rev.\
  Lett.} {\bfseries 113} no.~17, (2014) 171802},
  \href{http://arxiv.org/abs/1406.2698}{{\ttfamily arXiv:1406.2698 [hep-ph]}}.

\bibitem{Marsicano:2018krp}
L.~Marsicano, M.~Battaglieri, M.~Bondi', C.~R. Carvajal, A.~Celentano,
  M.~De~Napoli, R.~De~Vita, E.~Nardi, M.~Raggi, and P.~Valente, ``{Dark photon
  production through positron annihilation in beam-dump experiments},''
  \href{http://dx.doi.org/10.1103/PhysRevD.98.015031}{{\em Phys. Rev. D}
  {\bfseries 98} no.~1, (2018) 015031},
  \href{http://arxiv.org/abs/1802.03794}{{\ttfamily arXiv:1802.03794
  [hep-ex]}}.

\bibitem{Blumlein:2011mv}
J.~Blumlein and J.~Brunner, ``{New Exclusion Limits for Dark Gauge Forces from
  Beam-Dump Data},''
  \href{http://dx.doi.org/10.1016/j.physletb.2011.05.046}{{\em Phys. Lett. B}
  {\bfseries 701} (2011) 155--159},
  \href{http://arxiv.org/abs/1104.2747}{{\ttfamily arXiv:1104.2747 [hep-ex]}}.

\bibitem{Blumlein:2013cua}
J.~Blumlein and J.~Brunner, ``{New Exclusion Limits on Dark Gauge Forces from
  Proton Bremsstrahlung in Beam-Dump Data},''
  \href{http://dx.doi.org/10.1016/j.physletb.2014.02.029}{{\em Phys. Lett.}
  {\bfseries B731} (2014) 320--326},
\href{http://arxiv.org/abs/1311.3870}{{\ttfamily arXiv:1311.3870 [hep-ph]}}.

\bibitem{Gninenko:2012eq}
S.~Gninenko, ``{Constraints on sub-GeV hidden sector gauge bosons from a search
  for heavy neutrino decays},''
  \href{http://dx.doi.org/10.1016/j.physletb.2012.06.002}{{\em Phys. Lett. B}
  {\bfseries 713} (2012) 244--248},
  \href{http://arxiv.org/abs/1204.3583}{{\ttfamily arXiv:1204.3583 [hep-ph]}}.

\bibitem{Chang:2016ntp}
J.~H. Chang, R.~Essig, and S.~D. McDermott, ``{Revisiting Supernova 1987A
  Constraints on Dark Photons},''
  \href{http://dx.doi.org/10.1007/JHEP01(2017)107}{{\em JHEP} {\bfseries 01}
  (2017) 107}, \href{http://arxiv.org/abs/1611.03864}{{\ttfamily
  arXiv:1611.03864 [hep-ph]}}.

\bibitem{Pospelov:2008zw}
M.~Pospelov, ``{Secluded U(1) below the weak scale},''
  \href{http://dx.doi.org/10.1103/PhysRevD.80.095002}{{\em Phys. Rev.}
  {\bfseries D80} (2009) 095002},
\href{http://arxiv.org/abs/0811.1030}{{\ttfamily arXiv:0811.1030 [hep-ph]}}.

\bibitem{Kou:2018nap}
{\bfseries Belle-II} Collaboration, W.~Altmannshofer {\em et~al.}, ``{The Belle
  II Physics Book},'' \href{http://dx.doi.org/10.1093/ptep/ptz106,
  10.1093/ptep/ptaa008}{{\em PTEP} {\bfseries 2019} no.~12, (2019) 123C01},
  \href{http://arxiv.org/abs/1808.10567}{{\ttfamily arXiv:1808.10567
  [hep-ex]}}.
[Erratum: PTEP2020,no.2,029201(2020)].

\bibitem{Ilten:2016tkc}
P.~Ilten, Y.~Soreq, J.~Thaler, M.~Williams, and W.~Xue, ``{Proposed Inclusive
  Dark Photon Search at LHCb},''
  \href{http://dx.doi.org/10.1103/PhysRevLett.116.251803}{{\em Phys. Rev.
  Lett.} {\bfseries 116} no.~25, (2016) 251803},
\href{http://arxiv.org/abs/1603.08926}{{\ttfamily arXiv:1603.08926 [hep-ph]}}.

\bibitem{Ilten:2015hya}
P.~Ilten, J.~Thaler, M.~Williams, and W.~Xue, ``{Dark photons from charm mesons
  at LHCb},'' \href{http://dx.doi.org/10.1103/PhysRevD.92.115017}{{\em Phys.
  Rev.} {\bfseries D92} no.~11, (2015) 115017},
\href{http://arxiv.org/abs/1509.06765}{{\ttfamily arXiv:1509.06765 [hep-ph]}}.

\bibitem{NA62:dump}
E.~Cortina~Gil {\em et~al.}, ``{ADDENDUM I TO P326 Continuation of the physics
  programme of the NA62 experiment},''.
  \url{https://cds.cern.ch/record/2691873?ln=en}.

\bibitem{Feng:2017uoz}
J.~L. Feng, I.~Galon, F.~Kling, and S.~Trojanowski, ``{ForwArd Search
  ExpeRiment at the LHC},''
  \href{http://dx.doi.org/10.1103/PhysRevD.97.035001}{{\em Phys. Rev.}
  {\bfseries D97} no.~3, (2018) 035001},
\href{http://arxiv.org/abs/1708.09389}{{\ttfamily arXiv:1708.09389 [hep-ph]}}.

\bibitem{Berlin:2018pwi}
A.~Berlin, S.~Gori, P.~Schuster, and N.~Toro, ``{Dark Sectors at the Fermilab
  SeaQuest Experiment},''
  \href{http://dx.doi.org/10.1103/PhysRevD.98.035011}{{\em Phys. Rev.}
  {\bfseries D98} no.~3, (2018) 035011},
\href{http://arxiv.org/abs/1804.00661}{{\ttfamily arXiv:1804.00661 [hep-ph]}}.

\bibitem{Adrian:2018scb}
{\bfseries HPS} Collaboration, P.~Adrian {\em et~al.}, ``{Search for a dark
  photon in electroproduced $e^{+}e^{-}$ pairs with the Heavy Photon Search
  experiment at JLab},''
  \href{http://dx.doi.org/10.1103/PhysRevD.98.091101}{{\em Phys.\ Rev.\ D}
  {\bfseries 98} no.~9, (2018) 091101},
  \href{http://arxiv.org/abs/1807.11530}{{\ttfamily arXiv:1807.11530
  [hep-ex]}}.

\bibitem{Caldwell:2018atq}
A.~Caldwell {\em et~al.}, ``{Particle physics applications of the AWAKE
  acceleration scheme},'' \href{http://arxiv.org/abs/1812.11164}{{\ttfamily
  arXiv:1812.11164 [physics.acc-ph]}}.

\bibitem{Doria:2019sux}
L.~Doria, P.~Achenbach, M.~Christmann, A.~Denig, and H.~Merkel, ``{Dark Matter
  at the Intensity Frontier: the new MESA electron accelerator facility},'' in
  {\em {An Alpine LHC Physics Summit 2019 (ALPS 2019) Obergurgl, Austria, April
  22-27, 2019}}.
\newblock 2019.
\newblock
\href{http://arxiv.org/abs/1908.07921}{{\ttfamily arXiv:1908.07921 [hep-ex]}}.
\newblock

\bibitem{Doria:2018sfx}
L.~Doria, P.~Achenbach, M.~Christmann, A.~Denig, P.~Gulker, and H.~Merkel,
  ``{Search for light dark matter with the MESA accelerator},'' in {\em {13th
  Conference on the Intersections of Particle and Nuclear Physics (CIPANP 2018)
  Palm Springs, California, USA, May 29-June 3, 2018}}.
\newblock 2018.
\newblock
\href{http://arxiv.org/abs/1809.07168}{{\ttfamily arXiv:1809.07168 [hep-ex]}}.
\newblock

\bibitem{Curtin:2014cca}
D.~Curtin, R.~Essig, S.~Gori, and J.~Shelton, ``{Illuminating Dark Photons with
  High-Energy Colliders},''
  \href{http://dx.doi.org/10.1007/JHEP02(2015)157}{{\em JHEP} {\bfseries 02}
  (2015) 157}, \href{http://arxiv.org/abs/1412.0018}{{\ttfamily arXiv:1412.0018
  [hep-ph]}}.

\bibitem{Karliner:2015tga}
M.~Karliner, M.~Low, J.~L. Rosner, and L.-T. Wang, ``{Radiative return
  capabilities of a high-energy, high-luminosity $e^+e^-$ collider},''
  \href{http://dx.doi.org/10.1103/PhysRevD.92.035010}{{\em Phys.\ Rev.\ D}
  {\bfseries 92} no.~3, (2015) 035010},
  \href{http://arxiv.org/abs/1503.07209}{{\ttfamily arXiv:1503.07209
  [hep-ph]}}.

\bibitem{DOnofrio:2019dcp}
M.~D'Onofrio, O.~Fischer, and Z.~S. Wang, ``{Searching for Dark Photons at the
  LHeC and FCC-he},'' \href{http://dx.doi.org/10.1103/PhysRevD.101.015020}{{\em
  Phys. Rev. D} {\bfseries 101} no.~1, (2020) 015020},
  \href{http://arxiv.org/abs/1909.02312}{{\ttfamily arXiv:1909.02312
  [hep-ph]}}.

\bibitem{Feng:2016jff}
J.~L. Feng, B.~Fornal, I.~Galon, S.~Gardner, J.~Smolinsky, T.~M.~P. Tait, and
  P.~Tanedo, ``{Protophobic Fifth-Force Interpretation of the Observed Anomaly
  in $^8$Be Nuclear Transitions},''
  \href{http://dx.doi.org/10.1103/PhysRevLett.117.071803}{{\em Phys. Rev.
  Lett.} {\bfseries 117} no.~7, (2016) 071803},
  \href{http://arxiv.org/abs/1604.07411}{{\ttfamily arXiv:1604.07411
  [hep-ph]}}.

\bibitem{Feng:2016ysn}
J.~L. Feng, B.~Fornal, I.~Galon, S.~Gardner, J.~Smolinsky, T.~M.~P. Tait, and
  P.~Tanedo, ``{Particle physics models for the 17 MeV anomaly in beryllium
  nuclear decays},'' \href{http://dx.doi.org/10.1103/PhysRevD.95.035017}{{\em
  Phys. Rev. D} {\bfseries 95} no.~3, (2017) 035017},
  \href{http://arxiv.org/abs/1608.03591}{{\ttfamily arXiv:1608.03591
  [hep-ph]}}.

\bibitem{Aaij:2017rft}
{\bfseries LHCb} Collaboration, R.~Aaij {\em et~al.}, ``{Search for Dark
  Photons Produced in 13 TeV $pp$ Collisions},''
  \href{http://dx.doi.org/10.1103/PhysRevLett.120.061801}{{\em Phys. Rev.
  Lett.} {\bfseries 120} no.~6, (2018) 061801},
\href{http://arxiv.org/abs/1710.02867}{{\ttfamily arXiv:1710.02867 [hep-ex]}}.

\bibitem{Bergsma:1985qz}
{\bfseries CHARM} Collaboration, F.~Bergsma {\em et~al.}, ``{Search for Axion
  Like Particle Production in 400-{GeV} Proton - Copper Interactions},''
\href{http://dx.doi.org/10.1016/0370-2693(85)90400-9}{{\em Phys. Lett.}
  {\bfseries 157B} (1985) 458--462}.

\bibitem{Dent:2012mx}
J.~B. Dent, F.~Ferrer, and L.~M. Krauss, ``{Constraints on Light Hidden Sector
  Gauge Bosons from Supernova Cooling},''
\href{http://arxiv.org/abs/1201.2683}{{\ttfamily arXiv:1201.2683
  [astro-ph.CO]}}.

\bibitem{Dreiner:2013mua}
H.~K. Dreiner, J.-F. Fortin, C.~Hanhart, and L.~Ubaldi, ``{Supernova
  constraints on MeV dark sectors from $e^+e^-$ annihilations},''
  \href{http://dx.doi.org/10.1103/PhysRevD.89.105015}{{\em Phys. Rev.}
  {\bfseries D89} no.~10, (2014) 105015},
\href{http://arxiv.org/abs/1310.3826}{{\ttfamily arXiv:1310.3826 [hep-ph]}}.

\bibitem{Hardy:2016kme}
E.~Hardy and R.~Lasenby, ``{Stellar cooling bounds on new light particles:
  plasma mixing effects},''
  \href{http://dx.doi.org/10.1007/JHEP02(2017)033}{{\em JHEP} {\bfseries 02}
  (2017) 033},
\href{http://arxiv.org/abs/1611.05852}{{\ttfamily arXiv:1611.05852 [hep-ph]}}.

\bibitem{Aad:2014yea}
{\bfseries ATLAS} Collaboration, G.~Aad {\em et~al.}, ``{Search for long-lived
  neutral particles decaying into lepton jets in proton-proton collisions at $
  \sqrt{s}=8 $ TeV with the ATLAS detector},''
  \href{http://dx.doi.org/10.1007/JHEP11(2014)088}{{\em JHEP} {\bfseries 11}
  (2014) 088},
\href{http://arxiv.org/abs/1409.0746}{{\ttfamily arXiv:1409.0746 [hep-ex]}}.

\bibitem{Aad:2015sms}
{\bfseries ATLAS} Collaboration, G.~Aad {\em et~al.}, ``{A search for prompt
  lepton-jets in $pp$ collisions at $\sqrt{s}=$ 8 TeV with the ATLAS
  detector},'' \href{http://dx.doi.org/10.1007/JHEP02(2016)062}{{\em JHEP}
  {\bfseries 02} (2016) 062},
\href{http://arxiv.org/abs/1511.05542}{{\ttfamily arXiv:1511.05542 [hep-ex]}}.

\bibitem{Khachatryan:2015wka}
{\bfseries CMS} Collaboration, V.~Khachatryan {\em et~al.}, ``{A search for
  pair production of new light bosons decaying into muons},''
  \href{http://dx.doi.org/10.1016/j.physletb.2015.10.067}{{\em Phys. Lett.}
  {\bfseries B752} (2016) 146--168},
\href{http://arxiv.org/abs/1506.00424}{{\ttfamily arXiv:1506.00424 [hep-ex]}}.

\bibitem{Fradette:2014sza}
A.~Fradette, M.~Pospelov, J.~Pradler, and A.~Ritz, ``{Cosmological Constraints
  on Very Dark Photons},''
  \href{http://dx.doi.org/10.1103/PhysRevD.90.035022}{{\em Phys. Rev.}
  {\bfseries D90} no.~3, (2014) 035022},
\href{http://arxiv.org/abs/1407.0993}{{\ttfamily arXiv:1407.0993 [hep-ph]}}.

\bibitem{Bediaga:2018lhg}
{\bfseries LHCb} Collaboration, R.~Aaij {\em et~al.}, ``{Physics case for an
  LHCb Upgrade II - Opportunities in flavour physics, and beyond, in the HL-LHC
  era},'' \href{http://arxiv.org/abs/1808.08865}{{\ttfamily arXiv:1808.08865
  [hep-ex]}}.

\bibitem{Ariga:2018uku}
{\bfseries FASER} Collaboration, A.~Ariga {\em et~al.}, ``{FASER physics reach
  for long-lived particles},''
  \href{http://dx.doi.org/10.1103/PhysRevD.99.095011}{{\em Phys. Rev.}
  {\bfseries D99} no.~9, (2019) 095011},
\href{http://arxiv.org/abs/1811.12522}{{\ttfamily arXiv:1811.12522 [hep-ph]}}.

\bibitem{Strategy:2019vxc}
R.~K. Ellis {\em et~al.}, ``{Physics Briefing Book}: {Input for the European
  Strategy for Particle Physics Update 2020},''
  \href{http://arxiv.org/abs/1910.11775}{{\ttfamily arXiv:1910.11775
  [hep-ex]}}.

\bibitem{Adler:2001xv}
{\bfseries E787} Collaboration, S.~Adler {\em et~al.}, ``{Further evidence for
  the decay $K^+ \to \pi^+ \nu \bar \nu$},''
  \href{http://dx.doi.org/10.1103/PhysRevLett.88.041803}{{\em Phys. Rev. Lett.}
  {\bfseries 88} (2002) 041803},
\href{http://arxiv.org/abs/hep-ex/0111091}{{\ttfamily arXiv:hep-ex/0111091
  [hep-ex]}}.

\bibitem{Artamonov:2009sz}
{\bfseries BNL-E949} Collaboration, A.~V. Artamonov {\em et~al.}, ``{Study of
  the decay $K^+\to\pi^+\nu \bar\nu$ in the momentum region $140 < P_\pi < 199$
  MeV/c},'' \href{http://dx.doi.org/10.1103/PhysRevD.79.092004}{{\em Phys.
  Rev.} {\bfseries D79} (2009) 092004},
\href{http://arxiv.org/abs/0903.0030}{{\ttfamily arXiv:0903.0030 [hep-ex]}}.

\bibitem{CortinaGil:2019nuo}
{\bfseries NA62} Collaboration, E.~Cortina~Gil {\em et~al.}, ``{Search for
  production of an invisible dark photon in $\pi^0$ decays},''
  \href{http://dx.doi.org/10.1007/JHEP05(2019)182}{{\em JHEP} {\bfseries 05}
  (2019) 182}, \href{http://arxiv.org/abs/1903.08767}{{\ttfamily
  arXiv:1903.08767 [hep-ex]}}.

\bibitem{Lees:2017lec}
{\bfseries BaBar} Collaboration, J.~P. Lees {\em et~al.}, ``{Search for
  Invisible Decays of a Dark Photon Produced in ${e}^{+}{e}^{-}$ Collisions at
  BaBar},'' \href{http://dx.doi.org/10.1103/PhysRevLett.119.131804}{{\em Phys.
  Rev. Lett.} {\bfseries 119} no.~13, (2017) 131804},
\href{http://arxiv.org/abs/1702.03327}{{\ttfamily arXiv:1702.03327 [hep-ex]}}.

\bibitem{NA64:2019imj}
D.~Banerjee {\em et~al.}, ``{Dark matter search in missing energy events with
  NA64},'' \href{http://dx.doi.org/10.1103/PhysRevLett.123.121801}{{\em Phys.
  Rev. Lett.} {\bfseries 123} no.~12, (2019) 121801},
\href{http://arxiv.org/abs/1906.00176}{{\ttfamily arXiv:1906.00176 [hep-ex]}}.

\bibitem{Ambrosino:2019qvz}
{\bfseries KLEVER Project} Collaboration, F.~Ambrosino {\em et~al.}, ``{KLEVER:
  An experiment to measure BR($K_L\to\pi^0\nu\bar{\nu}$) at the CERN SPS},''
  \href{http://arxiv.org/abs/1901.03099}{{\ttfamily arXiv:1901.03099
  [hep-ex]}}.

\bibitem{Raggi:2015gza}
M.~Raggi, V.~Kozhuharov, and P.~Valente, ``{The PADME experiment at LNF},''
  \href{http://dx.doi.org/10.1051/epjconf/20159601025}{{\em EPJ Web Conf.}
  {\bfseries 96} (2015) 01025},
  \href{http://arxiv.org/abs/1501.01867}{{\ttfamily arXiv:1501.01867
  [hep-ex]}}.

\bibitem{Caputo:2020bdy}
A.~Caputo, H.~Liu, S.~Mishra-Sharma, and J.~T. Ruderman, ``{Dark Photon
  Oscillations in Our Inhomogeneous Universe},''
  \href{http://arxiv.org/abs/2002.05165}{{\ttfamily arXiv:2002.05165
  [astro-ph.CO]}}.

\bibitem{Garcia:2020qrp}
A.~A. Garcia, K.~Bondarenko, S.~Ploeckinger, J.~Pradler, and A.~Sokolenko,
  ``{Effective photon mass and (dark) photon conversion in the inhomogeneous
  Universe},'' \href{http://arxiv.org/abs/2003.10465}{{\ttfamily
  arXiv:2003.10465 [astro-ph.CO]}}.

\bibitem{Witte:2020rvb}
S.~J. Witte, S.~Rosauro-Alcaraz, S.~D. McDermott, and V.~Poulin, ``{Dark Photon
  Dark Matter in the Presence of Inhomogeneous Structure},''
  \href{http://arxiv.org/abs/2003.13698}{{\ttfamily arXiv:2003.13698
  [astro-ph.CO]}}.

\bibitem{McDermott:2019lch}
S.~D. McDermott and S.~J. Witte, ``{Cosmological evolution of light dark photon
  dark matter},'' \href{http://dx.doi.org/10.1103/PhysRevD.101.063030}{{\em
  Phys. Rev. D} {\bfseries 101} no.~6, (2020) 063030},
  \href{http://arxiv.org/abs/1911.05086}{{\ttfamily arXiv:1911.05086
  [hep-ph]}}.

\bibitem{Ehret:2010mh}
K.~Ehret {\em et~al.}, ``{New ALPS Results on Hidden-Sector Lightweights},''
  \href{http://dx.doi.org/10.1016/j.physletb.2010.04.066}{{\em Phys. Lett.}
  {\bfseries B689} (2010) 149--155},
\href{http://arxiv.org/abs/1004.1313}{{\ttfamily arXiv:1004.1313 [hep-ex]}}.

\bibitem{Betz:2013dza}
M.~Betz, F.~Caspers, M.~Gasior, M.~Thumm, and S.~Rieger, ``{First results of
  the CERN Resonant Weakly Interacting sub-eV Particle Search (CROWS)},''
  \href{http://dx.doi.org/10.1103/PhysRevD.88.075014}{{\em Phys. Rev. D}
  {\bfseries 88} no.~7, (2013) 075014},
  \href{http://arxiv.org/abs/1310.8098}{{\ttfamily arXiv:1310.8098
  [physics.ins-det]}}.

\bibitem{Redondo:2008aa}
J.~Redondo, ``{Helioscope Bounds on Hidden Sector Photons},''
  \href{http://dx.doi.org/10.1088/1475-7516/2008/07/008}{{\em JCAP} {\bfseries
  07} (2008) 008}, \href{http://arxiv.org/abs/0801.1527}{{\ttfamily
  arXiv:0801.1527 [hep-ph]}}.

\bibitem{An:2013yua}
H.~An, M.~Pospelov, and J.~Pradler, ``{Dark Matter Detectors as Dark Photon
  Helioscopes},'' \href{http://dx.doi.org/10.1103/PhysRevLett.111.041302}{{\em
  Phys. Rev. Lett.} {\bfseries 111} (2013) 041302},
  \href{http://arxiv.org/abs/1304.3461}{{\ttfamily arXiv:1304.3461 [hep-ph]}}.

\bibitem{Schwarz:2015lqa}
M.~Schwarz, E.-A. Knabbe, A.~Lindner, J.~Redondo, A.~Ringwald, M.~Schneide,
  J.~Susol, and G.~Wiedemann, ``{Results from the Solar Hidden Photon Search
  (SHIPS)},'' \href{http://dx.doi.org/10.1088/1475-7516/2015/08/011}{{\em JCAP}
  {\bfseries 08} (2015) 011}, \href{http://arxiv.org/abs/1502.04490}{{\ttfamily
  arXiv:1502.04490 [hep-ph]}}.

\bibitem{Danilov:2018bks}
M.~Danilov, S.~Demidov, and D.~Gorbunov, ``{Constraints on hidden photons
  produced in nuclear reactors},''
  \href{http://dx.doi.org/10.1103/PhysRevLett.122.041801}{{\em Phys. Rev.
  Lett.} {\bfseries 122} no.~4, (2019) 041801},
  \href{http://arxiv.org/abs/1804.10777}{{\ttfamily arXiv:1804.10777
  [hep-ph]}}.

\bibitem{Jaeckel:2010xx}
J.~Jaeckel and S.~Roy, ``{Spectroscopy as a test of Coulomb's law: A Probe of
  the hidden sector},''
  \href{http://dx.doi.org/10.1103/PhysRevD.82.125020}{{\em Phys. Rev.}
  {\bfseries D82} (2010) 125020},
\href{http://arxiv.org/abs/1008.3536}{{\ttfamily arXiv:1008.3536 [hep-ph]}}.

\bibitem{Redondo:2013lna}
J.~Redondo and G.~Raffelt, ``{Solar constraints on hidden photons
  re-visited},'' \href{http://dx.doi.org/10.1088/1475-7516/2013/08/034}{{\em
  JCAP} {\bfseries 1308} (2013) 034},
\href{http://arxiv.org/abs/1305.2920}{{\ttfamily arXiv:1305.2920 [hep-ph]}}.

\bibitem{An:2013yfc}
H.~An, M.~Pospelov, and J.~Pradler, ``{New stellar constraints on dark
  photons},'' \href{http://dx.doi.org/10.1016/j.physletb.2013.07.008}{{\em
  Phys. Lett.} {\bfseries B725} (2013) 190--195},
\href{http://arxiv.org/abs/1302.3884}{{\ttfamily arXiv:1302.3884 [hep-ph]}}.

\bibitem{An:2014twa}
H.~An, M.~Pospelov, J.~Pradler, and A.~Ritz, ``{Direct Detection Constraints on
  Dark Photon Dark Matter},''
  \href{http://dx.doi.org/10.1016/j.physletb.2015.06.018}{{\em Phys. Lett. B}
  {\bfseries 747} (2015) 331--338},
  \href{http://arxiv.org/abs/1412.8378}{{\ttfamily arXiv:1412.8378 [hep-ph]}}.

\bibitem{Angle:2011th}
{\bfseries XENON10} Collaboration, J.~Angle {\em et~al.}, ``{A search for light
  dark matter in XENON10 data},''
  \href{http://dx.doi.org/10.1103/PhysRevLett.107.051301}{{\em Phys.\ Rev.\
  Lett.} {\bfseries 107} (2011) 051301},
  \href{http://arxiv.org/abs/1104.3088}{{\ttfamily arXiv:1104.3088
  [astro-ph.CO]}}. [Erratum: Phys.Rev.Lett. 110, 249901 (2013)].

\bibitem{Aprile:2014eoa}
{\bfseries XENON100} Collaboration, E.~Aprile {\em et~al.}, ``{First Axion
  Results from the XENON100 Experiment},''
  \href{http://dx.doi.org/10.1103/PhysRevD.90.062009,
  10.1103/PhysRevD.95.029904}{{\em Phys. Rev.} {\bfseries D90} no.~6, (2014)
  062009}, \href{http://arxiv.org/abs/1404.1455}{{\ttfamily arXiv:1404.1455
  [astro-ph.CO]}}.
[Erratum: Phys. Rev.D95,no.2,029904(2017)].

\bibitem{Aprile:2019xxb}
{\bfseries XENON} Collaboration, E.~Aprile {\em et~al.}, ``{Light Dark Matter
  Search with Ionization Signals in XENON1T},''
  \href{http://dx.doi.org/10.1103/PhysRevLett.123.251801}{{\em Phys. Rev.
  Lett.} {\bfseries 123} no.~25, (2019) 251801},
\href{http://arxiv.org/abs/1907.11485}{{\ttfamily arXiv:1907.11485 [hep-ex]}}.

\bibitem{Aprile:2020tmw}
{\bfseries XENON} Collaboration, E.~Aprile {\em et~al.}, ``{Observation of
  Excess Electronic Recoil Events in XENON1T},''
  \href{http://arxiv.org/abs/2006.09721}{{\ttfamily arXiv:2006.09721
  [hep-ex]}}.

\bibitem{Aguilar-Arevalo:2016zop}
{\bfseries DAMIC} Collaboration, A.~Aguilar-Arevalo {\em et~al.}, ``{First
  Direct-Detection Constraints on eV-Scale Hidden-Photon Dark Matter with DAMIC
  at SNOLAB},'' \href{http://dx.doi.org/10.1103/PhysRevLett.118.141803}{{\em
  Phys. Rev. Lett.} {\bfseries 118} no.~14, (2017) 141803},
\href{http://arxiv.org/abs/1611.03066}{{\ttfamily arXiv:1611.03066
  [astro-ph.CO]}}.

\bibitem{Aralis:2019nfa}
{\bfseries SuperCDMS} Collaboration, T.~Aralis {\em et~al.}, ``{Constraints on
  Dark Photons and Axion-Like Particles from SuperCDMS Soudan},''
  \href{http://dx.doi.org/10.1103/PhysRevD.101.052008}{{\em Phys. Rev.}
  {\bfseries D101} no.~5, (2020) 052008},
\href{http://arxiv.org/abs/1911.11905}{{\ttfamily arXiv:1911.11905 [hep-ex]}}.

\bibitem{She:2019skm}
Z.~She {\em et~al.}, ``{Direct Detection Constraints on Dark Photons with
  CDEX-10 Experiment at the China Jinping Underground Laboratory},''
  \href{http://dx.doi.org/10.1103/PhysRevLett.124.111301}{{\em Phys. Rev.
  Lett.} {\bfseries 124} no.~11, (2020) 111301},
\href{http://arxiv.org/abs/1910.13234}{{\ttfamily arXiv:1910.13234 [hep-ex]}}.

\bibitem{Armengaud:2018cuy}
{\bfseries EDELWEISS} Collaboration, E.~Armengaud {\em et~al.}, ``{Searches for
  electron interactions induced by new physics in the EDELWEISS-III Germanium
  bolometers},'' \href{http://dx.doi.org/10.1103/PhysRevD.98.082004}{{\em Phys.
  Rev.} {\bfseries D98} no.~8, (2018) 082004},
\href{http://arxiv.org/abs/1808.02340}{{\ttfamily arXiv:1808.02340 [hep-ex]}}.

\bibitem{Abramoff:2019dfb}
{\bfseries SENSEI} Collaboration, O.~Abramoff {\em et~al.}, ``{SENSEI:
  Direct-Detection Constraints on Sub-GeV Dark Matter from a Shallow
  Underground Run Using a Prototype Skipper-CCD},''
  \href{http://dx.doi.org/10.1103/PhysRevLett.122.161801}{{\em Phys. Rev.
  Lett.} {\bfseries 122} no.~16, (2019) 161801},
\href{http://arxiv.org/abs/1901.10478}{{\ttfamily arXiv:1901.10478 [hep-ex]}}.

\bibitem{Abe:2018owy}
{\bfseries XMASS} Collaboration, K.~Abe {\em et~al.}, ``{Search for dark matter
  in the form of hidden photons and axion-like particles in the XMASS
  detector},'' \href{http://dx.doi.org/10.1016/j.physletb.2018.10.050}{{\em
  Phys. Lett.} {\bfseries B787} (2018) 153--158},
\href{http://arxiv.org/abs/1807.08516}{{\ttfamily arXiv:1807.08516
  [astro-ph.CO]}}.

\bibitem{Andrianavalomahefa:2020ucg}
{\bfseries FUNK Experiment} Collaboration, A.~Andrianavalomahefa {\em et~al.},
  ``{Limits from the Funk Experiment on the Mixing Strength of Hidden-Photon
  Dark Matter in the Visible and Near-Ultraviolet Wavelength Range},''
  \href{http://arxiv.org/abs/2003.13144}{{\ttfamily arXiv:2003.13144
  [astro-ph.CO]}}.

\bibitem{Bartlett:1970js}
D.~F. Bartlett, P.~E. Goldhagen, and E.~A. Phillips, ``{Experimental Test of
  Coulomb's Law},''
\href{http://dx.doi.org/10.1103/PhysRevD.2.483}{{\em Phys. Rev.} {\bfseries D2}
  (1970) 483--487}.

\bibitem{Deniz:2009mu}
{\bfseries TEXONO} Collaboration, M.~Deniz {\em et~al.}, ``{Measurement of
  Nu(e)-bar -Electron Scattering Cross-Section with a CsI(Tl) Scintillating
  Crystal Array at the Kuo-Sheng Nuclear Power Reactor},''
  \href{http://dx.doi.org/10.1103/PhysRevD.81.072001}{{\em Phys. Rev. D}
  {\bfseries 81} (2010) 072001},
  \href{http://arxiv.org/abs/0911.1597}{{\ttfamily arXiv:0911.1597 [hep-ex]}}.

\bibitem{Fixsen:1996nj}
D.~Fixsen, E.~Cheng, J.~Gales, J.~C. Mather, R.~Shafer, and E.~Wright, ``{The
  Cosmic Microwave Background spectrum from the full COBE FIRAS data set},''
  \href{http://dx.doi.org/10.1086/178173}{{\em Astrophys. J.} {\bfseries 473}
  (1996) 576}, \href{http://arxiv.org/abs/astro-ph/9605054}{{\ttfamily
  arXiv:astro-ph/9605054}}.

\bibitem{Mirizzi:2009iz}
A.~Mirizzi, J.~Redondo, and G.~Sigl, ``{Microwave Background Constraints on
  Mixing of Photons with Hidden Photons},''
  \href{http://dx.doi.org/10.1088/1475-7516/2009/03/026}{{\em JCAP} {\bfseries
  03} (2009) 026}, \href{http://arxiv.org/abs/0901.0014}{{\ttfamily
  arXiv:0901.0014 [hep-ph]}}.

\bibitem{Wadekar:2019xnf}
D.~Wadekar and G.~R. Farrar, ``{First direct astrophysical constraints on dark
  matter interactions with ordinary matter at very low velocities},''
  \href{http://arxiv.org/abs/1903.12190}{{\ttfamily arXiv:1903.12190
  [hep-ph]}}.

\bibitem{Bhoonah:2018gjb}
A.~Bhoonah, J.~Bramante, F.~Elahi, and S.~Schon, ``{Galactic Center gas clouds
  and novel bounds on ultralight dark photon, vector portal, strongly
  interacting, composite, and super-heavy dark matter},''
  \href{http://dx.doi.org/10.1103/PhysRevD.100.023001}{{\em Phys. Rev. D}
  {\bfseries 100} no.~2, (2019) 023001},
  \href{http://arxiv.org/abs/1812.10919}{{\ttfamily arXiv:1812.10919
  [hep-ph]}}.

\bibitem{Redondo:2008ec}
J.~Redondo and M.~Postma, ``{Massive hidden photons as lukewarm dark matter},''
  \href{http://dx.doi.org/10.1088/1475-7516/2009/02/005}{{\em JCAP} {\bfseries
  02} (2009) 005}, \href{http://arxiv.org/abs/0811.0326}{{\ttfamily
  arXiv:0811.0326 [hep-ph]}}.

\bibitem{Sikivie:1983ip}
P.~Sikivie, ``{Experimental Tests of the Invisible Axion},''
  \href{http://dx.doi.org/10.1103/PhysRevLett.51.1415}{{\em Phys. Rev. Lett.}
  {\bfseries 51} (1983) 1415--1417}. [Erratum: Phys.Rev.Lett. 52, 695 (1984)].

\bibitem{deNiverville:2011it}
P.~deNiverville, M.~Pospelov, and A.~Ritz, ``{Observing a light dark matter
  beam with neutrino experiments},''
  \href{http://dx.doi.org/10.1103/PhysRevD.84.075020}{{\em Phys.\ Rev.\ D}
  {\bfseries 84} (2011) 075020},
  \href{http://arxiv.org/abs/1107.4580}{{\ttfamily arXiv:1107.4580 [hep-ph]}}.

\bibitem{Aguilar-Arevalo:2018wea}
{\bfseries MiniBooNE DM} Collaboration, A.~Aguilar-Arevalo {\em et~al.},
  ``{Dark Matter Search in Nucleon, Pion, and Electron Channels from a Proton
  Beam Dump with MiniBooNE},''
  \href{http://dx.doi.org/10.1103/PhysRevD.98.112004}{{\em Phys.\ Rev.\ D}
  {\bfseries 98} no.~11, (2018) 112004},
  \href{http://arxiv.org/abs/1807.06137}{{\ttfamily arXiv:1807.06137
  [hep-ex]}}.

\bibitem{Angloher:2015ewa}
{\bfseries CRESST} Collaboration, G.~Angloher {\em et~al.}, ``{Results on light
  dark matter particles with a low-threshold CRESST-II detector},''
  \href{http://dx.doi.org/10.1140/epjc/s10052-016-3877-3}{{\em Eur.\ Phys.\ J.\
  C} {\bfseries 76} no.~1, (2016) 25},
  \href{http://arxiv.org/abs/1509.01515}{{\ttfamily arXiv:1509.01515
  [astro-ph.CO]}}.

\bibitem{Anelli:2015pba}
{\bfseries SHiP} Collaboration, M.~Anelli {\em et~al.}, ``{A facility to Search
  for Hidden Particles (SHiP) at the CERN SPS},''
  \href{http://arxiv.org/abs/1504.04956}{{\ttfamily arXiv:1504.04956
  [physics.ins-det]}}.

\bibitem{Battaglieri:2016ggd}
{\bfseries BDX} Collaboration, M.~Battaglieri {\em et~al.}, ``{Dark Matter
  Search in a Beam-Dump eXperiment (BDX) at Jefferson Lab},''
  \href{http://arxiv.org/abs/1607.01390}{{\ttfamily arXiv:1607.01390
  [hep-ex]}}.

\bibitem{Antonello:2015lea}
{\bfseries MicroBooNE, LAr1-ND, ICARUS-WA104} Collaboration, M.~Antonello {\em
  et~al.}, ``{A Proposal for a Three Detector Short-Baseline Neutrino
  Oscillation Program in the Fermilab Booster Neutrino Beam},''
  \href{http://arxiv.org/abs/1503.01520}{{\ttfamily arXiv:1503.01520
  [physics.ins-det]}}.

\bibitem{Battaglieri:2017aum}
M.~Battaglieri {\em et~al.}, ``{US Cosmic Visions: New Ideas in Dark Matter
  2017: Community Report},'' in {\em {U.S. Cosmic Visions: New Ideas in Dark
  Matter}}.
\newblock 7, 2017.
\newblock \href{http://arxiv.org/abs/1707.04591}{{\ttfamily arXiv:1707.04591
  [hep-ph]}}.

\bibitem{Agnese:2016cpb}
{\bfseries SuperCDMS} Collaboration, R.~Agnese {\em et~al.}, ``{Projected
  Sensitivity of the SuperCDMS SNOLAB experiment},''
  \href{http://dx.doi.org/10.1103/PhysRevD.95.082002}{{\em Phys. Rev. D}
  {\bfseries 95} no.~8, (2017) 082002},
  \href{http://arxiv.org/abs/1610.00006}{{\ttfamily arXiv:1610.00006
  [physics.ins-det]}}.

\bibitem{Ade:2015xua}
{\bfseries Planck} Collaboration, P.~Ade {\em et~al.}, ``{Planck 2015 results.
  XIII. Cosmological parameters},''
  \href{http://dx.doi.org/10.1051/0004-6361/201525830}{{\em Astron.\
  Astrophys.} {\bfseries 594} (2016) A13},
  \href{http://arxiv.org/abs/1502.01589}{{\ttfamily arXiv:1502.01589
  [astro-ph.CO]}}.

\bibitem{Caputo:2019xum}
A.~Caputo, A.~Esposito, E.~Geoffray, A.~D. Polosa, and S.~Sun, ``{Dark Matter,
  Dark Photon and Superfluid He-4 from Effective Field Theory},''
  \href{http://dx.doi.org/10.1016/j.physletb.2020.135258}{{\em Phys. Lett. B}
  {\bfseries 802} (2020) 135258},
  \href{http://arxiv.org/abs/1911.04511}{{\ttfamily arXiv:1911.04511
  [hep-ph]}}.

\bibitem{Tiffenberg:2017aac}
{\bfseries SENSEI} Collaboration, J.~Tiffenberg, M.~Sofo-Haro,
  A.~Drlica-Wagner, R.~Essig, Y.~Guardincerri, S.~Holland, T.~Volansky, and
  T.-T. Yu, ``{Single-electron and single-photon sensitivity with a silicon
  Skipper CCD},'' \href{http://dx.doi.org/10.1103/PhysRevLett.119.131802}{{\em
  Phys.\ Rev.\ Lett.} {\bfseries 119} no.~13, (2017) 131802},
  \href{http://arxiv.org/abs/1706.00028}{{\ttfamily arXiv:1706.00028
  [physics.ins-det]}}.

\bibitem{Barak:2020fql}
L.~Barak {\em et~al.}, ``{SENSEI: Direct-Detection Results on sub-GeV Dark
  Matter from a New Skipper-CCD},''
  \href{http://arxiv.org/abs/2004.11378}{{\ttfamily arXiv:2004.11378
  [astro-ph.CO]}}.

\bibitem{Peccei:1977hh}
R.~Peccei and H.~R. Quinn, ``{CP Conservation in the Presence of Instantons},''
  \href{http://dx.doi.org/10.1103/PhysRevLett.38.1440}{{\em Phys. Rev. Lett.}
  {\bfseries 38} (1977) 1440--1443}.

\bibitem{Ellis:1988er}
J.~R. Ellis, J.~Gunion, H.~E. Haber, L.~Roszkowski, and F.~Zwirner, ``{Higgs
  Bosons in a Nonminimal Supersymmetric Model},''
  \href{http://dx.doi.org/10.1103/PhysRevD.39.844}{{\em Phys. Rev. D}
  {\bfseries 39} (1989) 844}.

\bibitem{Derendinger:1983bz}
J.~Derendinger and C.~A. Savoy, ``{Quantum Effects and SU(2) x U(1) Breaking in
  Supergravity Gauge Theories},''
  \href{http://dx.doi.org/10.1016/0550-3213(84)90162-7}{{\em Nucl. Phys. B}
  {\bfseries 237} (1984) 307--328}.

\bibitem{Frere:1983ag}
J.~Frere, D.~Jones, and S.~Raby, ``{Fermion Masses and Induction of the Weak
  Scale by Supergravity},''
  \href{http://dx.doi.org/10.1016/0550-3213(83)90606-5}{{\em Nucl. Phys. B}
  {\bfseries 222} (1983) 11--19}.

\bibitem{Silveira:1985rk}
V.~Silveira and A.~Zee, ``{SCALAR PHANTOMS},''
  \href{http://dx.doi.org/10.1016/0370-2693(85)90624-0}{{\em Phys. Lett. B}
  {\bfseries 161} (1985) 136--140}.

\bibitem{Binoth:1996au}
T.~Binoth and J.~van~der Bij, ``{Influence of strongly coupled, hidden scalars
  on Higgs signals},'' \href{http://dx.doi.org/10.1007/s002880050442}{{\em Z.
  Phys. C} {\bfseries 75} (1997) 17--25},
  \href{http://arxiv.org/abs/hep-ph/9608245}{{\ttfamily arXiv:hep-ph/9608245}}.

\bibitem{Patt:2006fw}
B.~Patt and F.~Wilczek, ``{Higgs-field portal into hidden sectors},''
  \href{http://arxiv.org/abs/hep-ph/0605188}{{\ttfamily arXiv:hep-ph/0605188}}.

\bibitem{Minkowski:1977sc}
P.~Minkowski, ``{$\mu \to e\gamma$ at a Rate of One Out of $10^{9}$ Muon
  Decays?},'' \href{http://dx.doi.org/10.1016/0370-2693(77)90435-X}{{\em Phys.
  Lett. B} {\bfseries 67} (1977) 421--428}.

\bibitem{Yanagida:1980xy}
T.~Yanagida, ``{Horizontal Symmetry and Masses of Neutrinos},''
  \href{http://dx.doi.org/10.1143/PTP.64.1103}{{\em Prog. Theor. Phys.}
  {\bfseries 64} (1980) 1103}.

\bibitem{GellMann:1980vs}
M.~Gell-Mann, P.~Ramond, and R.~Slansky, ``{Complex Spinors and Unified
  Theories},'' {\em Conf. Proc. C} {\bfseries 790927} (1979) 315--321,
  \href{http://arxiv.org/abs/1306.4669}{{\ttfamily arXiv:1306.4669 [hep-th]}}.

\bibitem{Mohapatra:1980yp}
R.~N. Mohapatra and G.~Senjanovic, ``{Neutrino Masses and Mixings in Gauge
  Models with Spontaneous Parity Violation},''
  \href{http://dx.doi.org/10.1103/PhysRevD.23.165}{{\em Phys. Rev. D}
  {\bfseries 23} (1981) 165}.

\bibitem{Pati:1974yy}
J.~C. Pati and A.~Salam, ``{Lepton Number as the Fourth Color},''
  \href{http://dx.doi.org/10.1103/PhysRevD.10.275}{{\em Phys. Rev. D}
  {\bfseries 10} (1974) 275--289}. [Erratum: Phys.Rev.D 11, 703--703 (1975)].

\bibitem{Mohapatra:1974gc}
R.~Mohapatra and J.~C. Pati, ``{A Natural Left-Right Symmetry},''
  \href{http://dx.doi.org/10.1103/PhysRevD.11.2558}{{\em Phys. Rev. D}
  {\bfseries 11} (1975) 2558}.

\bibitem{Mohapatra:1974hk}
R.~N. Mohapatra and J.~C. Pati, ``{Left-Right Gauge Symmetry and an
  Isoconjugate Model of CP Violation},''
  \href{http://dx.doi.org/10.1103/PhysRevD.11.566}{{\em Phys. Rev. D}
  {\bfseries 11} (1975) 566--571}.

\bibitem{Senjanovic:1975rk}
G.~Senjanovic and R.~N. Mohapatra, ``{Exact Left-Right Symmetry and Spontaneous
  Violation of Parity},''
  \href{http://dx.doi.org/10.1103/PhysRevD.12.1502}{{\em Phys. Rev. D}
  {\bfseries 12} (1975) 1502}.

\bibitem{Plehn:2017fdg}
M.~Bauer and T.~Plehn, ``{Yet Another Introduction to Dark Matter},''
  \href{http://dx.doi.org/10.1007/978-3-030-16234-4}{{\em Lect. Notes Phys.}
  {\bfseries 959} (2019) pp.--},
\href{http://arxiv.org/abs/1705.01987}{{\ttfamily arXiv:1705.01987 [hep-ph]}}.

\bibitem{Braaten:1993jw}
E.~Braaten and D.~Segel, ``{Neutrino energy loss from the plasma process at all
  temperatures and densities},''
  \href{http://dx.doi.org/10.1103/PhysRevD.48.1478}{{\em Phys.\ Rev.\ D}
  {\bfseries 48} (1993) 1478--1491},
  \href{http://arxiv.org/abs/hep-ph/9302213}{{\ttfamily arXiv:hep-ph/9302213}}.

\end{thebibliography}\endgroup
}
%
%
 \end{document}